\documentclass{aa}  

\usepackage{graphicx}
\usepackage{txfonts}

\usepackage{hyperref}
\hypersetup{
     colorlinks = true,
     linkcolor = blue,
     anchorcolor = blue,
     citecolor = blue,
     filecolor = blue,
     urlcolor = blue
}

\usepackage{bm}
\usepackage{caption}
\usepackage{subcaption}
\usepackage{bigstrut}
\usepackage{xspace}

\usepackage{todonotes}

\mathchardef\mhyphen="2D  

\begin{document}

   \title{Shaping the CO snowline in protoplanetary disks}

\titlerunning{CO snowline in protoplanetary disks}

 \author {
       S. Gavino \inst{1}
       J. Kobus \inst{2}
       A. Dutrey \inst{3}
       S. Guilloteau \inst{3}
       S. Wolf \inst{2}
       J.K. Jørgensen \inst{1}
       R. Sharma \inst{1}
	 }      
 \institute{
 	Niels Bohr Institute and Centre for Star and Planet Formation, University of Copenhagen, Øster Voldgade 5-7, 1350, Copenhagen K, Denmark
		\and University of Kiel, Institute of Theoretical Physics and Astrophysics, Leibnizstrasse 15, 24118 Kiel, Germany 
  		\and Laboratoire d'Astrophysique de Bordeaux, Universit\'e de Bordeaux, CNRS, B18N, 
			  All\'ee Geoffroy Saint-Hilaire, F-33615 Pessac
}

\authorrunning{Gavino, S.}

\offprints{ Sacha Gavino\\
\email{sacha.gavino@nbi.ku.dk}}

   \date{Received September 15, 1996; accepted March 16, 1997}

\abstract
{Characterizing the dust thermal structure in protoplanetary disks is a fundamental task as the dust surface temperature can affect both the planetary formation and the chemical evolution. Since the temperature is dependent on many parameters, including the grain size, properly modeling the grain temperature structure can be challenging. Many chemistry disk models usually employ a sophisticated single dust structure designed to reproduce the effect of a realistic population presumably composed of a large diversity of sizes. This generally represents a good approximation in most cases. Nonetheless, this dilutes the effects of the complex radiative interactions between the different grain populations on the resulting dust temperature, and thus the chemistry.}
{We seek to show that the radiative interactions between dust grains of different sizes can induce a non-trivial dust temperature structure that cannot be reproduced by a single dust population and that can significantly affect the chemical outcome.}
{The disk thermal structures are computed using the Monte-Carlo radiative transfer code RADMC-3D. The thermal structures are post-processed using the gas-grain code NAUTILUS to calculate the evolution of the chemical abundance.}
{We find that simultaneously using at least two independent dust grain populations in disk models produces a complex temperature structure due to the starlight intercepted by the upper layers of the disk. In particular, we find that micron-sized dust grains are warmer than larger grains and can even show a radial temperature bump in some conditions. This dust temperature spread between the grains populations results in the segregation of the CO snowline and the presence of an unexpected CO gas hole along the midplane. We compare the results with observed close to edge-on class I/II disks.}
{Our study shows that the size-dependence of the dust temperature significantly impacts the chemistry, and that a minimum of two dust populations is required to account for this property of the thermal structure in protoplanetary disk models, over a wide range of disk masses and dust properties.}
 
 



   \maketitle

\section{Introduction}
\label{sec:intro}
Numerous studies have now confirmed the existence of grain growth in circumstellar disks, with grains of size up to millimeter detected \citep[e.g.,][]{Testi+etal_2014, Ohashi+etal_2023}. Some authors reported the presence of grain growth at a very early stage. For instance, \citet{Harsono+etal_2018} confirmed the presence of millimeter-sized grains in the very young disk ($\sim$ 100,000 years) around the protostar TMC1A. These observations are of prior importance as they suggest that planet formation can initiate at a very early stage of protostellar evolution.  

More specifically, it is known that the grain size distribution in protoplanetary disks is different from the one in the interstellar medium (ISM). Many authors have shown that the distribution can deviate from the typical \citet*{Mathis+Rumpl+Nordsieck_1977} model (hereafter MRN) found in the ISM \citep{Ricc+etal_2010}. In particular, the grain size range in disks is less narrow than in the ISM and is likely to be continuously distributed due to dust coagulation, fragmentation and radial drift \citep[e.g.,][]{Dullemond+Dominik_2005, Brauer+etal_2008, Birnstiel+etal_2010, Birnstiel+etal_2018}, resulting in a typical range from nanometer sizes to centimeter sizes. The spatial distribution is also more complex. Grains with a small Stokes number are coupled to the disk gas (small Stokes numbers typically apply to small grains, and vice versa) and can be observed in near-infrared at very large vertical extent whereas grains with a larger Stokes number are more efficiently settled toward the disk midplane and occupy a thinner vertical extent \citep{Fromang+etal_2009}. Moreover, inward radial drift can occur for coagulated grains \citep[e.g.,][]{Birnstiel+etal_2010, Lambrechts+etal_2014}, strongly reshuffling the local dust density and size distributions.

Consequently, the extinction opacity in disks is also much more difficult to characterize than in the ISM. Most authors assume size-averaged opacities from a given local distribution \citep[e.g.,][]{Pinte+etal_2016, Birnstiel+etal_2018}. The opacities are commonly averaged over a typical dust size distribution from the MRN model, with the maximum grain size chosen to align with multi-wavelength observations. More precisely, most recent studies consider a dust model composed of two-grain populations, a small dust population and a large dust population \citep[e.g.,][]{Du+Bergen_2014, Ballering+etal_2021, Schwarz+etal_2021, Zhang+etal_2021, Murillo+etal_2022} meant to fit observationally constrained evidence of grain growth and size-dependence to vertical and radial segregation \citep{Grafe+etal_2013, Perez+etal_2015, Tazzari+etal_2016, Villenave+etal_2020}. Typically, each population has its opacity averaged over a given size range, each with a different maximum grain size $a_\mathrm{max}$, the latter being a major parameter impacting the wavelength-dependent optical properties \citep{Birnstiel+etal_2018}. A single resulting temperature is then extracted (density and temperature), either by averaging-out the two dust structures or by forcing dust thermal coupling during the radiative transfer simulations. The structure is later used for chemistry post-processes. However, since surface chemistry is strongly sensitive to dust surface temperatures, this treatment may overlook possible chemical effects by not considering the two dust structures independently.

Indeed, we saw that real disks are most likely composed of grains of multiple sizes with independent optical properties, resulting in polydispersed dust grains \citep{Birnstiel+etal_2018}. Note that the term 'polydispersion' (as opposed to monodispersion) in the present study is used in the strict sense as defined in \citet{Deirmendjian_1969} and \citet{Bohren+Huffman_1998}, that is we speak of polydispersion when a population of scattering particles in a medium is uniform in shape and bulk material but not uniform in size. Grains are heated both by the external field and by the resulting scattered and re-emitted light from all the other grains. The absorption efficiency of grains is dependent on grain size and wavelengths which results in a) a complex interaction between the local radiation field and b) a size-dependent dust temperature structure \citep{Heese+etal_2017, Gavino+etal_2021}. It is therefore expected that utmost attention should be given to the selection of the dust population, dust absorption and scattering profiles in disk models. 

In this work, we reproduce similar dust models as in recent studies \citep[e.g.,][]{Zhang+etal_2021}. However, we keep the dust structures independent so we can analyze (i) the precise dust temperature structures resulting from polydispersed dust grains computed with dust continuum radiative transfer simulations and (ii) the resulting chemical structure by treating the various dust populations simultaneously and independently in chemistry post-processes. We compare the results with test models composed of a single dust population meant to depict the classical treatment mentioned above. In particular, we will investigate the effect of polydispersion in the regions around the disk midplane and observe the impact on the CO snowline. 

The paper is structured as follows: Section~\ref{sec:model} describes the disk models used in the present study. Section~\ref{sec:results} presents the resulting thermal structures and chemical post-process and, finally, in Sect.~ \ref{sec:discussion} we discuss in more details the origin of the temperature substructures and its potential implications on the CO gas distribution in observed disks. A summary is then provided in Sect.~\ref{sec:summary}.

\section{Model}\label{sec:model}
In the present study, the models are assumed to be static protoplanetary disks, consisting of a typical 2D parametric smooth, axisymmetric and geometrically flared structure. Additionally, only passive heating is assumed. The disk is assumed to be in Keplerian rotation. We provide hereafter a concise description of the parametrization but the model is similar to \citet{Gavino+etal_2021}. 


\subsection{Fiducial disk structure}\label{sec:model:disk}
\subsubsection{Spatial distribution of gas}\label{sec:model:gasdistrib} 

The radial gas surface density follows a simple truncated power-law,

\begin{equation}\label{eq:surdensg}
\Sigma_{g}  = \Sigma_{g,0} \bigg(\frac{r}{R_0}\bigg)^{-p},
\end{equation}	

\noindent where $p=3/2$ \citep{Shakura_Sunyaev_1973, Hersant+etal_2009, Guilloteau+etal_2011, LeGal+etal_2019} and $R_\mathrm{0}$ is the reference radius chosen to be 100 au in the study. 
 
The surface density at the given reference radius $R_\mathrm{0}$ is derived using 
 
\begin{equation}\label{eq:surdens0}
\Sigma_{g,0}  = \frac{M_\mathrm{disk} R_\mathrm{0}^{-(3/2)}}{4 \pi \left( R_\mathrm{out}^{1/2} - R_\mathrm{in}^{1/2} \right)},
\end{equation}	

\noindent where $M_\mathrm{disk}$ is the total disk mass and its inner radius  and outer radius are $R_\mathrm{in}$ and $R_\mathrm{out}$, respectively. 

Following \citet{Dartois+etal_2003}, the gas temperature is imposed by analytical laws.
The gas kinetic temperature in the mid-plane $T_{mid}$ is given by a power law 

\begin{equation}
\label{eq:Tmiddef}
    T_\mathrm{mid}(r) = T_\mathrm{mid,0} \bigg(\frac{r}{R_0}\bigg)^{-q}
\end{equation}

\noindent where $T_\mathrm{mid,0}$ is the gas temperature in the midplane at $R_0$, and we allow for a warmer disk atmosphere using the formulation of \citet{Williams+Best_2014}
\begin{equation}
\label{eq:verticalT}
       T_\mathrm{g}(r, z) = T_\mathrm{mid}(r) + (T_\mathrm{atm}(r) - T_\mathrm{mid}(r)) \sin\bigg({\frac{\pi z}{2 z_\mathrm{atm}}}\bigg)^{2\sigma}
\end{equation}

\noindent where $\sigma = 2$ is a stiffness parameter of the vertical profile. Above $z_\mathrm{atm}$, the temperature is vertically constant. $T_\mathrm{atm}(r)$ follows a power law similar to $T_\mathrm{mid}(r)$ and $z_\mathrm{atm}$ is taken as being 4 times the mid-plane gas scale height, $H_g$,
which at the the reference radius $R_0$ is given by, 

\begin{equation}\label{eq:H_ref}
H_{g,0}  = \sqrt{\frac{k_B T_{mid,0} R_0^3}{\mu_m m_H G M_\star}}.
\end{equation}

\noindent where $\mu$ is the mean molecular weight equals to 2.4, $m_\mathrm{H}$ is the proton mass, $k_\mathrm{B}$ the Boltzmann constant, and $G$ the gravitational constant. The mid-plane gas scale height then follows a radial power-law varying as $r^{3/2 - q/2}$, a typical flared geometry. 

Our assumed gas temperature and density profiles were selected for comparison with previous works and differ from those of a steady-state viscous disk with radially constant viscosity, which would have a $r^{-1.1}$ dependence. Note, however, that a self-consistent model would have a more
complex radial density profile even under the simplified viscosity assumption because of the regions of increased temperatures that result from the treatment of different dust populations.

We solve the gas density by considering the (non-isothermal) vertical hydrostatic pressure equilibrium, that we solve iteratively following
\citet{Hersant+etal_2009},

\begin{equation}
\label{eq:not-iso}
ln(\rho_g(z_i)) = ln(\rho_g(z_i-1)) - (\Omega^2\frac{\mu m_H}{k_BT_g(z_i)}+(ln(T_g(z_i))-ln(T_g(z_{i-i})))
\end{equation} 

\noindent where $\rho_\mathrm{g(z_i)}$ is the gas mass density at altitude $z_i$, $\Omega$ the Keplerian rotation at the given radius. 
The vertical gas density structure thus slightly deviates from a Gaussian profile at altitudes $z > 2-3$ H$_\mathrm{g}$.

Although this imposed density and gas temperature structure are not exactly self-consistent, we
verified that the gas temperature $T_\mathrm{g}(r,z)$ is not too different from the
area-weighted dust temperature (calculated from Monte-Carlo simulations as described in Sect.\,\ref{sec:model:rad}),  
\begin{equation}
\label{eq:ta}
        T_{a}(r, z) =  \frac{ \sum_j a_j^2 T_d(a_j, r, z) n_d(a_j, r, z) }{ \sum_j a_j^2 n_d(a_j, r, z) }
\end{equation}
as would be expected when the gas is mostly heated by thermal contact with the dust grains, rather than by the radiation field. Here, $a_j$ represents the grain size of bin $j$ and $n_d$ is the dust number density (details are given in the next section). Note that the chemical models can be computed using the gas temperature as given by Eq.\,\ref{eq:verticalT} or derived
from the area-weighted dust temperature (Eq.\,\ref{eq:ta}). We systematically verified that both cases lead to very similar abundances for the 
most common molecules, as expected since the rates of the main gas phase chemical reactions are not strong functions of the temperature\footnote{Differences could however appear for molecules that can only be formed through schemes involving reactions with an activation barrier between the two temperatures.}. We thus only present chemistry results from the model using the gas temperature as given by Eq.\,\ref{eq:verticalT},
the other choice resulting in practically identical figures.

\subsubsection{Spatial distribution of dust}\label{sec:model:dustdistrib} 
The vertical density distribution of any dust component is defined by the following equation:

\begin{equation}
\label{eq:rhod}
\rho_d(r,z, a) =  \frac{\sigma_d(r, a)}{\sqrt{2\pi}H_d(r, a)} \exp\left(- \frac{z^2}{2\,H_{d}(r, a)^2}\right).
\end{equation}
\noindent where $\sigma_\mathrm{d(r, a)}$ and $H_\mathrm{d(r, a)}$ are the surface density and scale height, respectively, of the grain population of size $a$ at radius $r$. The dust scale height is defined following \citep[see][]{Dubrulle+etal_1995, Youdin+Lithwick_2007, Fromang+etal_2009, Dong+etal_2015},  
\begin{equation}
\label{eq:settling-dong} 
	H_{d}(a,r) = H_g(r) \frac{1}{\sqrt{1 + \mathrm{St} \frac{S_c}{\alpha}}} 
\end{equation}
\noindent so that the dust scale height represents a fraction of the gas scale height, depending on the dimensionless stopping time, or Stokes number, $\mathrm{St}$. $S_\mathrm{c}$ is the Schmidt number and $\alpha$ the turbulent viscosity coefficient, set to 1 and 0.01, respectively, in the whole study. The coefficient $\alpha$ sets the degree of turbulent viscosity and therefore the degree of settling \citep{Shakura_Sunyaev_1973}. The settling factor depends on the dust grain size $a$ via the Stokes number, which can be defined near the midplane as 

\begin{equation}
\label{eq:setfact} 
	\mathrm{St} =  \frac{a \rho_m}{\Sigma_g} \frac{\pi}{2}
\end{equation}

\noindent where $\rho_m$ is the dust material density and under the assumption of Epstein regime and spherical particles \citep[see][]{Cuzzi+etal_2001, Birnstiel+etal_2012}.

The dust-to-gas mass ratio varies from an altitude to another due to the settling, but the dust-to-gas surface density ratio is $\zeta$ at all radii. The model does not include radial drift.

We point out here that dust settling is not expected to result in a simple Gaussian distribution. For instance, \citet{Fromang+etal_2009}'s simulations and analytical approach (their Eq.\,19) show that in a vertically isothermal disk, the dust settling results in a dust mass distribution that is approximately Gaussian near the mid-plane ($z < 2H$), but falls more steeply than a Gaussian at high heights. Nevertheless, Eq.\,19 of \citet{Fromang+etal_2009} is only valid for a Gaussian gas density vertical distribution. In our case, the increased temperature at high heights above $z/H > 2$, where the dust profile starts to be steeper, leads to a profile that has broader wings than a Gaussian (our Eq.\,\ref{eq:not-iso}). The two effects partially cancel and the settled dust distribution is thus quite likely better approximated by a simple Gaussian than in the vertically isothermal case. In order to make sure that the extra term in Eq.\,19 of \citet{Fromang+etal_2009} does not cause major differences, we performed additional simulations using their approach. These tests have showed very minor changes, we therefore only present the results using our Eq.\,\ref{eq:rhod}.

\subsubsection{Dust mass and size distribution}\label{sec:model:dustpop}
In all the paper, the size distributions follow the MRN distribution with a power-law $dn(a) \propto a^{-d}$ and with $d$ = 3.5. It is possible to divide the distribution into $N_\mathrm{d}$ intervals (logarithmically distributed in all the study). From this, a mass and a size can be defined for each grain bin $j$. Following \citet{Gavino+etal_2021}, the effective discretized size value of the $j$th interval is given by the mass-weighted average,

\begin{equation}
\label{eq:size_av} 
	a_j = \bigg(\frac{1-d}{4-d} \cdot \frac{a_{+,j}^{4-d} - a_{-,j}^{4-d}}{a_{+,j}^{1-d} - a_{-,j}^{1-d}}\bigg)^{\frac{1}{3}}.
\end{equation}

\noindent where $a_{+,j}$ and $a_{-,j}$ are the maximum and minimum cutoff values of the $j$th interval, respectively.

The total dust mass in the disk of the population $j$ of size $a_j$ is a fraction $x(a_j)$ of the total disk dust mass $M_\mathrm{dust}$:

\begin{equation}
\label{eq:dustmass} 
	m_d(a_j) = x(a_j) M_{dust} = x(a_j) \zeta M_{disk}
\end{equation}

\noindent with $\zeta$ the dust-to-gas mass ratio and $M_\mathrm{disk}$ the total disk mass (gas+dust). The term $x(a_j)$ derives from the MRN distribution and is a fraction of the full size distribution 

\begin{equation}
\label{eq:frac} 
	x(a_j)= \frac{\int_{a_{-,j}}^{a_{+,j}} m(a)dn(a)}{\int_{a_{min}}^{a_{max}}m(a)dn(a)}. 
\end{equation}

\noindent where $a_{min}$ and $a_{max}$ are the minimum and maximum cutoff values, respectively, of the full size distribution and such that 

\begin{equation}
\label{eq:sumfrac} 
	\sum_{j=1}^{N_d} x(a_j) = 1.
\end{equation}

The fiducial physical parameters used in all disk models are summarized in Table~\ref{tab:param}. We note that these parameters are chosen to provide an interpretative framework for the resulting thermal structure and are not intended to directly compare with observations. However, the models are chosen to be similar to a typical low-mass T Tauri disk.

\subsection{Radiation source and radiative transfer modeling}\label{sec:model:rad}
For the stellar radiation we consider a pre-main sequence star emitting as a blackbody with an effective temperature of $T_\mathrm{star}$ = 4100 K, a stellar luminosity of $L_\mathrm{star}$ = 1 $L_\odot$, and a stellar mass of $M_\mathrm{star}$ = 1 $M_\odot$. As for the interstellar radiation field (ISRF), we consider a typical distribution from \citet{Draine_1978} with an extension of \citet{Dishoeck+Black_1982} for wavelengths > 200 nm. We use these parameters for all models described in this study. The stellar irradiation will be the dominant source of thermal heating for the dust and gas. On the other hand, the photochemistry of the disk surface will be mainly controlled by the FUV spectra of both the ISRF and stellar radiation field. 

To compute the dust temperature we perform radiative transfer calculations using the RADMC-3D code \citep{Dullemond+etal_2012}. Each dust population has its own density distribution and are allowed to be thermally decoupled from each other. 
The radiative transfer calculations include the absorption opacity, scattering opacity, and the Henyey-Greenstein $g$ parameter for the treatment of anisotropic scattering.

The structure is a spherical grid centered at the star's location, with 300 logarithmically distributed radii, and 180 angles from 0 to $\pi$. We use 100 logarithmically distributed wavelengths from 0.1 $\mu$m to 2 mm for the radiative transfer calculation. We use 10$^7$ photon packages in each simulation.

\begin{table}
\centering
\caption{Fiducial disk parameters \label{tab:param}}
\begin{tabular}{p{0.68\linewidth}r}
\hline
\noalign{\smallskip}
\multicolumn{2}{c}{Disk Parameters}\\
\noalign{\smallskip}
\hline
 \noalign{\smallskip}
\bm{$T_\mathrm{\star,eff}$} (star temperature)            	& 4100 K \\
\bm{$L_\star$} (star luminosity)       & 1 $\mathrm{L_{\odot}}$  \\
\bm{$M_\star$} (star mass)        & 1 $\mathrm{M_{\odot}}$ \\
\bm{$\mathrm{M_{disk}}$} (disk mass (dust+gas)) & 7.5.10$^{-3}$ $\mathrm{M_{\odot}}$\\
\bm{$R_\mathrm{in}$} (innermost radius)  & 1 $\mathrm{au}$ \\
\bm{$R_\mathrm{out}$} (outermost radius) & 500 $\mathrm{au}$ \\
\bm{$R_0$} (reference radius)& 100 $\mathrm{au}$  \\
\bm{$\zeta$} (dust-to-gas mass ratio) & 0.01 \\
\bm{$\rho_\mathrm{m}$} (dust material density) & 1.675 g.cm$^{-3}$ \\
\bm{$\alpha$} (turbulence coefficient) & 10$^{-2}$ \\
\bm{$T_\mathrm{mid,0}$} (chosen gas temperature at $R_0$) & 25 $\mathrm{K}$\\
\bm{$q$} (radial temperature profile exponent) & 0.4 \\
\bm{$p$} (surface density exponent) & 1.5

\end{tabular}
\end{table}

\subsection{Chemical network and gas-grain simulation}\label{sec:model:chem}
We use the three-phase Nautilus Multi-Grain Code (NMGC) \citep{Ruaud+etal_2016, Iqbal+Wakelam_2018, Gavino+etal_2021}, derived from the NAUTILUS gas-grain code \citep{Hersant+etal_2009}. NMGC is specifically designed for chemistry simulations of models embedding multiple, independent grain populations. All dust populations are chemically active and all chemical compounds can interact with each population independently.
The chemistry model is a 1D+1 structure spanning from 4 au to 360 au from the star. All 1D structures are composed of 64 vertical spatial points from the disk midplane up to 4 H$_\mathrm{g}$, defined as the surface of the disk. 

NMGC uses the rate equation approach \citep{Hasegawa+etal_1992, Hasegawa+Herbst_1993b} in which the gas phase, the grain surfaces, and the mantles are chemically active \citep{Ruaud+etal_2016}. The gas phase can exchange species 
with the surfaces but not directly with the mantle. For surface chemistry, adsorption, thermal desorption, cosmic-ray induced desorption, photodesorption and chemical desorption, swapping from the 
mantle to the surface and vice-versa, are included. Physisorbed species can diffuse on the grain surface 
via tunneling or thermal hopping. For the gas phase, chemistry is driven by bi-molecular, ion-neutral, 
and neutral-neutral reactions as well as ionization and dissociation triggered by cosmic-ray-induced 
processes, and both interstellar and stellar radiation fields. 

Following the prescription of \citet{Gavino+etal_2021}, the H$_2$ formation rates are calculated by consistently interpolating rate curves which are initially solved by using a 
stochastic treatment that considers fluctuations of hydrogen atoms on the grain surfaces by \citet{Bron+etal_2014}. 

We use the KInetic Database for Astrochemistry network (\textit{kida.uva.2014})\footnote{\url{https://kida.astrochem-tools.org/networks.html}} \citep{Wakelam+etal_2015}, 
to which we include the additional reaction rates for the H$_2$ formation on grain surfaces. The chemical evolution is simulated over a period of $t_\mathrm{chem}$ = 2 Myrs, well above the CO freeze-out timescale in the whole disk. Considering the relatively large simulation run-time, 
we use atomic initial conditions (except for H$_2$) as presented in Table~\ref{tab:init_ab}. 
See \citet{Wakelam+etal_2019} for more information on the impact of initial conditions on the chemical evolution.

\begin{table}
\caption{Adopted elemental initial abundances relative to H. \label{tab:init_ab}}
\centering
\begin{tabular}{c c c}
\hline
\noalign{\smallskip}
Element & abundance (relative to H) & mass (amu) \\
\noalign{\smallskip}
\hline
\noalign{\smallskip}	
H$_2$  & 0.5 & 4.00	\\
He & 9.0(-2)  & 4.00	\\
C &	 1.7(-4) & 12.00	\\
N &	 6.2(-5) & 14.00\\
O &	 2.4(-4) & 16.00 \\
Si &	 8.0(-9) & 28.00	\\
S &	 8.0(-8) & 32.00	\\
Fe &	 3.0(-9) & 56.00	\\
Na &	 2.0(-9) & 23.00	\\
Mg &	  7.0(-9) & 24.00	\\
Cl &	 1.0(-9) & 35.00	\\
P &	 2.0(-10) & 31.00	\\
F &	 6.7(-9) & 19.00	\\

\end{tabular}
\end{table}

For the species other than H$_2$ and CO, the photoprocess rates are calculated using the following equation: 

\begin{equation}
\label{eq:photrate}
k = G_0 \alpha e^{-\gamma A_v}
\end{equation} 

\noindent Where G$_0$ is the unattenuated local UV flux at each radius relative to the standard ISM (Draine) value. Note that the $\gamma$ value are taken as the default values appropriate for the ISM and that for many species $\gamma = 1$ in the Kida network.

The $A_v$ is computed using the local UV flux calculated from the Monte-Carlo radiative transfer simulation.

For a given wavelength $\lambda$ and given coordinates ($r$, $z$), the local flux is the unattenuated flux $F_{0}$ reduced by the dust extinction:

\begin{equation}
\label{eq:reduction}
F_{local}(\lambda, r, z) = F_{0}(\lambda, r, 4\mathrm{H}) exp[-\tau_\lambda^{ext}].
\end{equation} 

\noindent where the opacity is related to the extinction by the usual relation 

\begin{equation}
\label{eq:av}
A_\lambda = 1.086 \tau_\lambda.
\end{equation} 

\subsection{Dust models and optical properties}\label{sec:model:dust}
To illustrate the importance of polydispersion,
we consider four dust models. Model S (for Single) is composed of a simple single dust distribution. Model C and D are models using two dust distributions. In Model D (for Dual), each dust population has its own temperature, 
while for Model C (for Common), an averaged temperature is derived from the two temperatures of Model D following Eq.~\ref{eq:ta}. The difference between Model C and D therefore lies entirely in the chemical computation, as we will see. Lastly, Model M16 (for Multi) is composed of 16 dust grain size bins. The four models are described below, while the thermal and chemical results are shown in Sects.\,\ref{sec:res:thermal} and \ref{sec:res:chemistry}, respectively. A summary of the four models is given in Table~\ref{tab:dustparam}. More specifically, the number of grain size bins used during radiative transfer and chemistry simulations as well as the effective grain size values are indicated in columns 2, 3, and 4, respectively.

\subsubsection{Model S, D, and C}\label{sec:model:SDC}
We adopt the opacities from the DSHARP collaboration by using the standard size-averaged smoothed optical properties of the \textit{dsharp\_opac} package\footnote{\url{https://github.com/birnstiel/dsharp_opac}} as described in \citet{Birnstiel+etal_2018} with the mixed mie coefficients used for the DSHARP project. 

\begin{figure*}
\begin{subfigure}{.33\linewidth}
  \centering
  \includegraphics[width=1.0\linewidth]{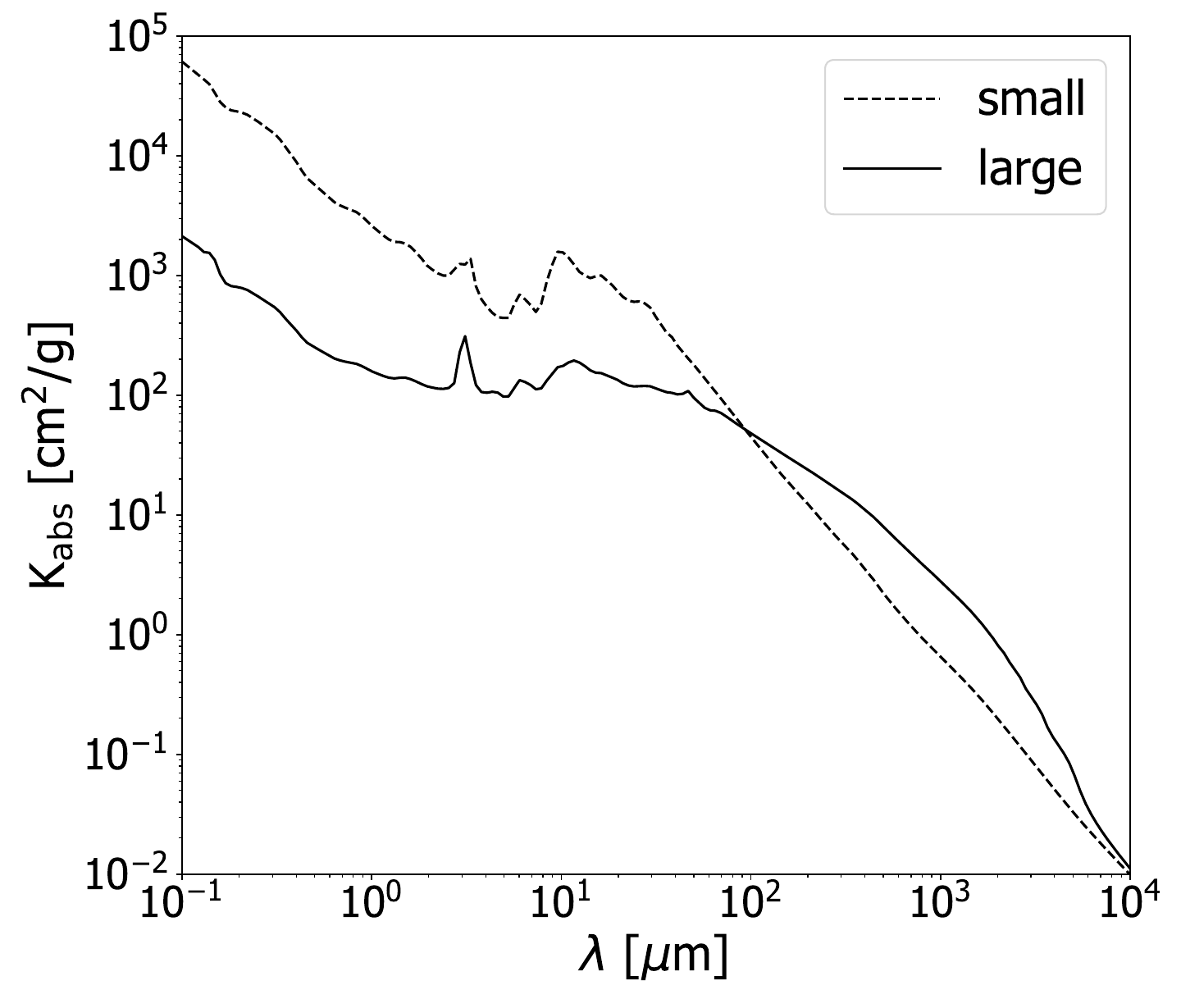}
\end{subfigure}
\begin{subfigure}{.33\linewidth}
  \centering
  \includegraphics[width=1.0\linewidth]{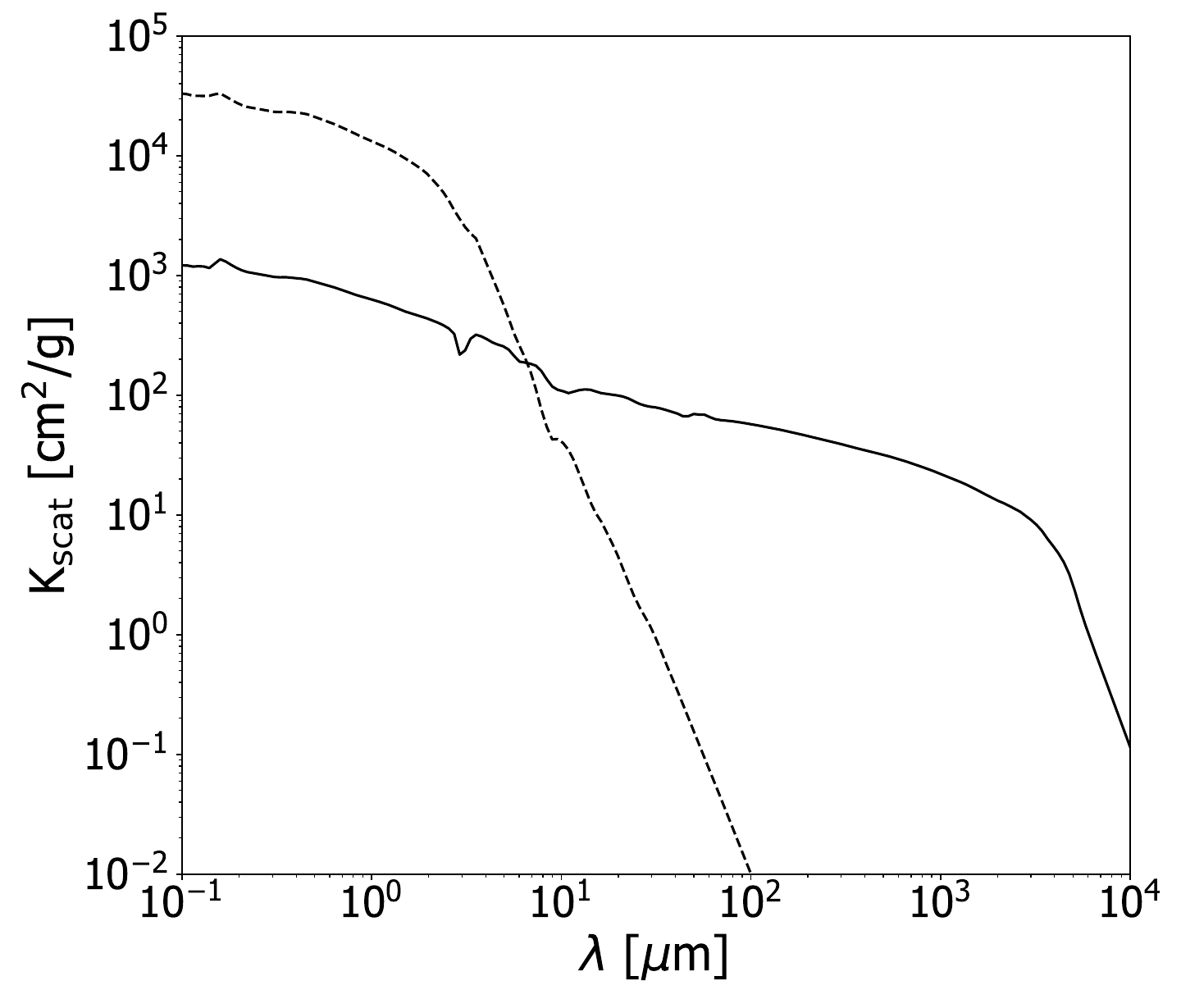}
\end{subfigure}
\begin{subfigure}{.33\linewidth}
  \centering
  \includegraphics[width=1.0\linewidth]{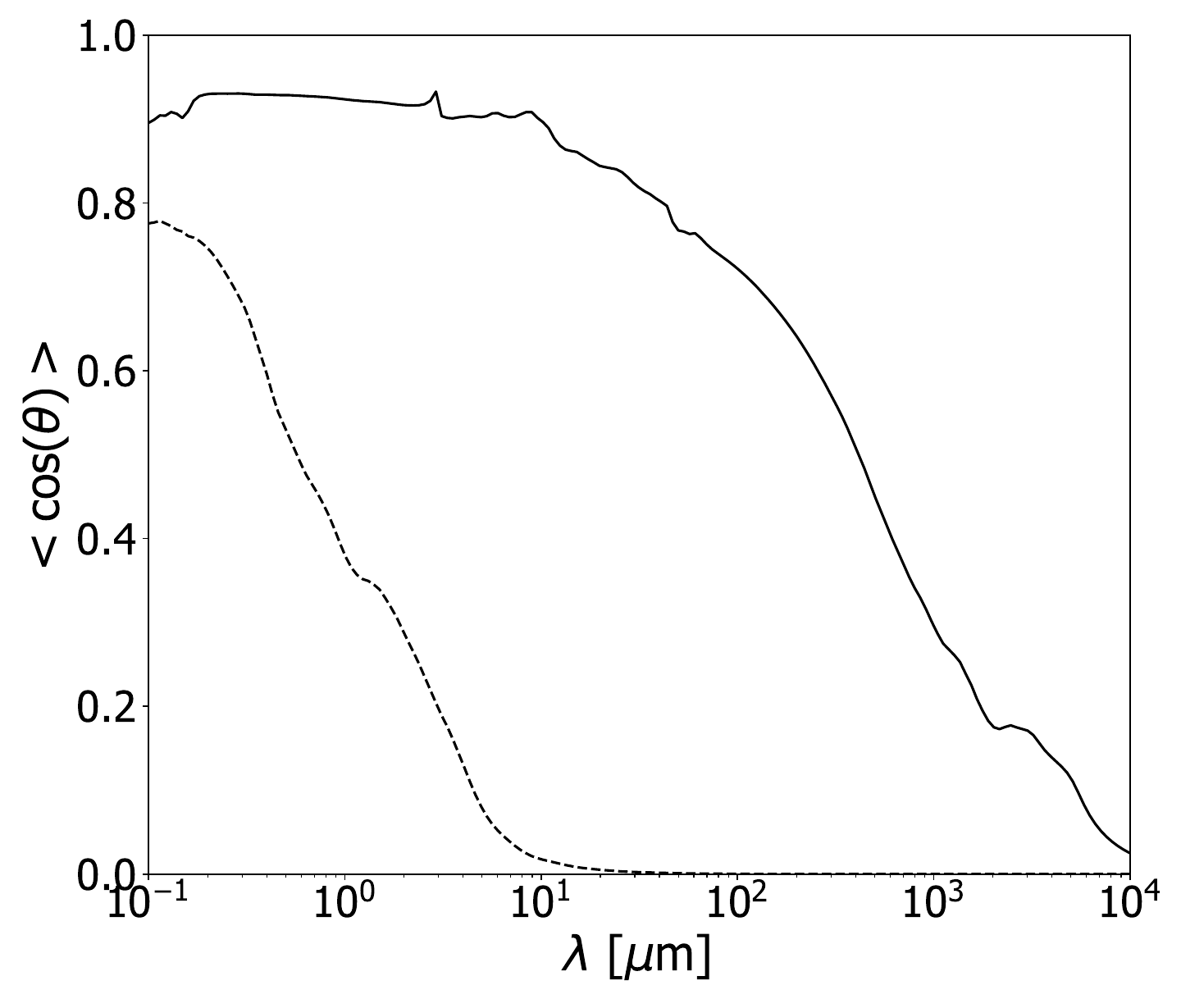}
\end{subfigure}
\caption{Dust optical properties of Model D (dashed and solid lines) and S (dashed lines only). Left: absorption opacity. Middle: scattering opacity. Right: Henyey-Greenstein g parameter of anisotropy (g = $<$cos($\theta$)$>$). }
\label{fig:compare-opa}
\end{figure*}

For Model D, we reproduce the same distribution as in \citet{Zhang+etal_2021} (who uses the DSHARP opacities), for each of the two populations we use a size-averaged opacity derived from a dust grain size distribution following the MRN distribution with a power-law index $q = 3.5$. The dust grains are composed of a mixture of water ice \citep{Warren+Brandt_2008}, astronomical silicates \citep{Draine_2003}, troilite and refractory organics \citep{Henning+Stognienko_1996}. We consider two populations, a small population $D_{\bm small}$ with an upper grain radius a$_\mathrm{max, s}$ = 1 $\mu$m, and a large population $D_{\bm large}$ with an upper grain radius a$_\mathrm{max, l}$ = 1 mm. Both populations have a lower grain radius of a$_\mathrm{min}$ = 0.005 $\mu$m \citep[e.g.,][]{Schwarz+etal_2018, Ballering+etal_2021, Zhang+etal_2021}. The wavelength range is a logarithmic distribution of 210 values between $\lambda_\mathrm{min} = 0.1$~$\mu$m and $\lambda_\mathrm{max} = 10^4$~$\mu$m. Because the small population is globally less protected from photodesorption, we set its ice mass fraction to zero (following the prescription in \citet{Zhang+etal_2021}). The resulting mass-weighted dust absorption opacities $\kappa_\mathrm{abs}$, dust scattering $\kappa_\mathrm{sca}$ and the $g$ parameter of anisotropy are shown in Fig.\,\ref{fig:compare-opa}.
These profiles are to reproduce as closely as possible the ones in \citet{Zhang+etal_2021}. The thermal calculations as well as the chemistry post-process are performed using the two populations simultaneously.

For Model S, we only use the small population opacity $D_{\bm small}$ (dashed lines in Fig.~\ref{fig:compare-opa}) for the computation of the dust temperature. A single dust structure is thus considered at all steps, including the chemistry post-process.

Model C is the same as Model D as it uses the two dust size distributions and optical properties simultaneously for the dust temperature computation, but differs during the chemistry post-process step. Indeed, the two resulting dust temperatures are averaged over the dust surface area using Eq.~\ref{eq:ta} so a single dust structure is used as input to solve the chemical evolution. Model C is therefore close to the prescription that has been commonly used in most disk models so far.

\subsubsection{Model M16}\label{sec:model:M} 
It is expected, based on models of dust coagulation and fragmentation, that a dust population is not composed of only two distinct populations but rather of a continuous size distribution \citep{Birnstiel+etal_2010, Birnstiel+etal_2018}. To address this, we consider Model M16, which mimics a continuous grain size distribution by splitting the MRN distribution into 16 logarithmically distributed grain size intervals, each with its own size (Eq.\,\ref{eq:size_av}) and associated single-grain-size opacities. Model M16 is therefore meant to be more realistic than the other models. 

Similarly to the other models, we adopt a classical power-law grain size distribution ranging from $a_\mathrm{min}$ = 5 nm to $a_\mathrm{max}$ = 1 mm with a size distribution index $q=3.5$. However, instead of having a total absorption opacity averaged over the total size distribution (as it is done in the other models), the resulting opacities are discretized into the 16 bins (see Table~\ref{tab:16sizes}).

To do so, we perform calculations for each grain bins using the Fortran subroutine included in the \textit{dsharp\_opac} package \citep{Birnstiel+etal_2018}, itself derived from the original Mie code from \citet{Bohren+Huffman_1998}. To avoid unrealistic interference effects due to the spherical treatment of the Mie theory, we calculate the opacity for 30 linearly distributed grain sizes around each of the 16 bin sizes.

To simplify the discussion in the following, we differentiate three sub-populations from the 16 sizes, each of which characterized by similar optical properties. We call small grains all grains of size < 0.1 $\mu$m (4 bins), intermediate grains all grains of sizes between 0.1 and 10 $\mu$m (6 bins) and large grains all grains of size > 10 $\mu$m (6 bins).

\begin{figure*}
\begin{subfigure}{.5\linewidth}
  \centering
  \includegraphics[width=1.0\linewidth]{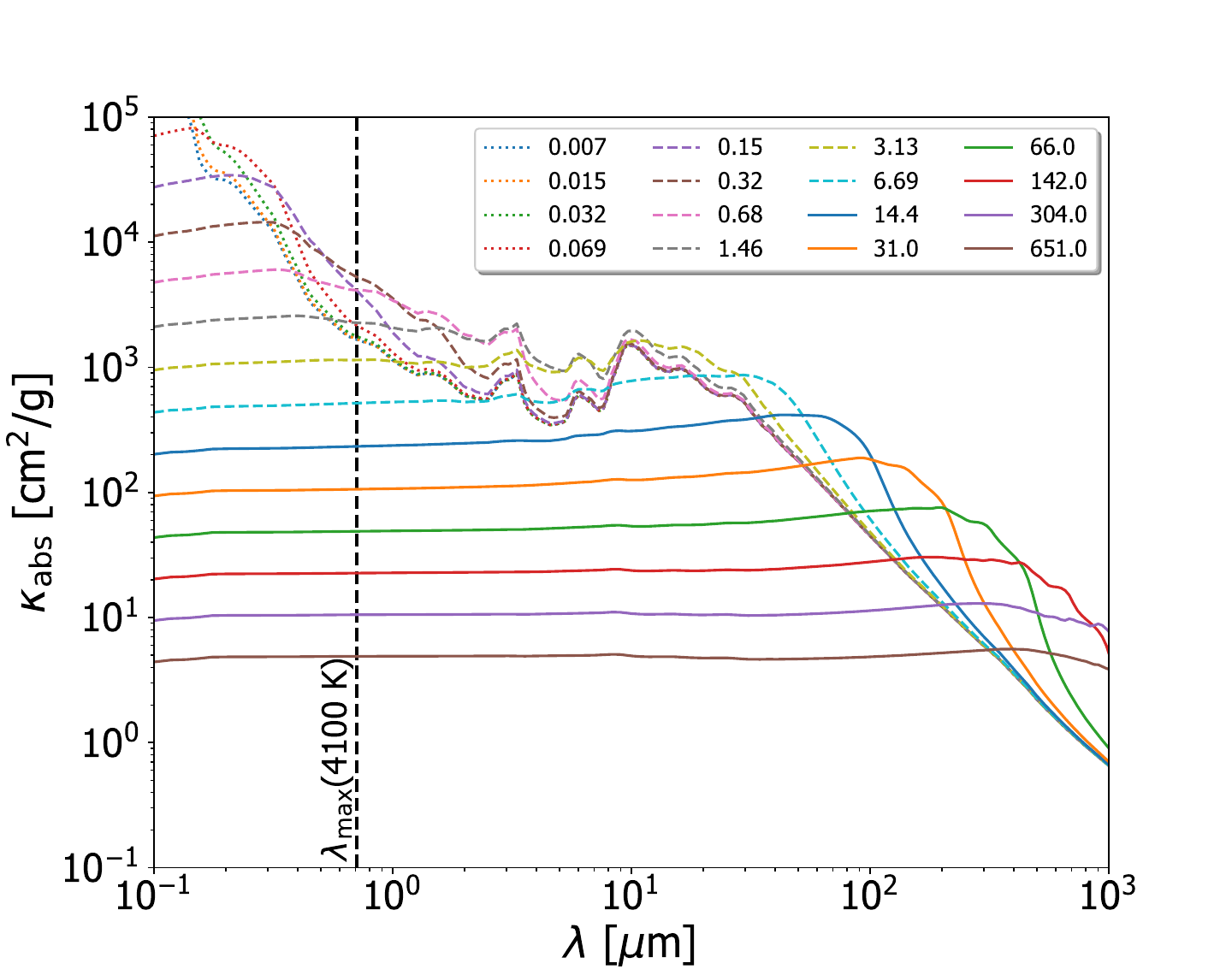}
  \caption{Absorption opacity \label{fig:cabs16}}
\end{subfigure}
\begin{subfigure}{.5\linewidth}
  \centering
  \includegraphics[width=1.0\linewidth]{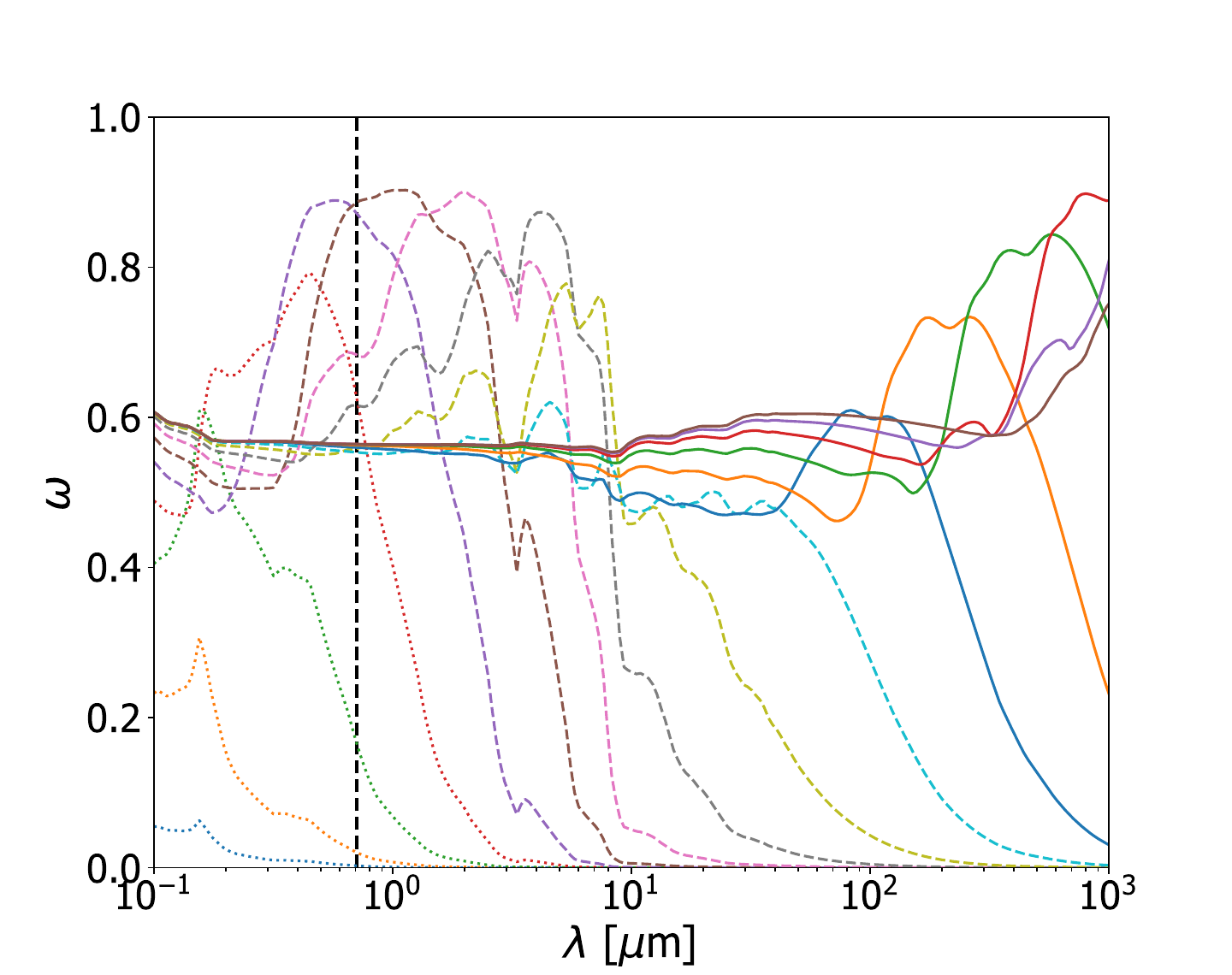}
  \caption{Scattering albedo \label{fig:albedo16}}
\end{subfigure}
\caption{Absorption opacity and scattering albedo profiles for the 16 bin sizes in Model M16. The dotted, dashed, and solid lines represent the small, intermediate, and large grains, respectively. The dashed black vertical line indicates the wavelength $\lambda_\mathrm{max}$ at which the star radiates according to Wien's law. The grain sizes are indicated in $\mu$m. Refer to Table~\,\ref{tab:16sizes} for description of bin sizes.}
\label{fig:dsharp16}
\end{figure*}

The resulting mass-weighted absorption opacities $\kappa_\mathrm{abs}(a) =  2 \pi a^2 Q_\mathrm{abs}(a)/ m(a)$ [g/cm$^2$] and scattering albedos $\omega$ are shown in Fig.\,\ref{fig:dsharp16}. The absorption opacities (Fig.\,\ref{fig:cabs16}) of the large grains (solid lines) are roughly flat in short wavelengths (< 100 $\mu$m) whereas they all asymptotically follow the same decreasing slope beyond. The large grains also prefer to emit at wavelengths comparable to their size. Intermediate (dashed lines) and small grain (dotted lines) opacities are strongly wavelength-dependent and show structured patterns due to their material composition. These various profiles result in various equilibrium grain temperatures and thermal emissions.

The small grains and intermediate grains show little albedo  or are completely opaque (Fig.\,\ref{fig:albedo16}) at wavelengths $\lambda$ > 10 $\mu$m. The albedo $\omega$ is at its maximum when $2\pi a \sim \lambda$ \citep{Birnstiel+etal_2018} and can be as high as 0.9 between $\sim$ 0.1 to 10 $\mu$m in the case of the intermediate grains. The albedo of the largest grains (bin 12 to bin 16) remains globally flat for $\lambda < 100$ $\mu$m at a fairly large albedo (0.5 < $\omega$ < 0.6). However, since their relative surface area is very small (< 1 \% in the midplane), their contribution to scattering events leading to significant effects on the thermal structure is assumed to be negligible.

\begin{table*}  
\centering
 \caption{Summary of the dust disk models. The grain sizes in column 4 and the mass fraction in column 5 are calculated using Eq.~\ref{eq:size_av} and Eq.~\ref{eq:dustmass}, respectively.}
 \begin{tabular}{l  c  c  c  c  c}
   \hline
    disk model & \# bins in RADMC3D & \# bins in NMGC & grain sizes [$\mu$m] & mass fraction [\%] & opacity \bigstrut \\
   \hline
    Model S & 1 & 1 & 0.02 & 100 & D$_\mathrm{small}$ \bigstrut \\
    \hline
    & & & 0.02 & 4.50 & \bigstrut\\[-3.5ex]
    \raisebox{-2ex}{Model D} & \raisebox{-2ex}{2} & \raisebox{-2ex}{2} &&& \raisebox{-2ex}{D} \\
     & & & 10.4 & 95.5 & \\
   \hline
    Model C & 2 & 1 & 0.02 & 100 & D \bigstrut \\
    \hline
    Model M16 & 16 & $\dagger$ & $[0.007 - 651]^{\ast}$  & $[0.10-32]^{\ast}$ & M16 \bigstrut \\
    \hline
\multicolumn{6}{l}{$\ast$ The values are given in Table~\ref{tab:16sizes}. $\dagger$ No chemistry computed.}\\
    \end{tabular}
\label{tab:dustparam}
\end{table*}

\begin{table}
\caption{Dust distribution in Model M16. \label{tab:16sizes}}
\centering
\begin{tabular}{c c c c}
\hline
\noalign{\smallskip}
bin & size [$\mathrm{\mu m}$] & $x^{\ast}$ [\%] & $\tau_\mathrm{drift}$(100 AU) [yr]\\
\noalign{\smallskip}
\hline	
\noalign{\smallskip}
1 &	 0.007 	& 0.10 &	1.04$\times10^9$\\
2 &	 0.015  & 0.15 &	4.84$\times10^8$\\
3 &	 0.032 	& 0.22 &	2.27$\times10^8$\\
4 &	 0.069 	& 0.33 &	1.05$\times10^8$\\
5 &	 0.15	& 0.48 &	4.84$\times10^7$\\
6 &	 0.32 	& 0.70 &	2.27$\times10^7$\\
7 &	 0.68 	& 1.03 &	1.05$\times10^7$\\
8 &	 1.46 	& 1.50 &	4.97$\times10^6$\\
9 &	 3.13 	& 2.20 &	2.32$\times10^6$\\
10 & 6.69 	& 3.22 &	1.09$\times10^6$\\
11 & 14.4 	& 4.72 &	5.04$\times10^5$\\
12 & 31.0 	& 6.91 &	2.34$\times10^5$\\
13 & 66.0 	& 10.1 &	1.10$\times10^5$\\
14 & 142 	& 14.8 &	5.11$\times10^4$\\
15 & 304 	& 21.7 &	2.39$\times10^4$\\
16 & 651 	& 31.8 &	1.12$\times10^4$\\
\multicolumn{4}{l}{$\ast$ Mass fraction.} \\
\end{tabular}
\end{table}

\subsubsection{Comparing models}\label{sec:model:caveat} 

The total disk mass is kept constant in all models. Thus, when comparing models, their respective optical depths can differ since the dust mass is not distributed among the same dust populations. Typically, Model D and C are expected to exhibit a slightly smaller vertical optical depth than Model S due to the presence of the large dust population in the former. However, these opacity differences are rather small
and the differences between models can be mainly attributed to their different dust temperature structures. Note that we tested various dust models and disk physical parameters (disk mass, size distribution, opacities, viscosity) and we show the results in Appendix\,\ref{app:sensitivity}.

\section{Results}\label{sec:results}
In this section, we evaluate the effects of different assumptions on the dust models on the thermal and chemical outcomes. First, we describe the 2D dust structures. 

\subsection{Dust structures}\label{sec:res:structure}
\begin{figure*}
\begin{subfigure}{.5\linewidth}
  \centering
  \includegraphics[width=1.0\linewidth]{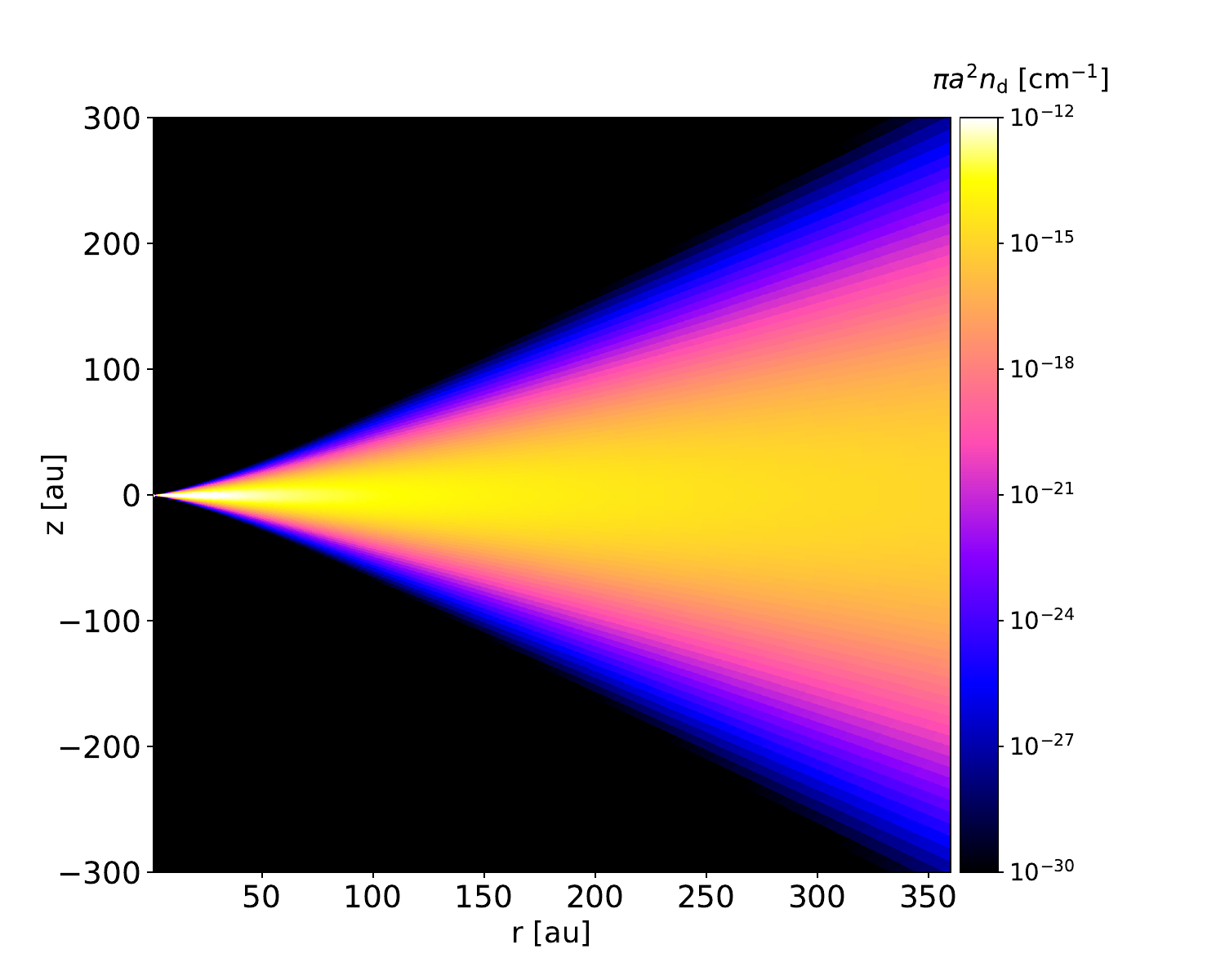}
  \caption{Small grains}\label{fig:D-dens-small}
\end{subfigure}
\begin{subfigure}{.5\linewidth}
  \centering
  \includegraphics[width=1.0\linewidth]{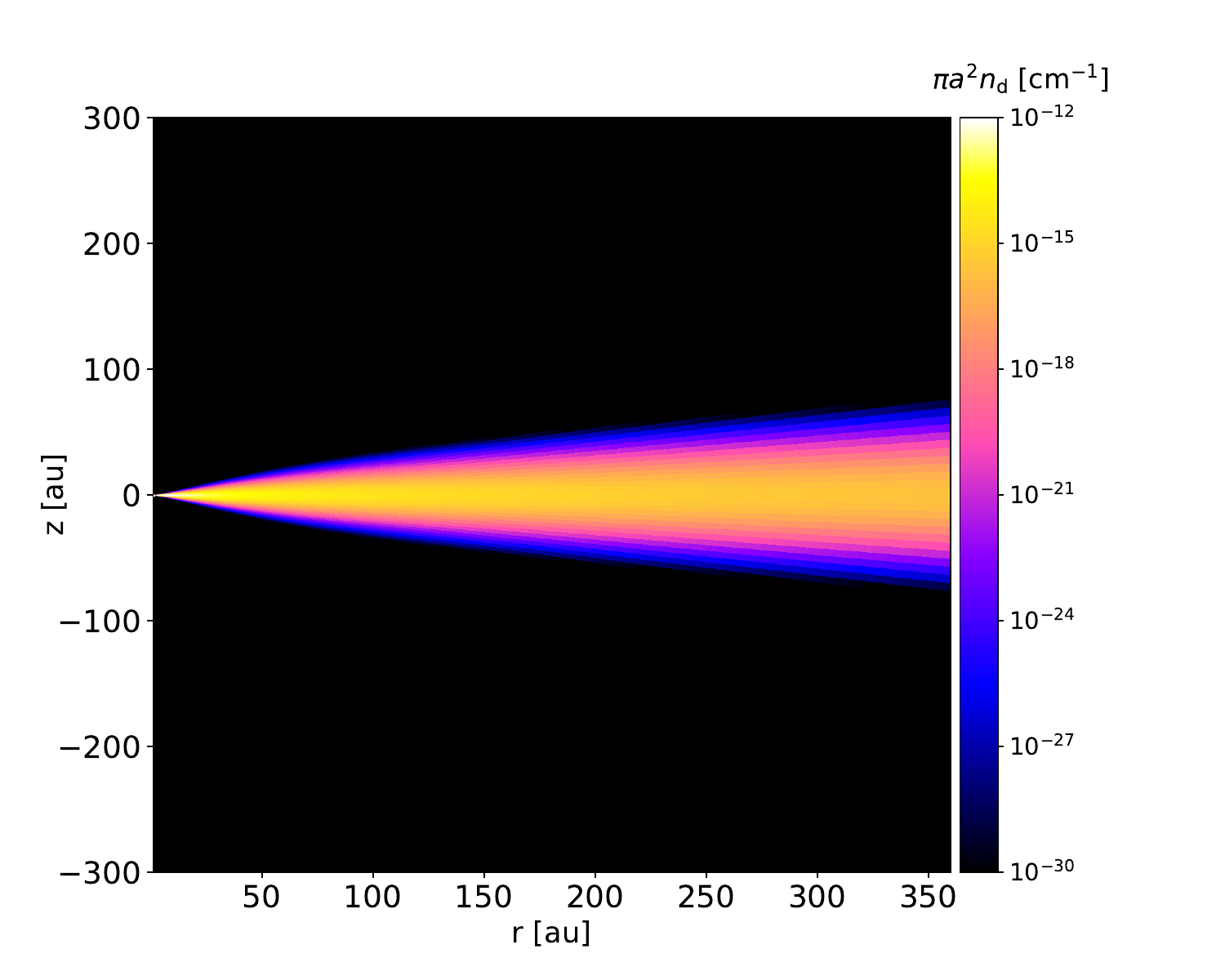}
  \caption{Large grains}\label{fig:D-dens-large}
\end{subfigure}
\begin{subfigure}{.5\linewidth}
  \centering
  \includegraphics[width=1.0\linewidth]{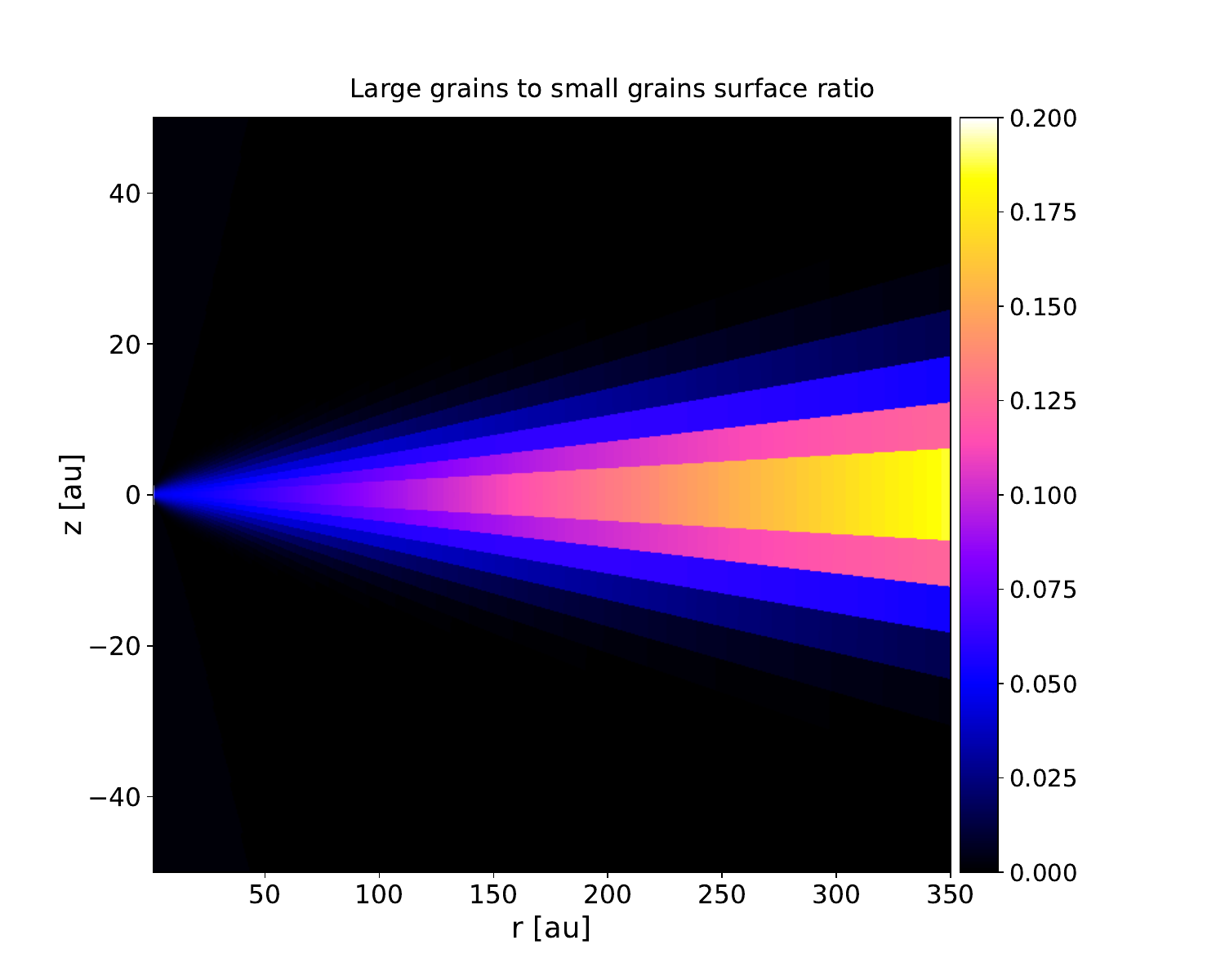}
  \caption{Surface ratio.}\label{fig:surf_ratio}
\end{subfigure}
\begin{subfigure}{.5\linewidth}
  \centering
  \includegraphics[width=1.0\linewidth]{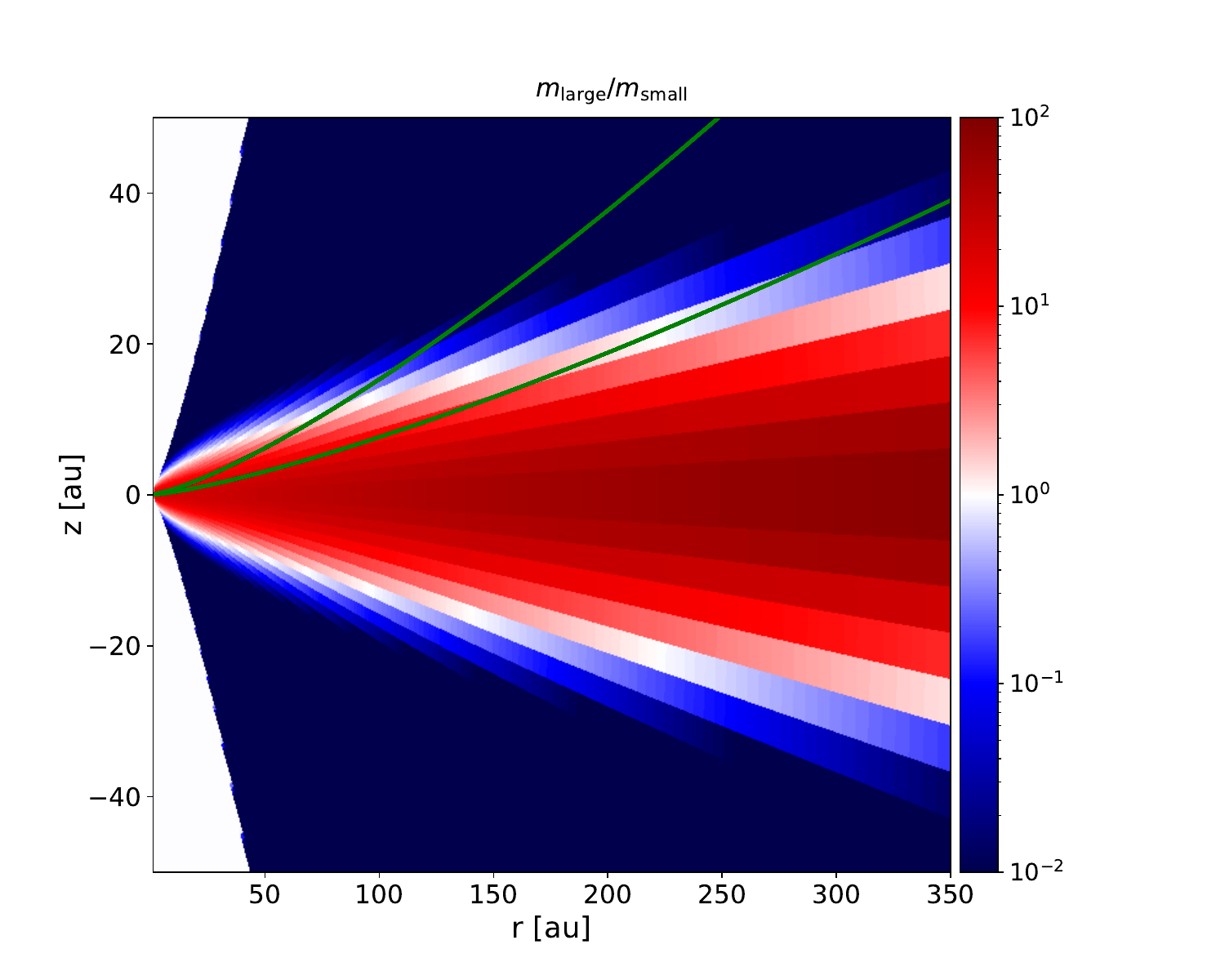}
  \caption{Mass ratio.}\label{fig:mass_ratio}
\end{subfigure}
\caption{Panels a and b: 2D dust area density [cm$^{-1}$] of Model D. Panel c: ratio of the large grains area density to the small grains area density (note the different $z$ scale). Panel d: mass ratio between the large and small grains. The green lines mark $1\times H_\mathrm{g}$ and $2\times H_\mathrm{g}$.
\label{fig:D-dens}}
\end{figure*}

Figure. \ref{fig:D-dens} shows the 2D dust area densities of Model D. The choice of showing the area density is motivated by the fact that the dust surface area is one of the main parameters that drives the gas-grain interactions during chemistry. As detailed in Sec.\,\ref{sec:model}, the dust populations are computed independently to one another so the two dust populations are to be shown separately. 

Using Eq.\,\ref{eq:size_av} and Eq.\,\ref{eq:dustmass}, we can define two discretized subranges representing two populations with their own size and mass fraction values. The small dust population (Fig.\,\ref{fig:D-dens-small}) is composed of grains of size 
$a_\mathrm{small} = 0.02$ $\mathrm{\mu}$m and, despite representing less than 5 \% of the total dust mass, is the larger contributor to the extinction as the grains represent most of the dust surface area. The large dust population is composed of grains of size 
$a_\mathrm{large} = 10.4$ $\mathrm{\mu}$m and represents 95.5 \% of the total dust mass (Fig.\,\ref{fig:D-dens-large}). 

The settling factor is defined using Eq.~\ref{eq:settling-dong}. The large population is more confined toward the midplane than the small population due to its significantly larger settling factor. Indeed, the Stokes number ratio of the two populations is $S_\mathrm{t, large}$/$S_\mathrm{t, small}$ = $a_\mathrm{large}$/$a_\mathrm{small}$ = 5.2$\times$10$^2$. While the settling factor of the small population is close to unity at all radii (the small dust scale height follows that of the gas), the ratio between the large dust and small dust scale height equals to 0.92 at 10 au, 0.35 at 100 au and 0.20 at 500 au.

\subsection{Thermal structures}\label{sec:res:thermal}
\subsubsection{2D temperature structures of Model D}\label{sec:res:thermal_D}

\begin{figure*} 
\centering
\includegraphics[width=1.05\linewidth]{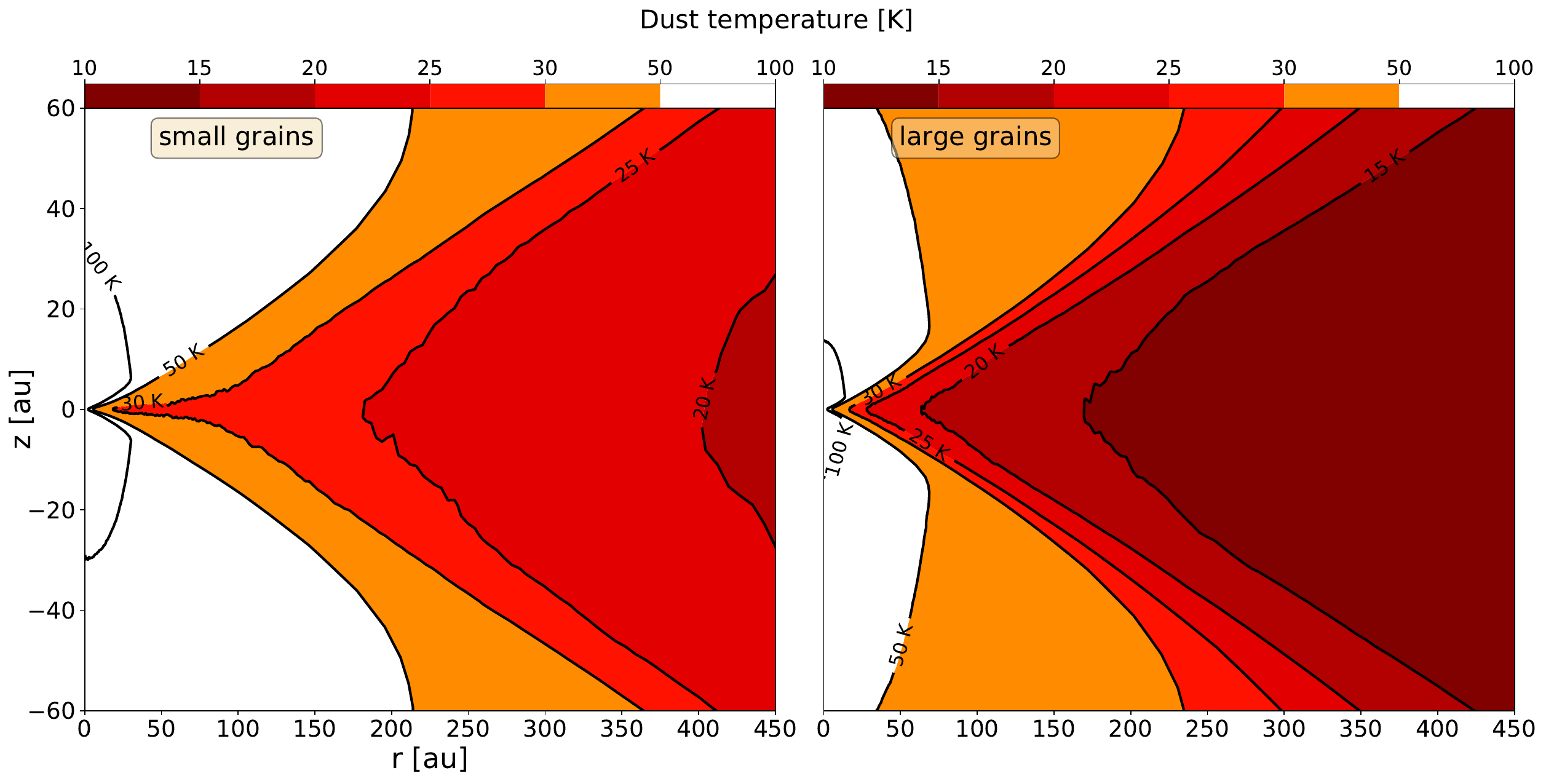}
\caption{2D dust temperature structures of Model D. Left: temperature of the small grains. Right: temperature of the large grains.  \label{fig:D-temp}}
\end{figure*}

Figure. \ref{fig:D-temp} shows the dust temperature structures of Model D computed with the thermal Monte Carlo simulation in RADMC3D. The two populations exhibit notable differences. In particular, the 25 K isothermal is located  at $\sim$ 30 au from the star (in the midplane) for the large grains (Fig.\,\ref{fig:D-temp}, right), which is a typical expected distance where the CO snowline is located in a T-Tauri disk. On the other hand, the small grain temperature drops below 25 K near 180 au only (Fig.\,\ref{fig:D-temp}, left). There is therefore a significant temperature spread with radial distance of 150 au between the two 25 K isothermal. 


\subsubsection{Comparison between Model D, S, and C}\label{sec:res:thermal_SDC}
\begin{figure} 
\centering
\includegraphics[width=1.05\linewidth]{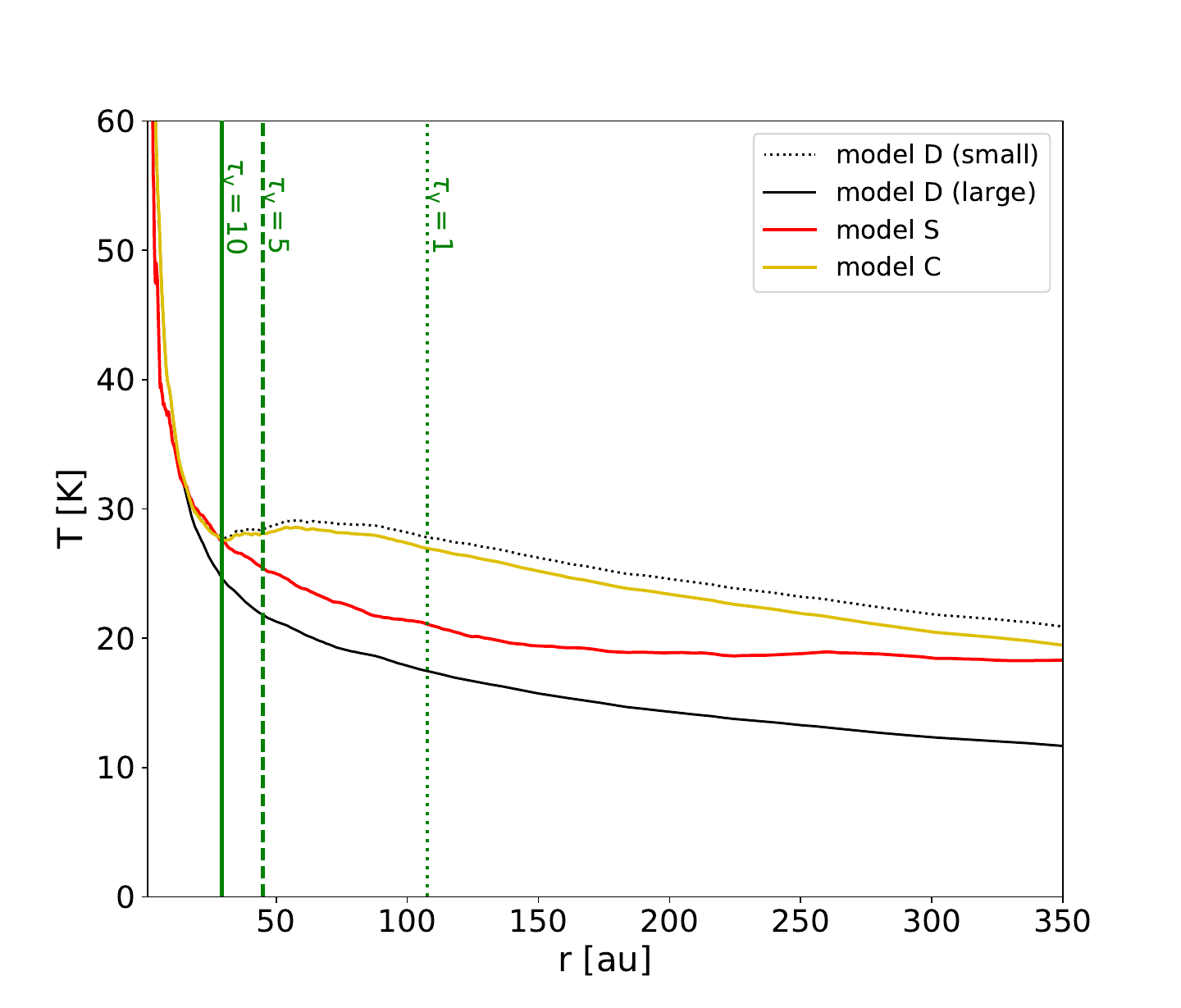}
\caption{Disk midplane radial profiles of dust temperatures. The black curves are the temperature profiles of Model D. The solid red curve is the temperature profile of Model S whereas the yellow curve is that of Model C. The vertical green solid, dashed, and dotted lines represent the radial distances in Model D from which the vertical optical depth (at 1 $\mu$m) becomes smaller than 10, 5, and 1, respectively. \label{fig:mid_SDC}}
\end{figure}

To have a better understanding of the temperature structures we further investigate the disk midplane by comparing the radial dust temperatures of Model D with that of Model S and C. 
Figure~\ref{fig:mid_SDC} shows the radial profiles in the midplane of the three models. In Model D, the two populations are thermally decoupled from $r \sim 30$~au and the small grains are warmer than the large grains of approximately 13 K at all radii. Moreover, the small dust population temperature shows a surprising bump between $\sim$ 30 au and $\sim$ 80 au. This structure therefore deviates from an expected monotonously decreasing temperature along the radial distance.

In Model S, the dust optical properties are the same as the small population of Model D. We can therefore expect that their respective dust temperature exhibit similar profiles but Fig.\, \ref{fig:mid_SDC} shows that it is not the case. The temperature profiles of the two models overlap in the optically thick ($\tau >> 1$) inner region ($r < 30$~au) due to the very effective temperature coupling. Further away, the two populations in Model D become spread out as they become thermally decoupled. Such temperature spread does not exist in Model S since the dust structure is unique. This also implies that there can be no bump in Model S.

The optical depth at both visible and infrared (IR) domains between the star and any radial distance in the midplane is too large ($\tau > 10^4$ at 1 $\mu$m), thus the re-heating of the small dust population cannot be generated by direct starlight emission. Since this bump does not exist in Model S, we can assume that it originates from the interaction of the re-emission and scattered light between the two dust populations. This assumption is also supported by the vertical optical depth. The vertical solid, dashed, and dotted green lines represent the radial distances from which the vertical optical depth at 1 $\mu$m becomes smaller than 10, 5, and 1, respectively (in Model D). We see that the dust temperature starts to increase as the vertical optical depth goes down to values close to 10, where dust radiative interactions between upper layers and lower layers start to be possible. 

As for Model C, the temperature profile follows closely the profile of the small dust population of Model D because most of the dust surface-area is from the small grain populations (> 95 \%).

\subsubsection{Model M16}\label{sec:res:thermal_M16}
\begin{figure*}
\begin{subfigure}{.5\linewidth}
  \centering
  \includegraphics[width=1.0\linewidth]{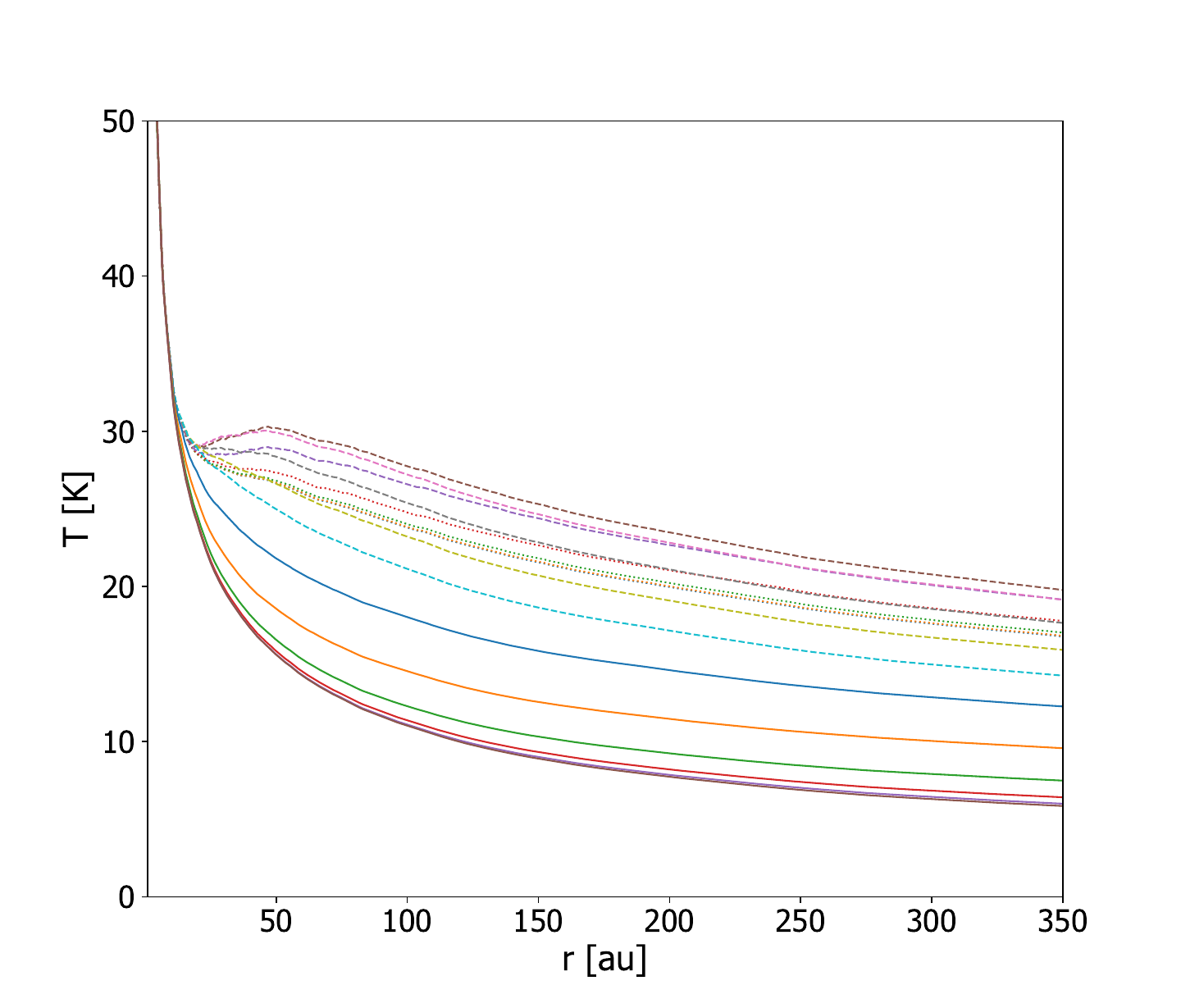}
  \caption{With Scattering \label{fig:radial_scat}}
\end{subfigure}
\begin{subfigure}{.5\linewidth}
  \centering
  \includegraphics[width=1.0\linewidth]{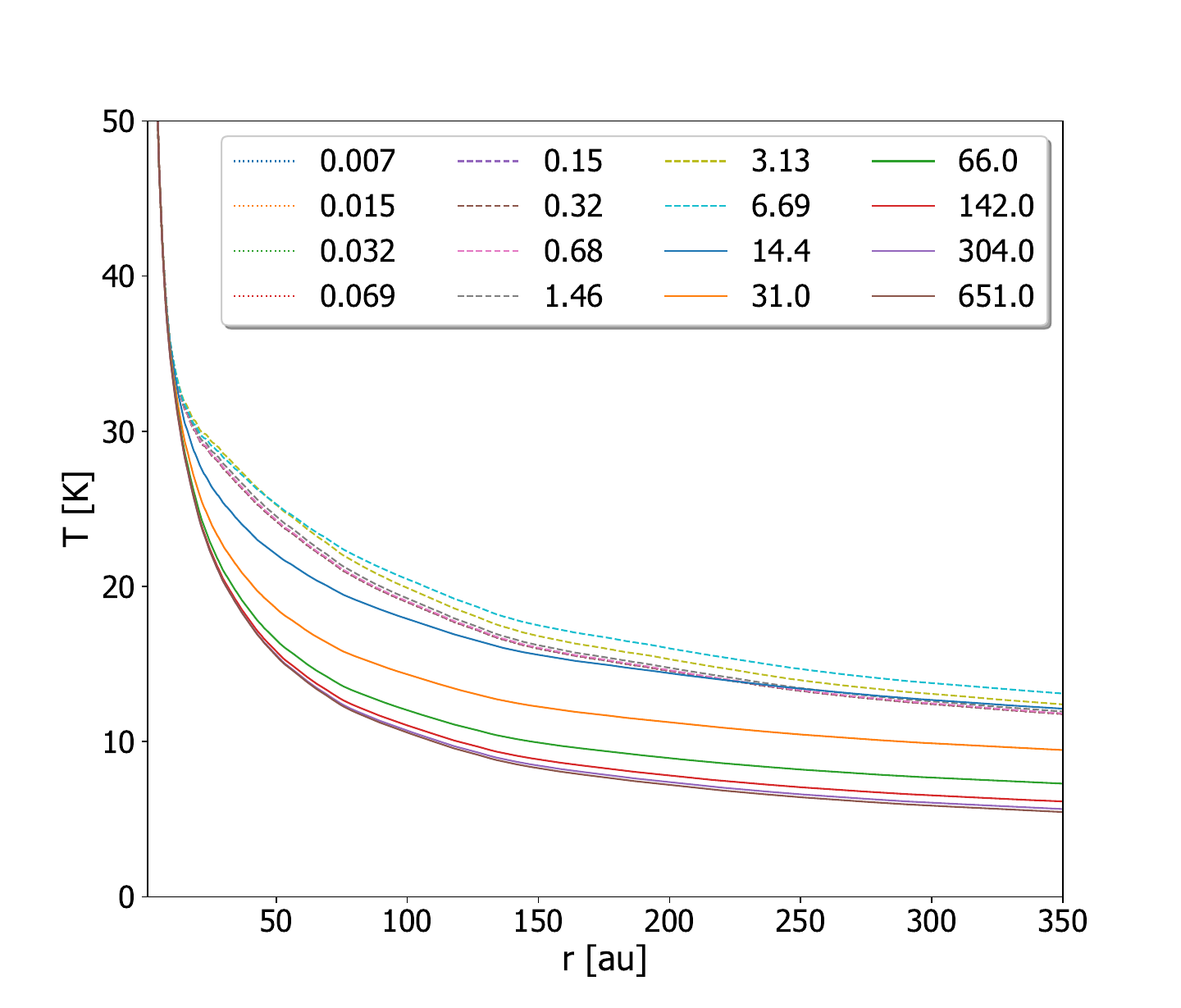}
  \caption{Without scattering \label{fig:radial_noscat}}
\end{subfigure}
\caption{Midplane radial dust temperature profiles of Model M16. The dotted curves represent the dust temperatures of the small dust grains, the dashed curves represent the temperature of the intermediate grains and the solid curves are the dust temperatures of the large dust grains. The grain sizes are given in $\mu$m. Left: temperatures with scattering. Right: temperatures with scattering opacity of the intermediate grains set to zero.}
\label{fig:compareM16}
\end{figure*}

Let us go further and test the effect of polydispersion in a more realistic structure by considering Model M16.
Figure.\,\ref{fig:radial_scat} shows the resulting radial temperature profiles of the 16 grain species in the midplane. There is a temperature spread just like in Model D, with the intermediate grains being warmer than the small grains and the large grains, and with a difference of about 17 K between the warmest grain and the coldest one. The model also shows a temperature bump inside 100~au. It is clear that the grain populations can be divided into two groups, one group that undergoes a bump effect (small and intermediate grains) and one that does not (large grains). The division in two groups is in fact analogous to the one in Model D. This suggests that a two-population Model Such as Model D could be sufficient to properly approximate the effect of polydispersion taking place in a realistic dust grain population.   
In order to help us understand which of the absorption, emission, or scattering processes as well as which of the grain populations contribute to the bump effect, we successively set the scattering opacity to zero inside wavelength intervals for each grain population and run radiative transfer simulations. We find that the radiation of wavelengths between 0.7 and 2 $\mu$m scattered by the intermediate grains (5 $\leq$ bin $\leq$ 10 i.e., grain sizes between 0.1 and 10 $\mu$m) seems to be the major reason behind the temperature bump. Figure.\,\ref{fig:radial_noscat} shows the result of the dust temperature structure computed with the albedo between 0.7 $\mu$m and 2 $\mu$m of the intermediate grains artificially set to zero. All grain temperature profiles are now monotonously decreasing with the distance. The spread of temperatures is smaller and all small and intermediate grains roughly exhibit the same temperature. 

We can conclude that when scattering is included (Fig.\,\ref{fig:radial_scat}), the temperature degeneracy of small and intermediate grains is raised, the spread is larger, and a bump is generated. Note that a similar test with Model D shows that setting the albedo to zero also suppresses the temperature bump. This effect will be discussed in more details in Sect.~\ref{sec:disc:bump}.

Another look at the temperatures shows that this temperature spread is located right in the vicinity of the CO freeze-out temperature (typically around 20-30 K in the disk midplane). We test the effect of the temperature spread on the chemical evolution in the following sections.

\subsection{Chemistry post-process: the case of CO}\label{sec:res:chemistry}
\subsubsection{Limitations}
It is easy to anticipate the effects of the temperature bump on the chemical abundance of light species, particularly that of CO. However, computation of the chemical evolution using as many grain sizes as in Model M16 can be highly cpu-consuming (a single 1D structure needs $\sim$ 60 h in cpu-time to be solved on a typical laptop compared to around $\sim$ 1 h for a model using a single grain) and can generate convergence issues due to the sizes and very small abundances of the largest grains. To duck this obstacle, we choose to use Model D instead of M16 to compute chemistry, based on the assumption previously made on the equivalent thermal structure in models D and M16. Said otherwise, we assume that the thermal spread from these models will have similar chemical outcomes. In addition, to fully appreciate the effect of the temperature spread, we also compute chemistry in Model S and C.

\subsubsection{CO in the gas phase}\label{sec:res:gasphase}

\begin{figure} 
\centering
\includegraphics[width=1.1\linewidth]{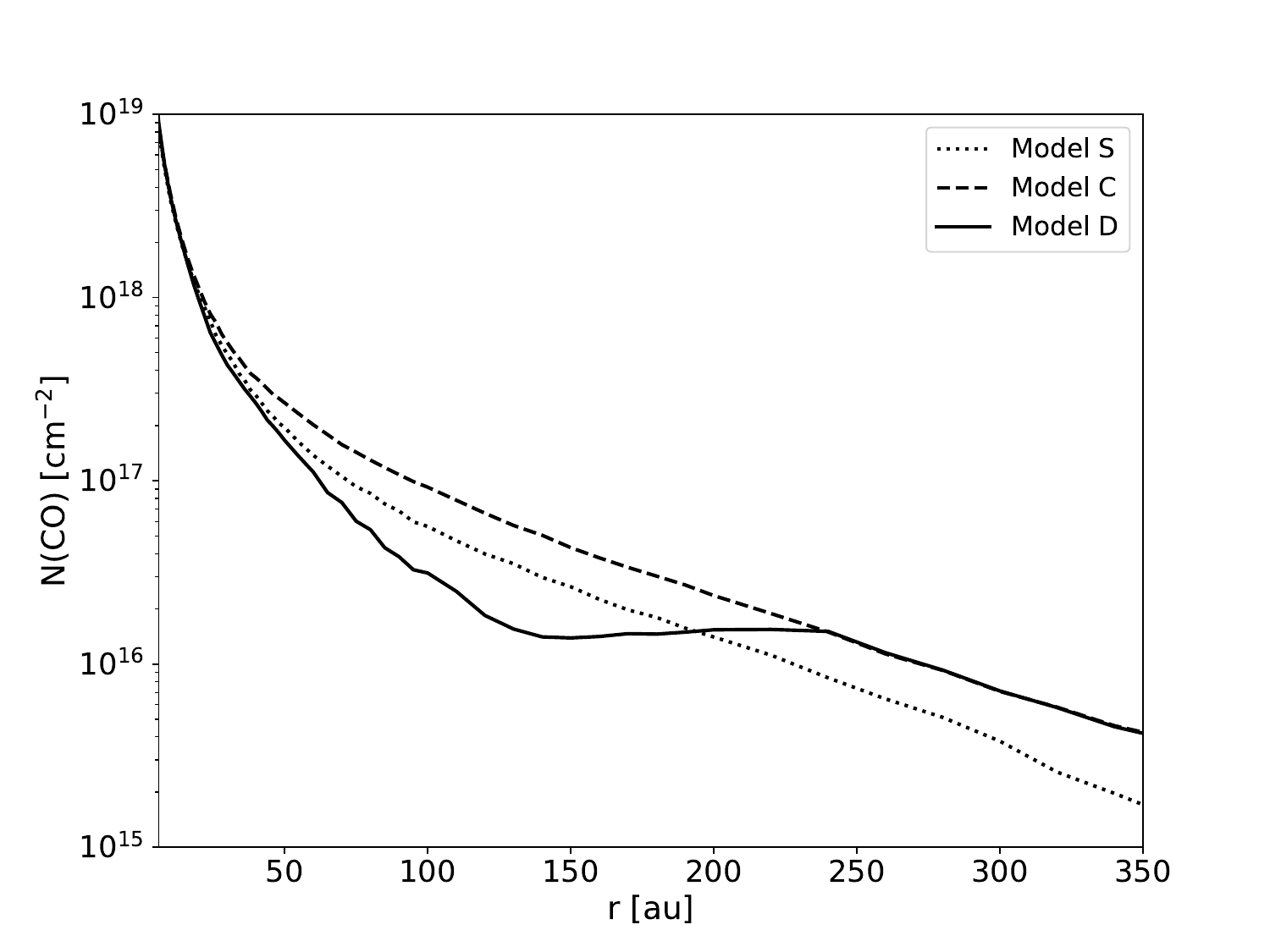}
\caption{Surface densities of CO in the gas phase for Model S (dotted line), C (dashed line), and D (solid line) as a function of the radius.} \label{fig:surfdens}
\end{figure}

Figure~\ref{fig:surfdens} shows the integrated surface density of CO in all 3 models. While Model S and C result in similar profiles, Model D shows
a clear depression in CO near 150 au, as a result of the combined effects of the small and large grains.

\begin{figure*}
\begin{subfigure}{.33\linewidth}
  \centering
  \includegraphics[width=1.0\linewidth]{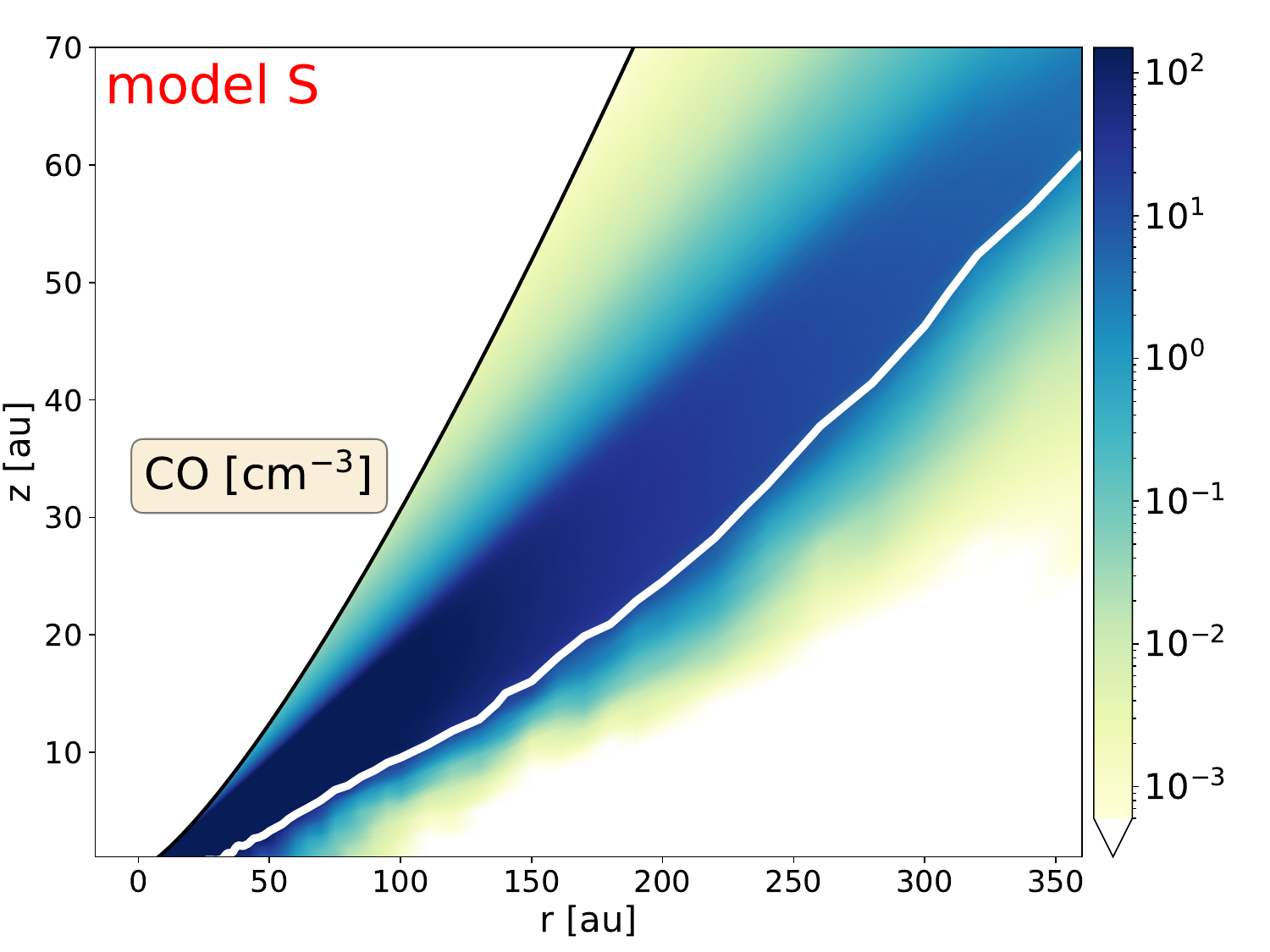}
  \caption{CO in Model S \label{fig:nCOs}}
\end{subfigure}
\begin{subfigure}{.33\linewidth}
  \centering
  \includegraphics[width=1.0\linewidth]{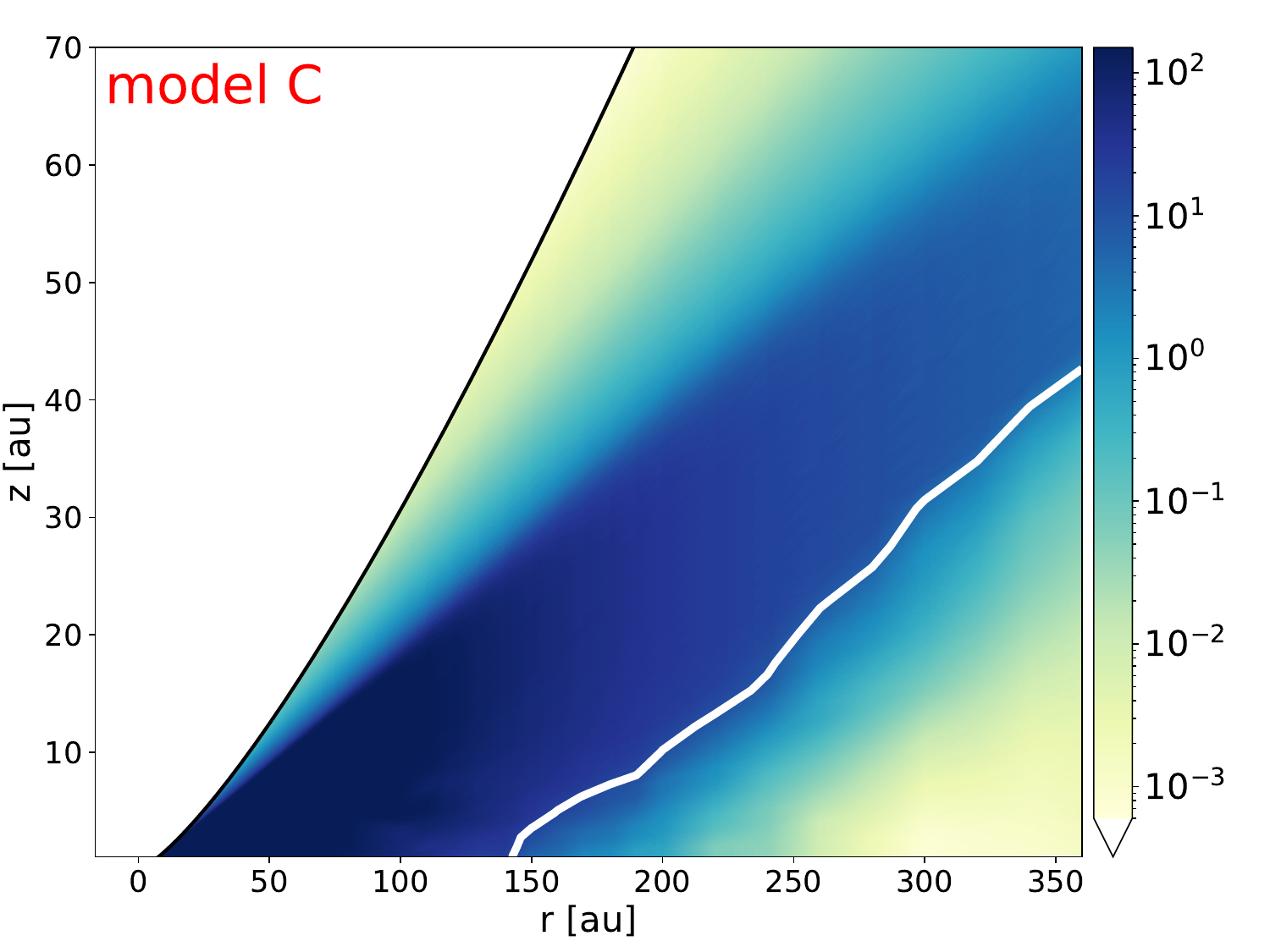}
  \caption{CO in Model C \label{fig:nCOc}}
\end{subfigure}
\begin{subfigure}{.33\linewidth}
  \centering
  \includegraphics[width=1.0\linewidth]{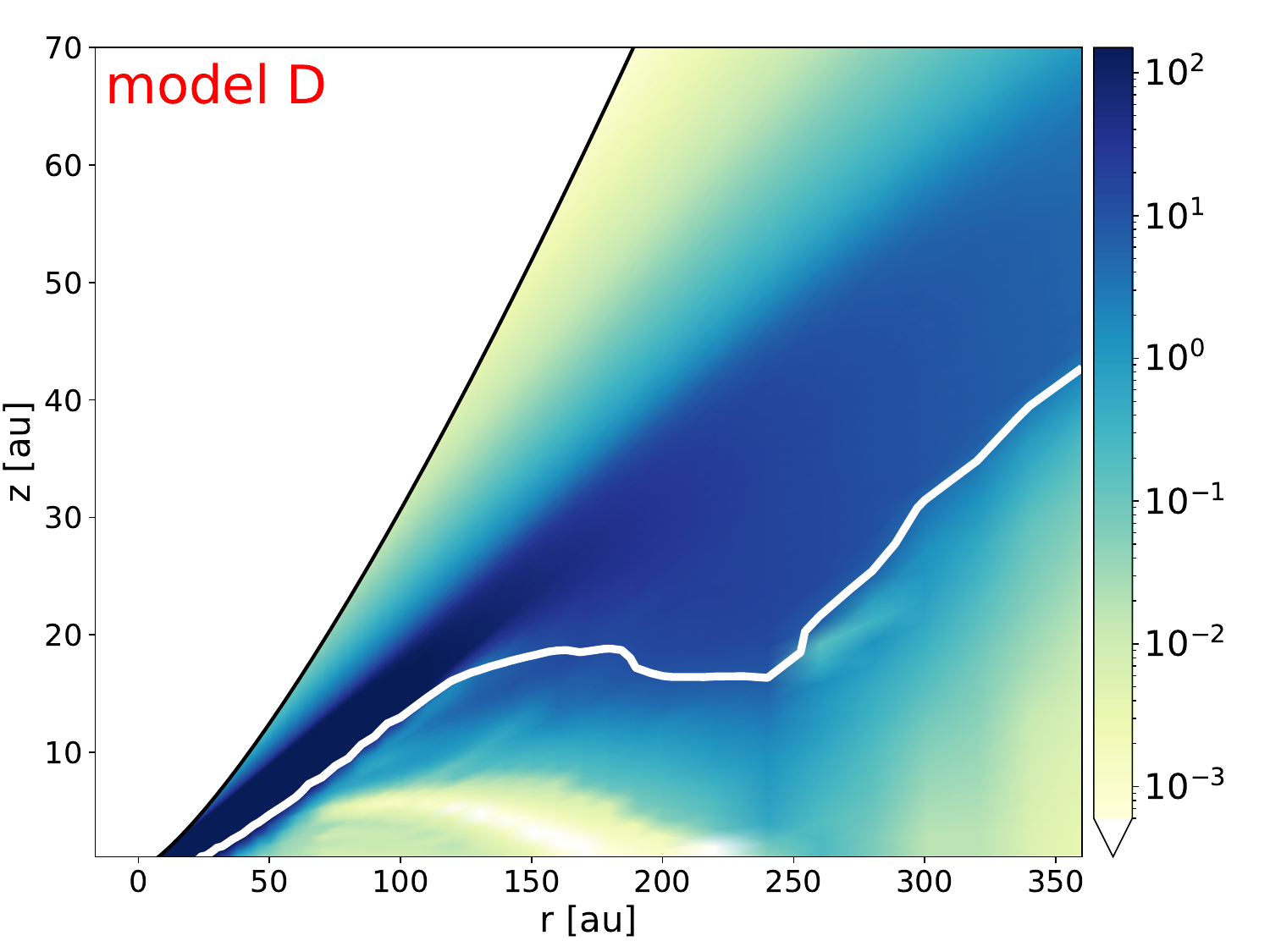}
  \caption{CO in Model D \label{fig:nCOd}}
\end{subfigure}
\caption{2D maps of CO gas number density [cm$^{-3}$]. Left: CO in Model S. Middle: same in Model C. Right: same in Model D. The solid white line corresponds to the CO snowline where n$_\mathrm{g}$(CO)/(n$_\mathrm{gas}$(CO)+n$_\mathrm{ice}$(CO)) = 0.5. The black line is the maximum scale height below which chemistry is computed.}
\label{fig:nCO_maps}
\end{figure*} 

In order to understand where those differences in column densities come from, we show in Fig.\,\ref{fig:nCO_maps} the resulting 2D CO gas number density [cm$^{-3}$] structures in ($r$,$z$) coordinates, from $r$ = 2 au to 360 au using Model S (Fig.\,\ref{fig:nCOs}), Model C (Fig.\,\ref{fig:nCOc}), and Model D (Fig.\,\ref{fig:nCOd}). The solid white lines marks the transition where the ratio n$_\mathrm{g}$(CO)/(n$_\mathrm{g}$(CO)+n$_\mathrm{s}$(CO)) becomes smaller than 0.5 or said otherwise the CO snowline. 

The interpretation of the CO gas structure in Model S is straightforward. The disk surface is vulnerable to the external UV radiation and the photodissociation rate of CO is large enough to prevent efficient formation. In the molecular layers the H$_2$ shielding and CO self-shielding become sufficiently protective and CO gas becomes the most abundant molecule (after H$_2$) with an abundance close to the interstellar value 10$^{-4}$. The grain surfaces are still too warm for effective CO condensation. Closer to the midplane, the grain surface temperatures drop and CO becomes massively depleted onto the grain surfaces. The CO snowline delimits the molecular layer and the cold midplane. Model C shows a similar pattern with a larger molecular layers. This is not surprising since the dust temperature is globally larger than in Model S. This 'V-shape' structure is typically observed in protoplanetary disks \citep[see][for a detailed description of the CO gas phase distribution in disks]{vanthoff+etal_2020}.

The CO gas map in Model D shows a more intriguing structure. Indeed, although the structure in the upper layers is similar to Model S and C, we note the presence of a CO 'hole' from $\sim$ 50 au to $\sim$ 200 au and extending up to $\sim$ 10 au vertically. We then observe a CO gas rise between 200 au and 300 au before diminishing beyond 300 au. The CO snowline exhibits a total different shape as well, with a gas-ice transition displaced toward the midplane between 150 au and 250 au.

By looking at Fig.\,\ref{fig:mid_SDC}, it appears that the CO gas rise is located well outside the temperature bump (around $\sim$ 70 au) and the CO depression is located right where the dust temperature is at its maximum. This striking discrepancy between CO depression and dust temperature is due to the dust distribution producing multiple condensation fronts, as shown in the following section.

\subsubsection{CO ice and snowline}\label{sec:res:snowline}

\begin{figure*}
\begin{subfigure}{.33\linewidth}
  \centering
  \includegraphics[width=1.0\linewidth]{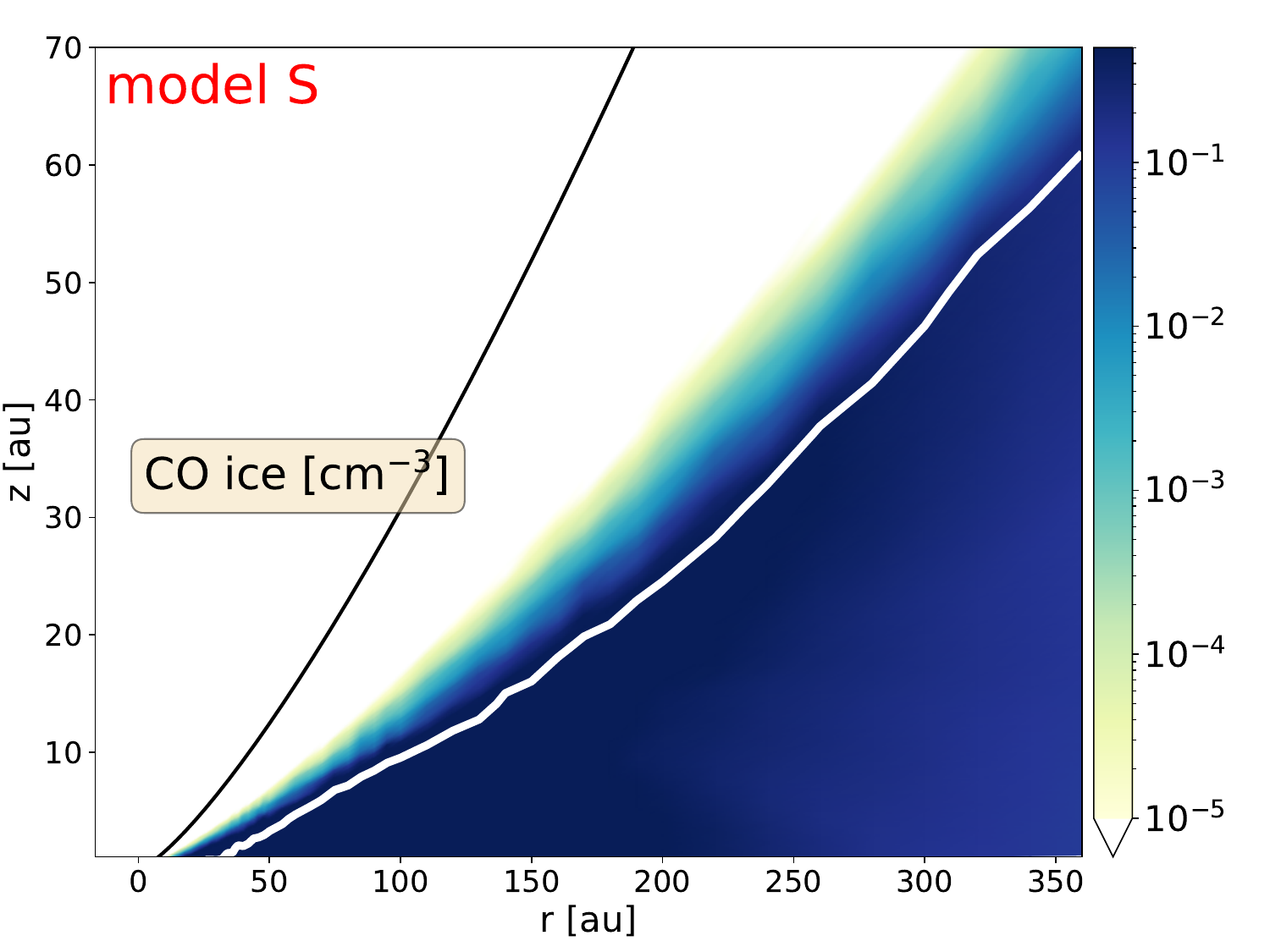}
  \caption{CO ice in Model S \label{fig:coice_s}}
\end{subfigure}
\begin{subfigure}{.33\linewidth}
  \centering
  \includegraphics[width=1.0\linewidth]{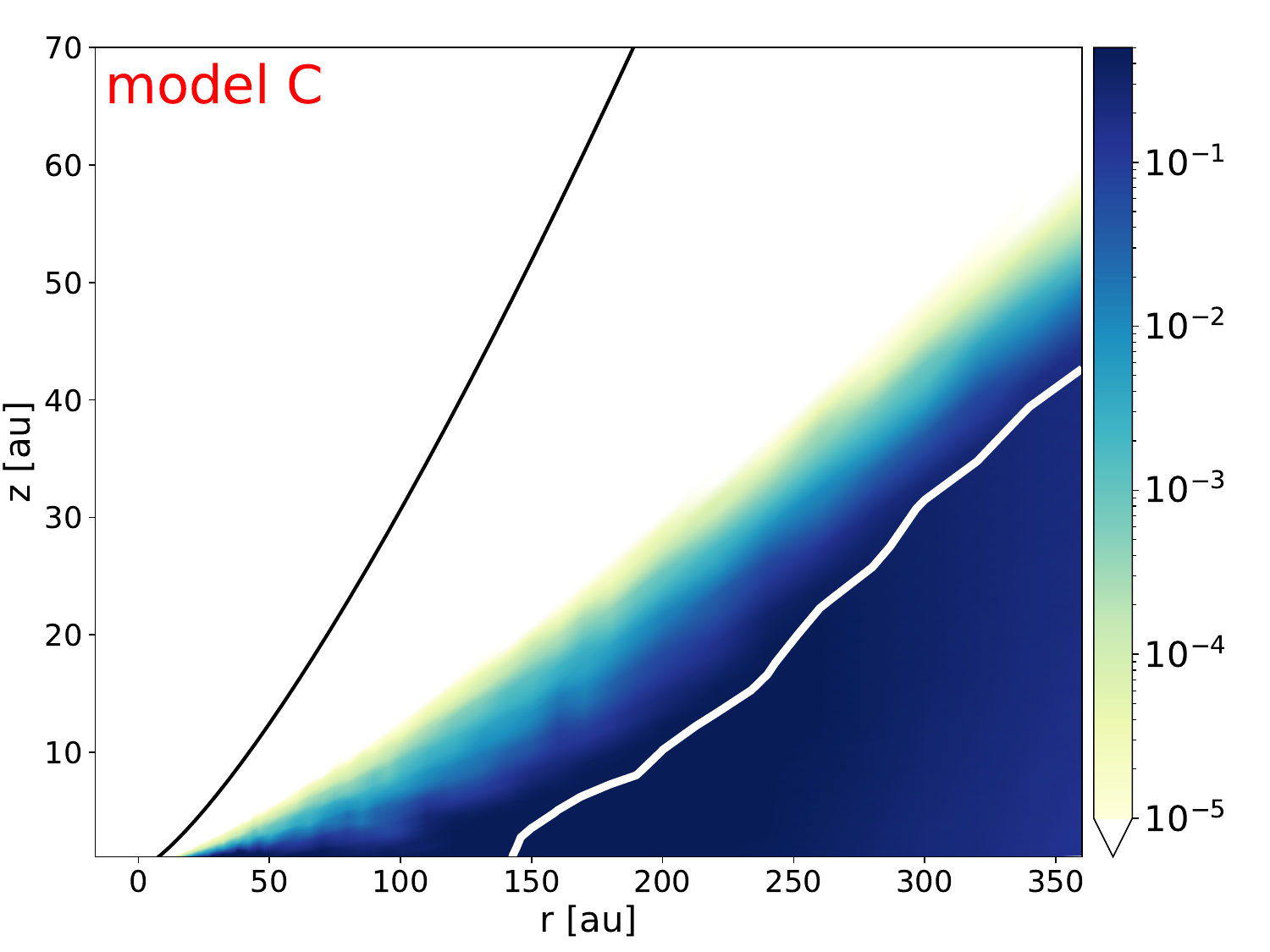}
  \caption{CO ice in Model C \label{fig:coice_c}}
\end{subfigure}
\begin{subfigure}{.33\linewidth}
  \centering
  \includegraphics[width=1.0\linewidth]{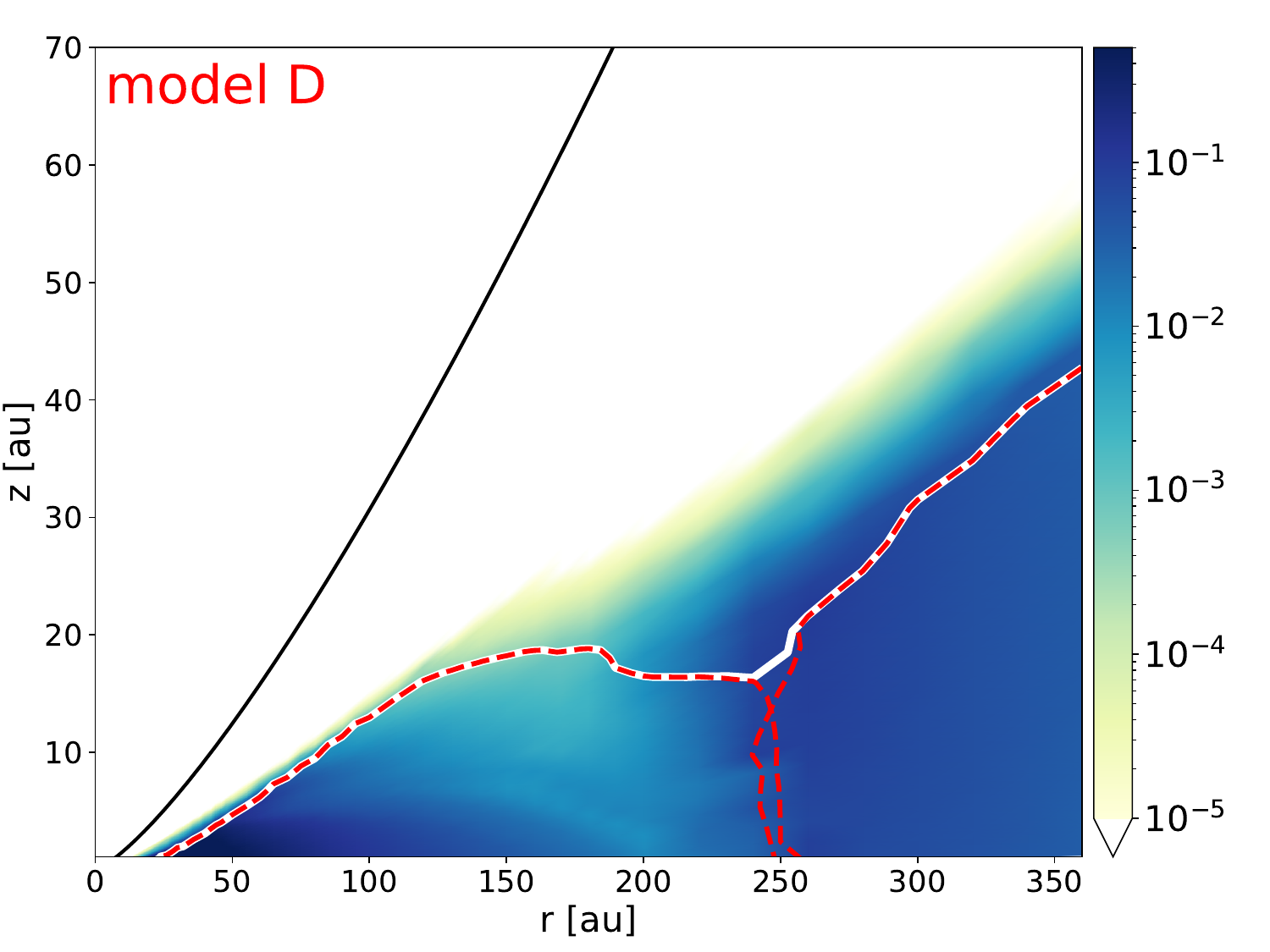}
  \caption{CO ice in Model D\label{fig:coice_d}}
\end{subfigure}
\caption{2D maps of number density CO ice. Left: CO adsorbed onto dust grains in Model S. Middle: same in Model C. Right: CO adsorbed both onto the small and large dust populations in Model D. The solid white line corresponds to the CO snowline. The dashed red lines correspond to the snowline of each population taken independently. The black line is the maximum scale height below which chemistry is computed.}
\label{fig:coice}
\end{figure*}

Figure~\ref{fig:coice} shows the 2D number density maps of the total CO ice in Model S, C, and D. All models present the typical dual structure split by the snowline, below which CO is massively depleted onto the grains. With a global higher dust temperature, the snowline of Model C is shifted toward lower layers of the disks as compared to Model S.

In Model D, the total CO ice content is the sum of CO ice adsorbed onto the small and large grain populations. It is clear that CO ice lies under the snowline profile, and although the snowline is clearly different than in the other models, the CO ice distribution seems to follow a similar trend as the other models. Additionally, we also present the snowlines of each population taken separately (red
dashed lines). The snowlines appear to be divided into two individual surfaces, overlapping near 250 au. To better understand, let us provide a formal definition of the snowline in the case of multiple discretized dust grain sizes. The freeze-out temperature is tantamount to solving the balance between the desorption and adsorption rate, and depends on multiple parameters among which the dust temperature and dust and gas densities \citep[see for instance][]{vanthoff+etal_2017}. Each grain population having its own density and temperature, we can define a snowline for each of them. More precisely, if $n_\mathrm{small}(CO)$ and $n_\mathrm{large}(CO)$ represent the CO ice number density on the small and large population, respectively, and $n_\mathrm{g}(CO)$ the number density of CO gas, the two dashed white lines thus correspond to the location where the ratios $n_\mathrm{g}(CO)/(n_\mathrm{g}(CO)+n_\mathrm{small}(CO))$ and $n_\mathrm{g}(CO)/(n_\mathrm{g}(CO)+n_\mathrm{large}(CO))$ equal 0.5. On the other hand, the total snowline (white line) is the location where

\begin{equation}
\label{eq:snowline} 
	\frac{n_\mathrm{g}(CO)}{n_\mathrm{g}(CO)+n_\mathrm{ice_\mathrm{tot}}(CO)} = 0.5,
\end{equation}

\noindent where 

\begin{equation}
\label{eq:snowline2} 
	n_\mathrm{ice_\mathrm{tot}}(CO) = \sum_{bin = 1}^{N} n_\mathrm{ice_\mathrm{bin}}(CO).
\end{equation}

\noindent In the case of Model D, the number of grain bins $N = 2$ and Eq.~\ref{eq:snowline2} writes $n_\mathrm{ice_\mathrm{tot}}(CO) = n_\mathrm{small}(CO) + n_\mathrm{large}(CO)$.

While Model S and Model C exhibit an expected structure of the snowline, the shape of the snowline in Model D is more complex. This is discussed in more details in Sect.~\ref{sec:disc:snowline}.

\section{Discussion}\label{sec:discussion}

 \subsection{Thermal decoupling and the 'Bump effect'}
 \subsubsection{Thermal decoupling in the disk midplane}\label{sec:disc:decoupling}
\begin{figure} 
\centering
\includegraphics[width=1.1\linewidth]{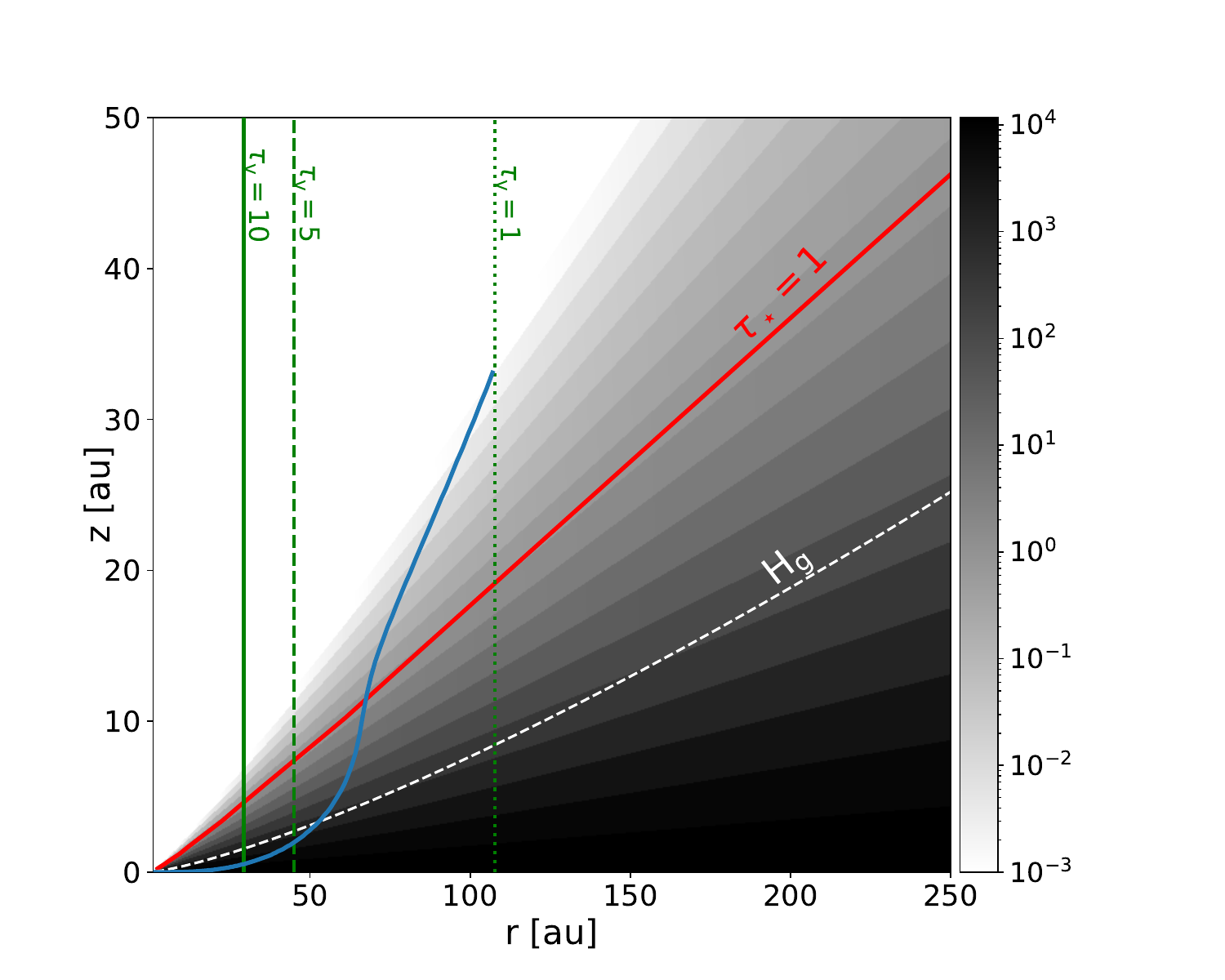}
\caption{2D map of the disk. The grey scale represent the values of the optical depth computed from the star to all points in the disk $\tau_\star(r,z)$. The red lines represent the surface where $\tau_\star(r,z)$ = 1. The green lines mark the location where the total vertical optical depth at the midplane becomes smaller than 10, 5, and 1. The solid blue line marks the altitude at which the optical depth in the z-direction toward the upper layers becomes 1. The line diverges to infinity beyond the line $\tau_v$ = 1. All opacities are given at 1 $\mu$m. \label{fig:taustar}}
\end{figure}

Figure\,\ref{fig:taustar} shows the 2D map (grey scale) of the optical depth at 1 $\mu$m in Model D computed from the star to all points in the disk $\tau_\star(r,z)$. We see that the optical depth values located below one scale height $z <$ H$_\mathrm{g}$ (white dashed curve) are very large ($\tau_\star$ >> 10$^2$) such that the disk midplane is always optically thick in regards to the star. The two dust populations are therefore expected to be strongly thermally coupled: the warmer population is cooled by the other one, and vice versa. However, outside 30~au, the grains become thermally decoupled (the small grains become warmer), even though the optical depth remains very large.

To take a closer look, the blue line marks the altitude where the vertical optical depth calculated from the midplane equals 1. This way we can investigate from which altitude radiation contributes to the heating of the midplane. We see that at small radii ($r$ < 30 au), the radiation reaching the midplane is exclusively due to thermal emission from nearby dust with very similar temperature, hence the temperature coupling. At larger radii ($r$ > 30 au), we see that the altitude of the blue line suddenly increases to reach the uppermost layers of the disk at around $r$ = 100 au. On the other hand, the red lines corresponds to the surface where the optical depth from the star $\tau_\star$ equals to 1. The layers above that surface are therefore directly heated by the star. At the coordinates ($r$ = 70 au, $z$ = 11 au), the blue line crosses the red line, meaning that the optical depth from the star to this point and from there to the midplane equals 2, a value much smaller than the optical depth calculated from the star through the midplane. The midplane can thus be effectively heated by stellar light coming from the disk surface. 

\begin{figure} 
\centering
\includegraphics[width=1.05\linewidth]{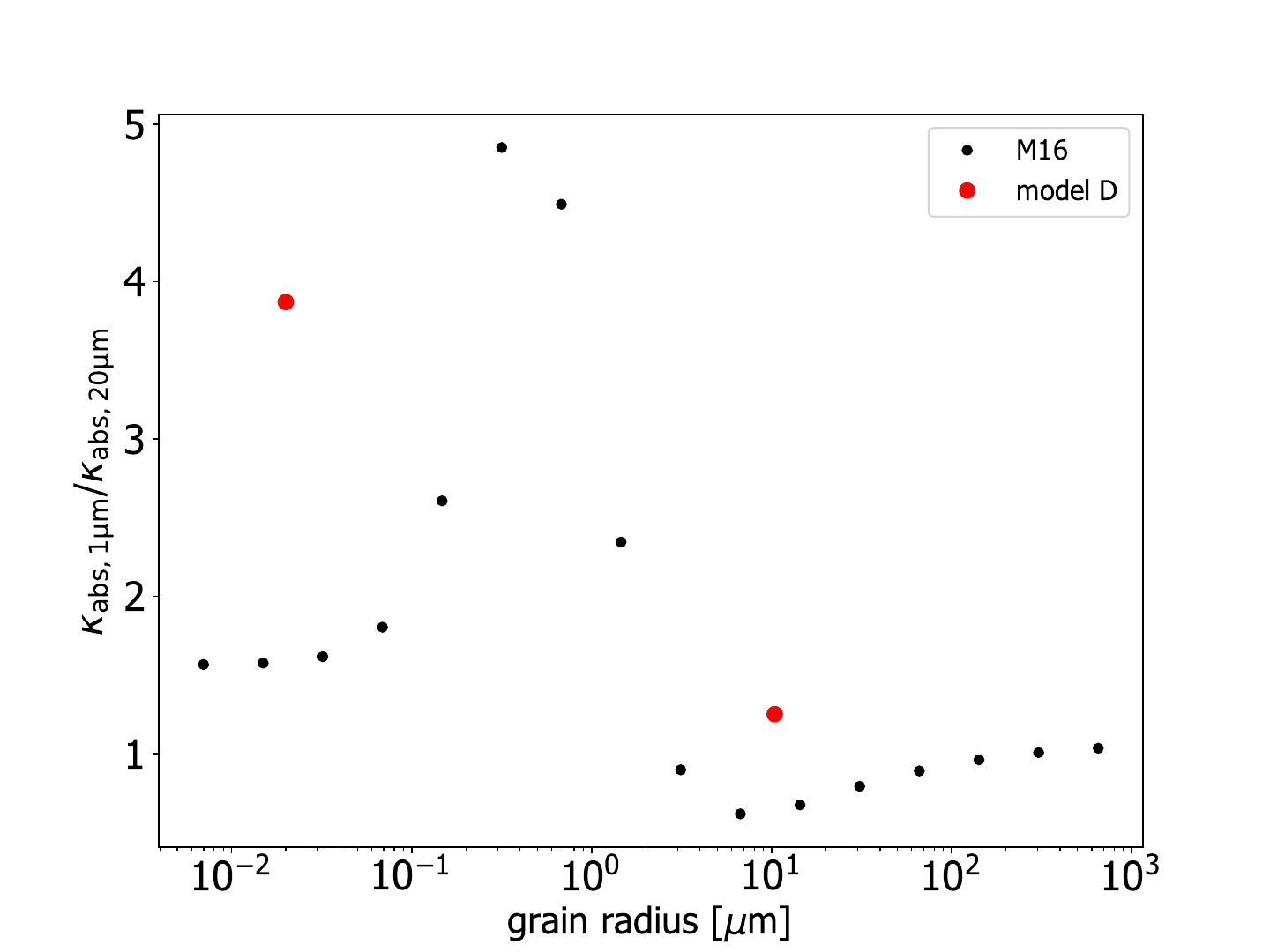}
\caption{ratios of the absorption cross section
for wavelengths at which the star radiates ($\sim$ 1 $\mu$m) and
the absorption cross section for wavelengths at which the dust
emits ($\sim$ 20 $\mu$m) for the grains in Model D (red points) and M16 (black points). \label{fig:ratio}}
\end{figure}

However, since each grain has a different absorption/emission opacity, they are not heated with the same efficiency. Figure\,\ref{fig:ratio} shows the ratios of the absorption cross section for wavelengths at which the star typically radiates to the absorption cross section for wavelengths at which the dust typically emits ($\sim$ 20 $\mu$m) as a function of the grain size. In Model D, the small population has a ratio close to 4 whereas the large population has a ratio close to 1.2. The starlight at 1 $\mu$m is thus more effective in heating the small grains than the large grains. Likewise, in Model M16, the intermediate grains have the largest ratio values, so the heating is maximum for these grains.
 
\subsubsection{The 'Bump effect'}\label{sec:disc:bump}

On top of that, we described a surprising effect where the temperature re-increases, reaches a maximum and then monotonously decreases again. This is what we called the bump effect.
 
In Sec.\,\ref{sec:res:thermal} we demonstrated that scattering is at the origin of the bump effect in the dust temperatures.

To illustrate the importance of the contribution of scattering of stellar radiation, we indicated the wavelength $\lambda_\mathrm{max}$ for the temperature at which the star emits according to Wien's law in Fig.\,\ref{fig:cabs16}.

In Model M16, we saw that the bump effect was caused by the scattering of stellar radiation by the intermediate grains where a fraction of the starlight intercepted in the upper layers is scattered vertically toward the midplane. In the case of Model D, this is the small grains that scatter the starlight vertically (we note that the large grains are strongly settled so they cannot effectively intercept the starlight in the upper layers. They can, however, interact with the starlight already scattered by the small dust population in the intermediate layers). 

More generally, the crucial combination for a bump to occur is the simultaneous presence of at least one scattering dust grain population with a large scattering albedo accompanied with the presence of at least one other grain population with a large absorption to emission opacity ratio.

In Model S, no bump effect is visible, even in the optically thin regions. This comes obvious given the mechanisms needed to generate a bump, as we saw that at least two dust species with different opacities are needed, so that the various dust populations can have different temperatures in regions where temperature coupling is not relevant. Therefore, a bump effect cannot occur in a dust model using a single opacity profile, regardless of how sophisticated the model is.

\subsubsection{Sensitivity to other parameters}
Thermal decoupling and the bump effect originate from different mechanisms. While thermal decoupling always occurs in optically thin environments and does not require radiative interaction between different grains, a bump effect is generated on certain grains by the scattered light coming from other grains in their vicinity. It is therefore sensitive to the chosen optical properties and in particular to parameters that affect the albedo: the material composition, the size range, etc. To illustrate this, we performed again dust continuum radiative transfer simulations for Model D but this time by using the DIANA standard opacities in Appendix\,\ref{app:diana}, and by using another size distribution in Appendix\,\ref{app:size}. We also tested the sensitivity to different disk masses (Appendix\,\ref{app:mass}) and different turbulence parameters (Appendix\,\ref{app:turbulence}). As expected, while the magnitude of the bump is sensitive to the opacity properties,
a temperature dichotomy exists in all cases, though at different radii.

Finally, the reader is referred to Appendix\,\ref{app:number} where we address the question of the possible numerical origin of the bump effect arising in optically thick regions.

\subsection{Snowline shape in Model D} \label{sec:disc:snowline}

\begin{figure*}
\begin{subfigure}{.5\linewidth}
  \centering
  \includegraphics[width=1.0\linewidth]{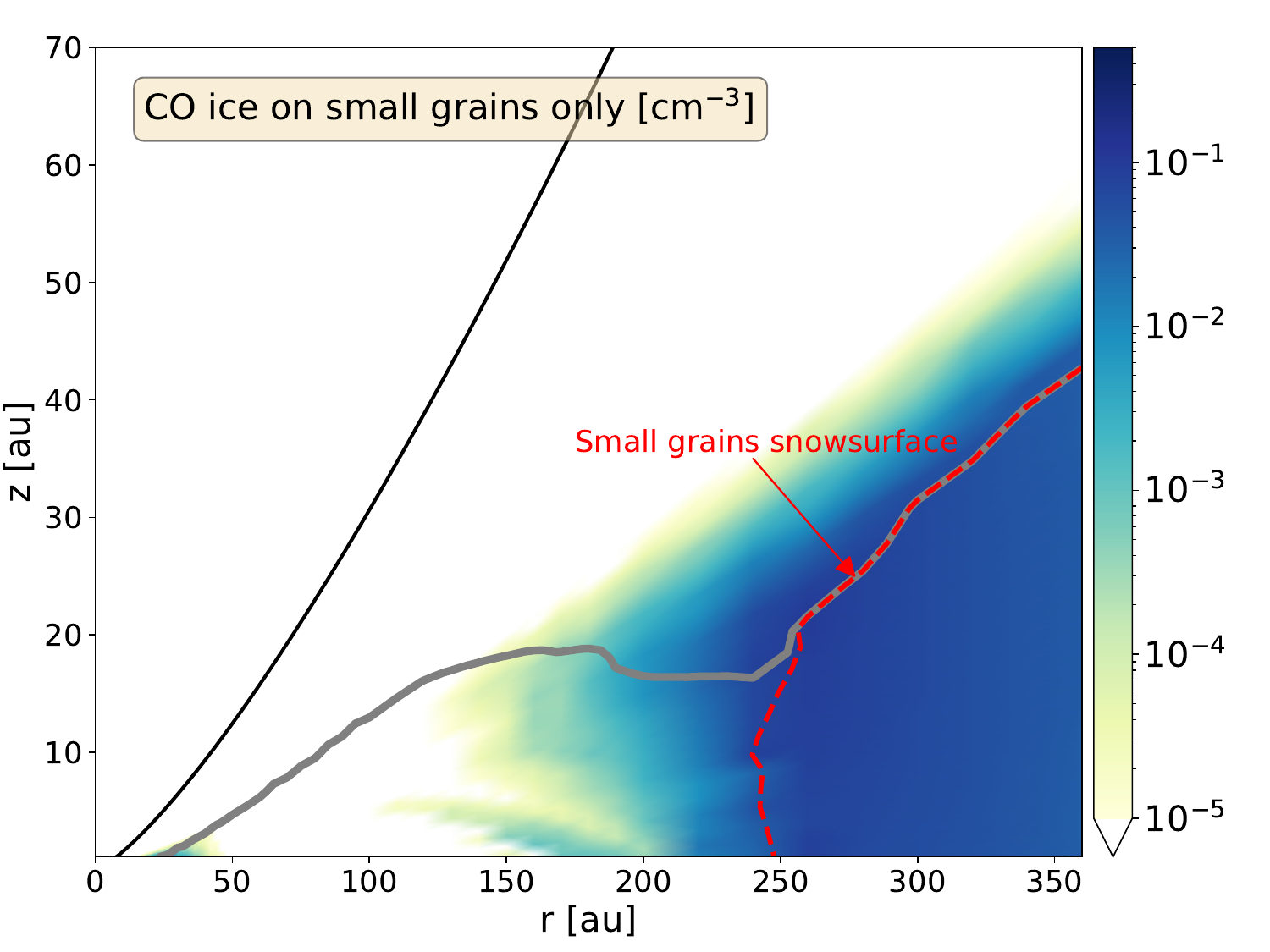}
  \caption{CO ice on small grains \label{fig:coice-small}}
\end{subfigure}
\begin{subfigure}{.5\linewidth}
  \centering
  \includegraphics[width=1.0\linewidth]{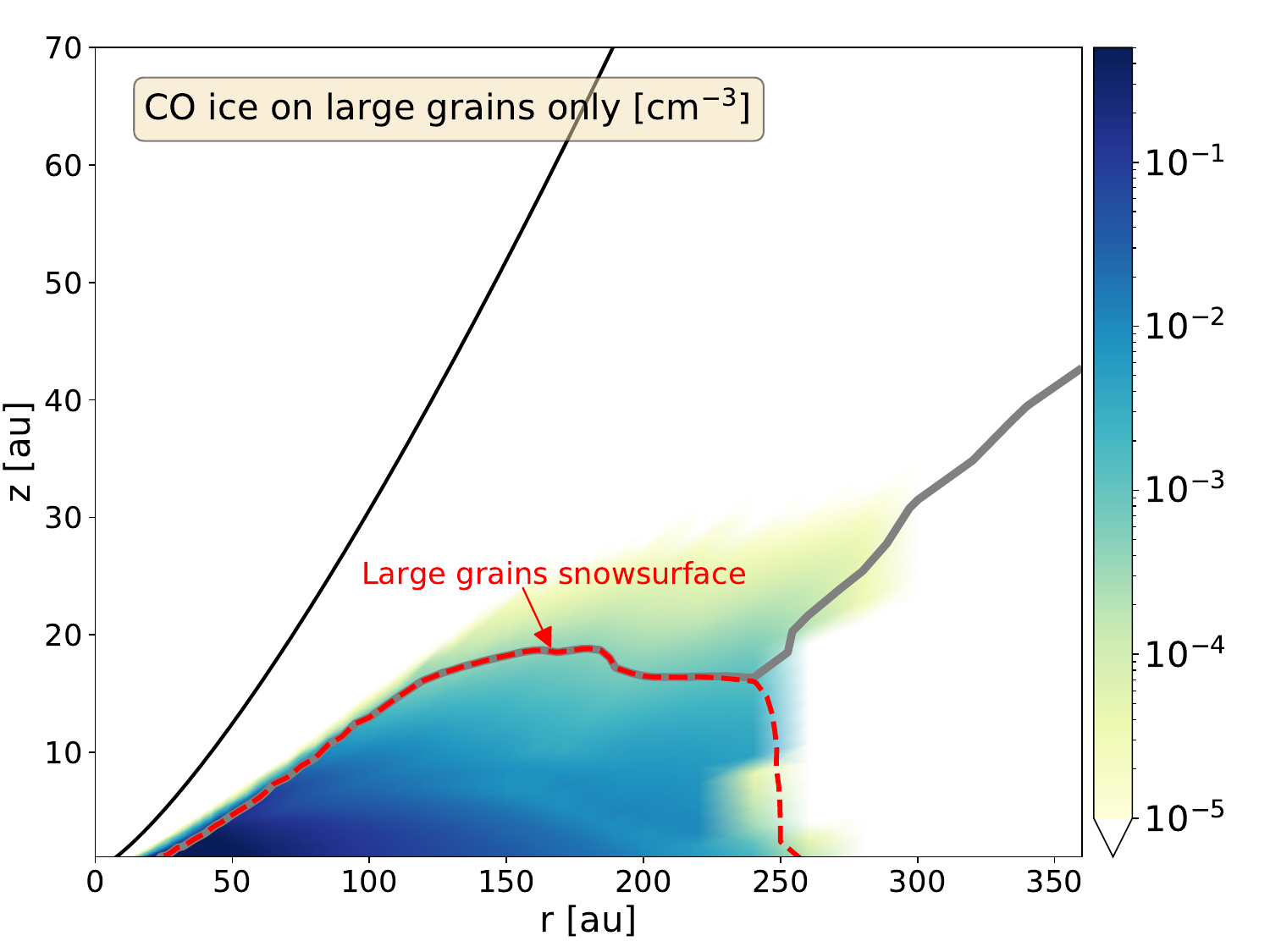}
  \caption{CO ice on large grains\label{fig:coice-large}}
\end{subfigure}
\caption{2D maps of CO ice number density [cm$^{-3}$] in Model D. Left: CO ice adsorbed onto the small dust population only. Right: CO ice adsorbed onto the large dust population only. The solid grey line corresponds to the total CO snowline. The dashed red lines correspond to the snowline of each population independently. The black line is the maximum scale height below which chemistry is computed.}
\label{fig:coice_independent}
\end{figure*} 

To understand the snowline shape requires to first consider the separate contributions of the two grain populations of Model D.

Figure~\ref{fig:coice_independent} shows the CO ice on each population in Model D. Figure~\ref{fig:coice_independent}a represents the distribution of CO adsorbed onto the small grains only whereas Fig.~\ref{fig:coice_independent}b represents CO ice adsorbed onto the large grains only. The grey line is the total snowline (Eq.~\ref{eq:snowline}) whereas the red dashed lines are the snowlines of the individual populations.

In the midplane, the CO snowline for the small dust population is located close to 250 au where the dust temperature has sufficiently decreased. Inside 250 au, CO ice on the small grains is virtually non-existent because the temperature is above the freeze-out threshold for this population. For the large grains, the CO snowline in the midplane is located at a much inner radius ($\sim$ 25 au) where the dust temperature drops below the freeze-out temperature for this population. We can straightforwardly conclude then that only the large dust population is at the origin of the CO gas hole visible in Fig.\,\ref{fig:nCOd}.

On the other hand, we see that the large grain population also exhibits another midplane snowline (near  $\sim$ 250 au, also  where the midplane snowline of the small grains is located), so that the overall snowline of the large grains forms a closed 'bubble'. Beyond that bubble, we can notice that CO is virtually non-existent on the large grains (Fig.~\ref{fig:coice_independent}b), even though their surface temperature is low enough for CO to adsorb. This behavior can be explained by multiple factors, but the most crucial one is the competition of interactions with CO molecules between the large and small grains. Indeed, the dust surface-areas (and therefore the number of accessible sites) is never the same between the two populations. For instance, at 250 au, the ratio between the small grain surface-area to the large grain surface-area is $n_\mathrm{small}(CO) a_\mathrm{small}^2 / n_\mathrm{large}(CO) a_\mathrm{large}^2 \sim 6.5$, meaning that a CO molecule has a higher probability (about 6.5 times higher) to interact with a small grain than a large grain. Consequently, Once the small grains are below the freeze-out temperature, they retain most of CO at the expense of the large grains. Hence, the sharp delimitation between the two snowlines.

The rise in CO gas observed at the delimitation (Fig.\,\ref{fig:nCOd}) is also a consequence of this competition. Between $r=150$\,au to $r=250$\,au, the desorption to adsorption rate ratio progressively increases, and reaches the value 0.5 close to 250 au (for the small grains). At that location, CO interacts more with the small grains, but the latter still being slightly too warm, is returned to the gas phase ultimately. In other words, the freeze-out timescale around 250 au is longer than elsewhere, and the snowline assumes this peculiar S-shape with a minimum height at 250 au.

Note the presence of a small isolated 'island' of CO ice on small grains (Fig.~\ref{fig:coice-small}) at $\sim$ 30 au around the midplane, which corresponds to the location of the temperature dip, where the large grains have cooled enough the smaller grains. This happens because of the specific dust mass and structure of the disk. Should the disk be slightly less massive, this small island would likely not exist.

Lastly, with two grain populations and therefore two temperatures, Model D is an efficient simplification of the dust distribution to understand the impact of a proper dust temperature onto the chemistry and in particular the behavior of snowlines. Nevertheless, a more realistic distribution with more grain populations and more related temperature would necessarily produce more "individual" (one per grain population and temperature) snowlines, making the resulting CO snowline even more complex. Although the resulting CO ice segregation should be hard to probe in observed protoplanetary disks, it requires an interpretation on its own that may have significant implications for the chemical evolution and planet formation.

\subsection{Radial drift and planet formation} \label{sec:disc:drift}
Our model does not include any radial drift so the CO snowline shape is
solely due to the different dust temperatures, though its shape and precise location is also
affected by dust settling.

\citet{Cleeves_2016} has shown, using a simple parametric model, that radial drift can reshape the CO distribution and create multiple snowlines. In particular, the removal of a large fraction of dust grains from the outer disk can allow the region to become warmer by reprocessed stellar radiation from the upper layers. We showed that such reprocessing can occur even without a significant change in dust-to-gas mass ratio and only requires the presence of different dust populations. We can therefore expect that the combination of the two mechanisms (i.e., temperature spread and radial drift) should build up the radial segregation of the snowlines. 

Radial drift occurs when gas drag acts on the dust grains (the gas disk 'feels' its own pressure and is therefore sub-Keplerian). As a consequence, the dust grains drift inward. The drift efficiency increases with the Stokes number $\mathrm{St} < 1$ under Epstein drag law \citep{Birnstiel+etal_2012}. If, for instance, we consider the grain distribution in Model M16, the large grains will drift more efficiently than the intermediate grains, whereas the small grains remain mostly coupled to the gas and will barely drift. Although the drift time is expected to be relatively large for the whole grain size range considered ($\mathrm{St} < 10^{-2}$ for all grains of size $<$ 1 mm at 10 au in our model), 
radial size-dependent segregation can occur with enough orbital periods. We can make an estimate of the radial segregation between all grain populations by considering the following drift timescale equation \citep{Birnstiel+etal_2012}, 

\begin{equation}
\label{eq:drift}
        \tau_\mathrm{drift} = \frac{r V_\mathrm{k}}{\mathrm{St} c_\mathrm{s}^2}\gamma^{-1}  
\end{equation}

\noindent where $V_\mathrm{k}$ is the Keplerian velocity at radius $r$, $c_\mathrm{s}$ is the isothermal sound speed, and $\gamma = |-p - \frac{q}{2} - \frac{3}{2}|$ is the absolute value of the power-law index of the gas pressure. The drift timescales at 100 au of all 16 grains are given in the fourth column of Table\,\ref{tab:16sizes}. These timescales can be compared with our chemical evolution time $t_\mathrm{chem}$ = 2~Myr. We can see that all grains larger than about 10~$\mu$m  have a drift timescale $< t_\mathrm{chem}$, whereas the small grain populations virtually do not drift  ($\tau_\mathrm{drift}(100$~AU) $>> t_\mathrm{chem}$). We can therefore expect a strong radial size-dependent segregation within the chemical evolution time. The colder and larger grains should be found within a smaller radial extent than the smaller grains, such that the surface-area ratios become even more in favor of the smaller grains at large radii.
 
Accordingly, we expect that the effect found in Model D, which depends on a competition between small and large dust grain populations, will be moved inward by the radial drift of large grains, and one
might expect an even more complex shape for the 2D snowline curve than in our Model D. Likewise, this effect should be enhanced by inward drift in a more realistic distribution like in Model M16. This could have observational implications (see Sect. ~\ref{sec:disc:cocaps}).

Our study ignores grain-grain collisions, which may lead to more homogeneous ice
compositions for grains of different size.  However, grain-grain collisions are a complex process. While it is conceivable that slow velocity collisions between a small grain and a large one effectively results in the ice coverage of the small grain being incorporated in the ice coverage of the larger one,  faster collisions, those which replenish the small
size distribution from the large size one, are unlikely to have such a simple outcome.
Grain growth has been shown by \citet{vanClepper+etal_2022} to affect the chemistry, with an impact on the C/O gas phase ratio as a function of disk age, but their
model did not consider disruptive collisions.  Our chemical model also ignores diffusion, in particular vertical mixing, which can homogeneize the gas phase abundances \citet{Hersant+etal_2009}.

In all cases, however, accounting for the size dependence of dust temperatures
leads to higher temperatures than using a common temperature, so that this process should
be considered when evaluating the chemical evolution.

On another hand, our mechanism also has implications for the chemical evolution during planet formation. Indeed, Model D shows that the small grains are virtually 'naked' from CO ice inside the large grain snowline whereas the large grains are naked from CO inside the small grain snowline. This suggests that the process of CO ice toward complex organic molecules (COMs) takes place only on the large grains in the inner regions where planet formation takes place, whereas it occurs only on the small grains in the outermost regions, which later act as a reservoir of solid particles after they slowly grew toward drifting larger grains \citep{Youdin+Shu_2002, Lambrechts+etal_2014}.

Additional chemistry models which take into account size-dependent radial drift are going to be investigated in future work.  

\subsection{CO holes in observed disks} \label{sec:disc:cocaps}

We found that Model D shows a CO depression 
of up to 3 orders of magnitude in density  and a factor of 10
in surface density, forming a 'hole' between 50 au and 200 au, followed by a rise beyond. Such behavior is reported in the case of the observations of edge-on disks. 

\citet{Dutrey+etal_2017} analyzed the temperature structure of the edge-on Class II disk 2MASS J16281370-2431391 (the 'Flying Saucer'). They found a CO and CS temperature drops inside 200 au followed by a rise between 200 and 300 au. They argued that an effective rise of UV penetration can explain the heat up of the disk in the outer region (200 au being the outer radius of the 'mm' dust disk).
Similarly, \citet{Flores+etal_2021, Villenave+etal_2022} reported in the  edge-on Class II disk around SSTC2D J163131.2-242627 (Oph 16313) a rise in the CO temperature in the outermost region. They also argued that an external UV field can heat the outer disk where UV photons can penetrate more effectively.

More recently, \citet{Lin+edisk_2023} detected a similar behavior in both the $^{12}$CO and $^{13}$CO emissions in the close-to edge-on Class I disk IRAS 04302+2247 \citep[the 'Butterfly'][]{Wolf+etal_2003} which closes off the freeze-out region, forming a hole. However, contrary to the two other disks, the rise does not coincide with the outer radius of millimeter-continuum emission. Furthermore, \citet{Grafe+etal_2013} showed that in this disk the population of large grains has drifted inward whereas the modeled scattered light indicates that the small grain distribution has a larger radial extent invisible to mm-wavelength.  

Hence, in these objects, the rise of CO appears beyond a more or less important drop of the dust-to-gas ratio (although it is less obvious in the case of the younger Butterfly star where the dust is less evolved). Such a behavior can be qualitatively explained by \citet{Cleeves_2016}, where a CO rise occurs naturally because the dust-to-gas ratio drops quickly beyond the 'mm' dust disk edge allowing for a raise of temperature.

The \citet{Cleeves_2016} model requires an effective change in the (small grain) dust-to-gas ratio, due to radial drift. On the contrary, our model
shows that even for a radially relatively constant value of this ratio,
the temperature difference between small and large grains can result
in a similar feature. This will be investigated in a future work.

Finally, Fig.~\ref{fig:surfdens} shows that the CO surface density, although of similar amplitude, clearly differs for Model D compared to the others. Model D, with two-grain temperatures, shows a dip between 70 and 230 au which reflects the CO depression observed in  Fig.\,\ref{fig:coice_independent}. Observing a disk with a moderate inclination, this feature can apparently mimic an unresolved gap which may be then erroneously attributed to planet formation.

In any case, we note that the observation of the CO snowline shape is only feasible for disks seen at high inclination, ideally close to edge-on. In less inclined disks, the substantial CO opacity from the top layer will hide the empty region behind the upper layers.
The surface density effect shown in Fig.\,\ref{fig:surfdens} should however be visible using isotopologues,
principally $^{13}$CO.

Other manifestations of the temperature gradient may be found with other molecular probes. The differences in temperature and timescales will also affect the surface chemistry because of the temperature dependence of the atoms and molecule mobility on the dust surfaces, and thus, in return, the chemical composition of the gas phase.
Grids of models may be required to quantitatively evaluate the impact and possibly find out more sensitive indicators of this behavior than the rotational lines of CO.

\section{Summary}\label{sec:summary}
We tested the impact of polydispersion in a medium composed of multiple independent grain populations on the resulting dust temperatures and chemical evolution in protoplanetary disk models. More specifically, we compared disk models using 1, 2, and a more realistic 16 individual grain populations with size-dependent vertical settling, densities, and optical properties and showed the effect on the CO distribution in the disks. The main results can be summarized as follows:

\begin{itemize}

\item The presence of multiple dust populations with different optical properties generates a systematic dust temperature spread, where the smaller grains are always warmer than the larger grains, in regions of the disk where dust can be thermally decoupled. This is true even in the midplane of massive disks. Interestingly, the temperature of micron-sized grains can show a radial bump: it slightly re-increases with the radial distance due to stellar scattered light coming from the upper layers when the vertical optical depth decreases.

\item Our work shows that opacities derived from a size distribution are not well-suited to reproduce the temperature spread that exists in a dust population composed of multiple sizes because these opacities dilute the effects of polydispersion, regardless of how sophisticated the opacity models are. In particular, they underestimate the temperature of the small grains. This makes impossible the choice of a 'fit-all' dust opacity model to compute a realistic dust temperature. However, given the computational limitations of a dust opacity model using single-grain size opacities like Model M16, we argue that a satisfactory compromise is to use two (maybe three) independent size-averaged dust opacities to approximate the dust thermal spread of a realistic dust grain population, providing a good choice of composition and size range.

\item The grains of different sizes, having different temperatures and different coupling timescales to the gas phase, do not interact in the same way with the gas phase species, in particular when CO condensation/evaporation balance is considered.

\item The multiple temperatures significantly affects the CO gas phase distribution. Our model using two dust populations shows the presence of a CO depletion from 50 au up to 250 au, followed by a CO gas rise at larger radii. This rise does not require additional external UV irradiation nor a change of dust-to-gas mass ratio. The effect cannot be reproduced in chemical models that use a single dust temperature structure.

\item The CO ice distribution is also affected and the snowline is reshaped. The dust temperature spread creates multiple CO condensation fronts, inducing a radial snowline splitting. This does not require radial drift, although inward drift will affect the snowline even more. A more physical model would need a larger number of grain sizes and related temperature, implying a more complex 2D CO snowline.

\item These results have potential implications for observations and planet formation. On one hand, the presence of CO holes followed by a rise in observed Class I/II disks has been recently reported. We speculate that the origin of these CO holes could be linked to the combination of dust temperature spreads and size-dependent dust radial segregation. On another hand, the CO snowline splitting has implications on the chemical evolution during planet formation. Indeed, the formation of COMs via CO surface processes is in turn also dust-size segregated such that it affects the way COMs are brought toward planetary cores.

\end{itemize}

Overall, this work illustrates the necessity for modelers to consider simultaneously multiple dust grain populations with independent optical properties, both for the thermal and chemical simulations.

\begin{acknowledgements}
We thank the referee for their valuable feedback. S.G., J.K.J, and R.S acknowledge support from the Independent Research Fund Denmark (grant No. 0135-00123B).
This work was supported by 
A. Dutrey and S. Guilloteau thank the French CNRS programs PNP, PNPS and PCMI.
J. Kobus and S. Wolf acknowledge support from the DFG grant WO 857/19-1.
\end{acknowledgements}

\bibliography{ref}

\bibliographystyle{bibtex/aa}

\begin{appendix} 

\section{Sensitivity to different physical parameters}\label{app:sensitivity}

\subsection{Grain properties and dust opacities}\label{app:diana}

\begin{figure*}
\begin{subfigure}{.33\linewidth}
  \centering
  \includegraphics[width=1.0\linewidth]{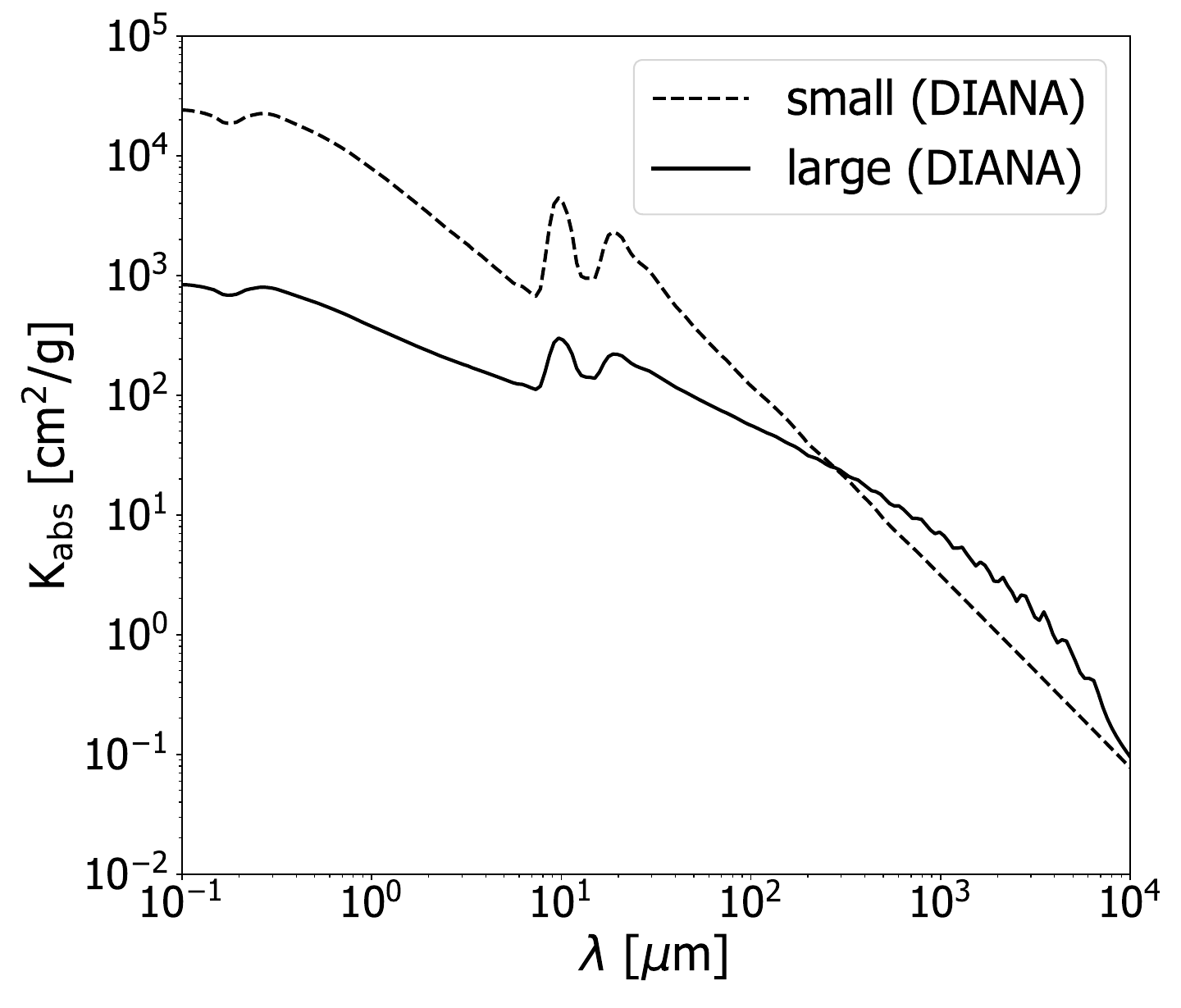}
\end{subfigure}
\begin{subfigure}{.33\linewidth}
  \centering
  \includegraphics[width=1.0\linewidth]{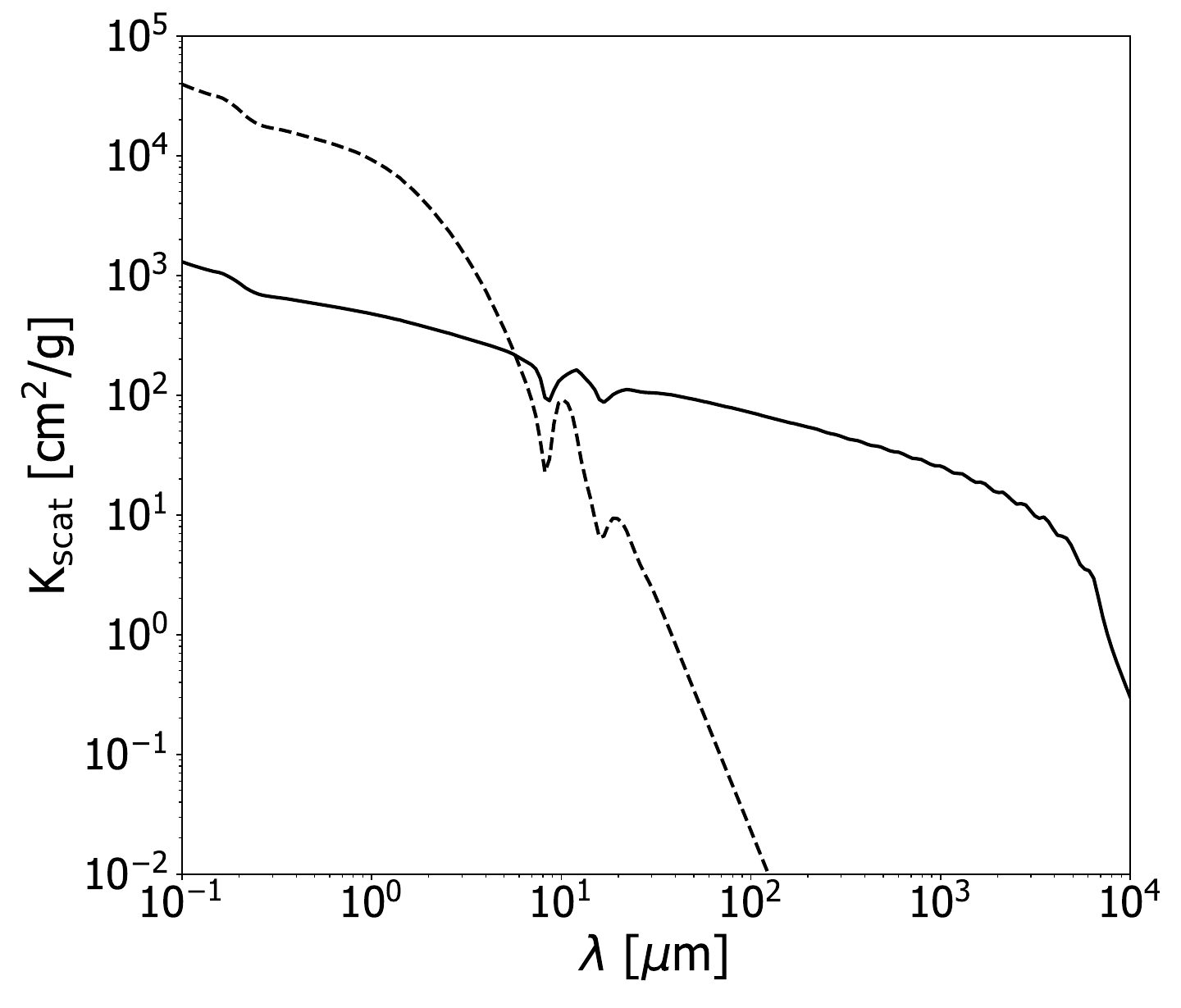}
\end{subfigure}
\begin{subfigure}{.33\linewidth}
  \centering
  \includegraphics[width=1.0\linewidth]{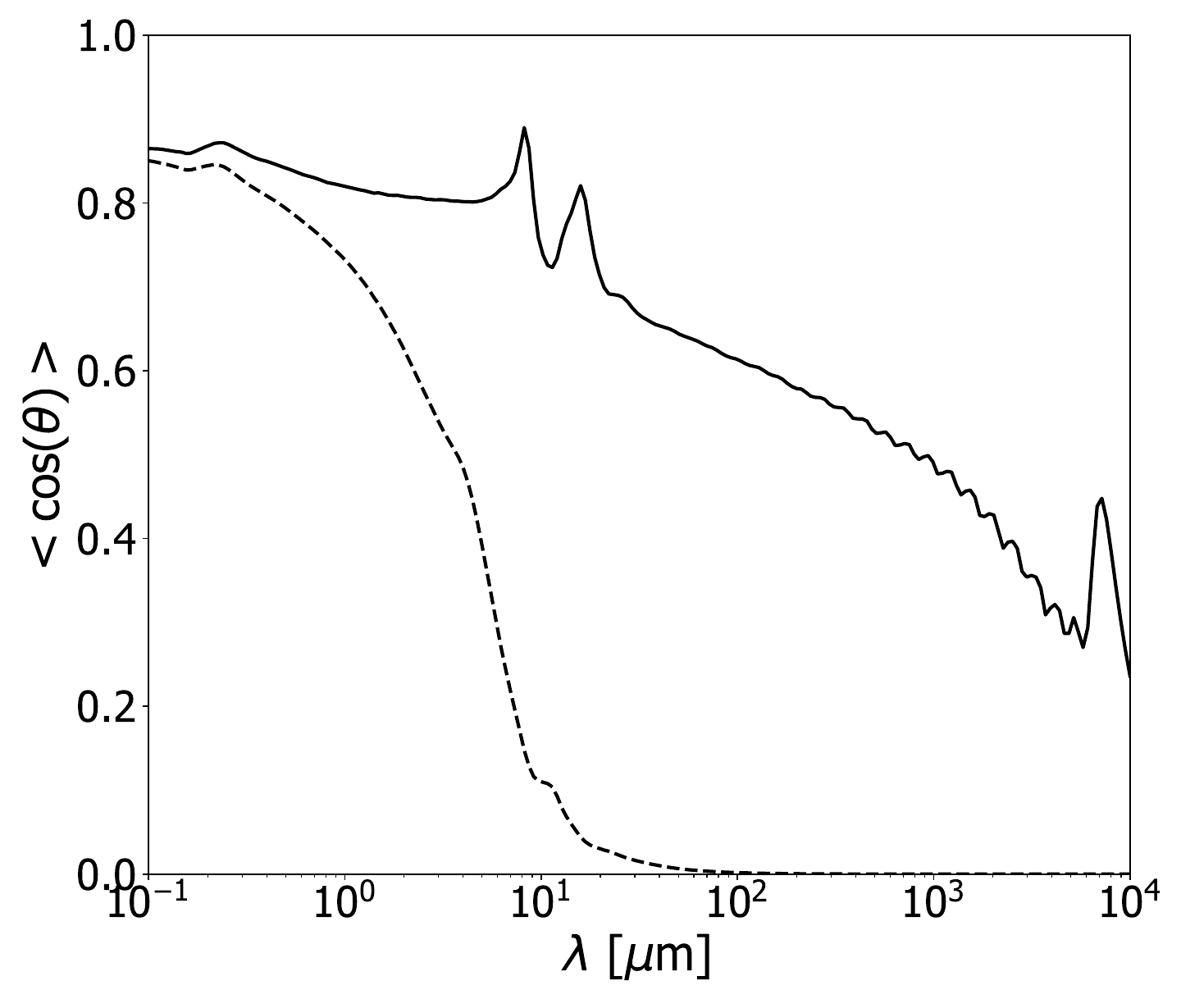}
\end{subfigure}
\caption{Dust optical properties from DIANA opacities for Model D. Left: absorption coefficients. Middle: scattering coefficients. Right: Henyey-Greenstein g parameter of anisotropy (g = $<$cos($\theta$)$>$). }
\label{fig:compare-opa-diana}
\end{figure*}

\begin{figure} 
\centering
\includegraphics[width=1.05\linewidth]{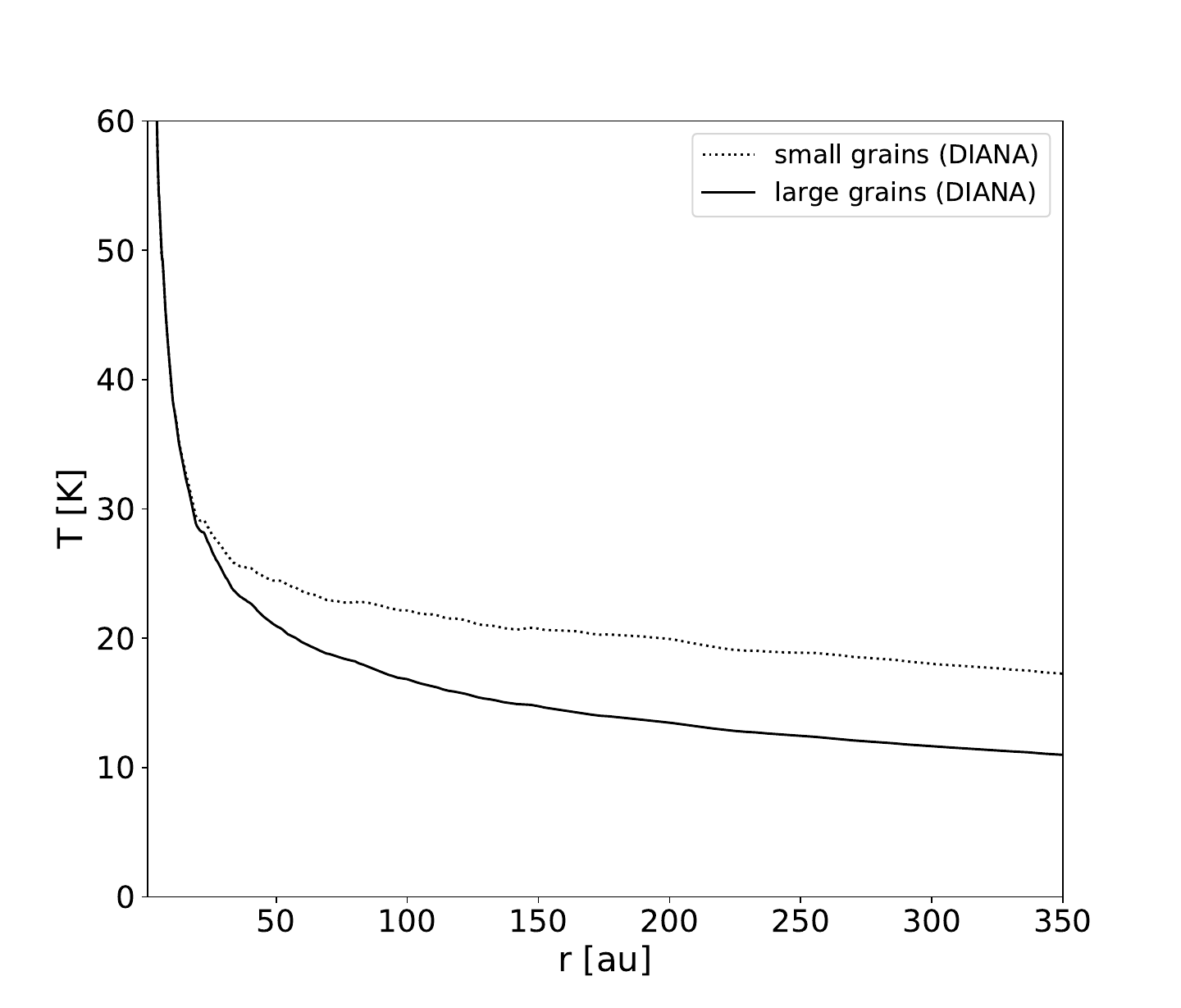}
\caption{Midplane radial dust temperature profiles of Model D with the DIANA standard opacities. \label{fig:mid_D_diana}}
\end{figure}

Here, we utilize another dust opacity model in order to check the impact of grain properties onto the thermal disk structure of Model D. We choose the standard DIANA opacities \citep{Woitke+etal_2016} in the \textit{optool} code \citep{optool} and use the same dust size and wavelength distribution as described in Sec.\,\ref{sec:model:SDC}. The opacity profiles are shown in Fig.\,\ref{fig:compare-opa-diana}. The composition is a mixture of amorphous laboratory silicate \citep{Diana+silicate} with amorphous carbon \citep{Diana+carbon} and 25 \% porosity. The difference with the DSHARP opacities seems minor, but they are not exactly similar. 
Figure\,\ref{fig:mid_D_diana} shows how 
the dust temperatures in the midplane are affected. The bump effect disappears but the small grain temperature profile always remains warmer than that of large grains and reaches 20K at about 250 au. Therefore, the CO chemistry should behave similarly as with the DSHARP opacities. Note that this result holds for Model M16 as well.

\subsection{Size range}\label{app:size}
In all the models shown above, both grain populations have the same minimum size value a$_\mathrm{min}$ = 0.005 $\mu$m, as in many other studies \citep[e.g.,][]{Cleeves_2016, Schwarz+etal_2018, Ballering+etal_2021, Zhang+etal_2021, Calahan+etal_2023}. This choice is partly justified by the fact that in a distribution following a power-law, the opacities are mostly sensitive to a$_\mathrm{max}$ because most of the mass is in the larger grains \citep{Draine_2006, Birnstiel+etal_2018}. 

\begin{figure*}
\begin{subfigure}{.5\linewidth}
  \centering
  \includegraphics[width=1.0\linewidth]{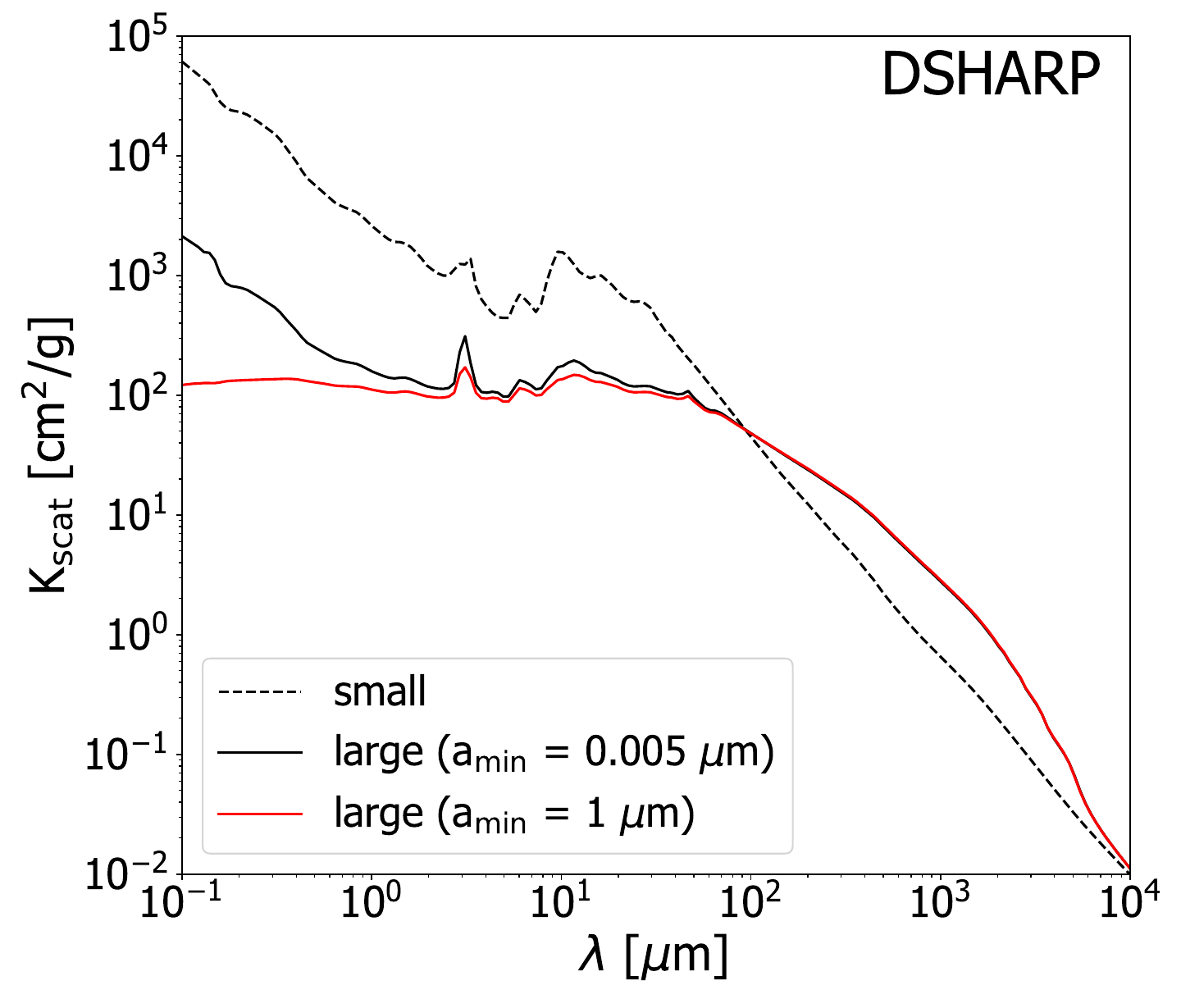}
  \caption{Absorption opacity (DSHARP) \label{fig:scat_dsharp_amin}}
\end{subfigure}
\begin{subfigure}{.5\linewidth}
  \centering
  \includegraphics[width=1.0\linewidth]{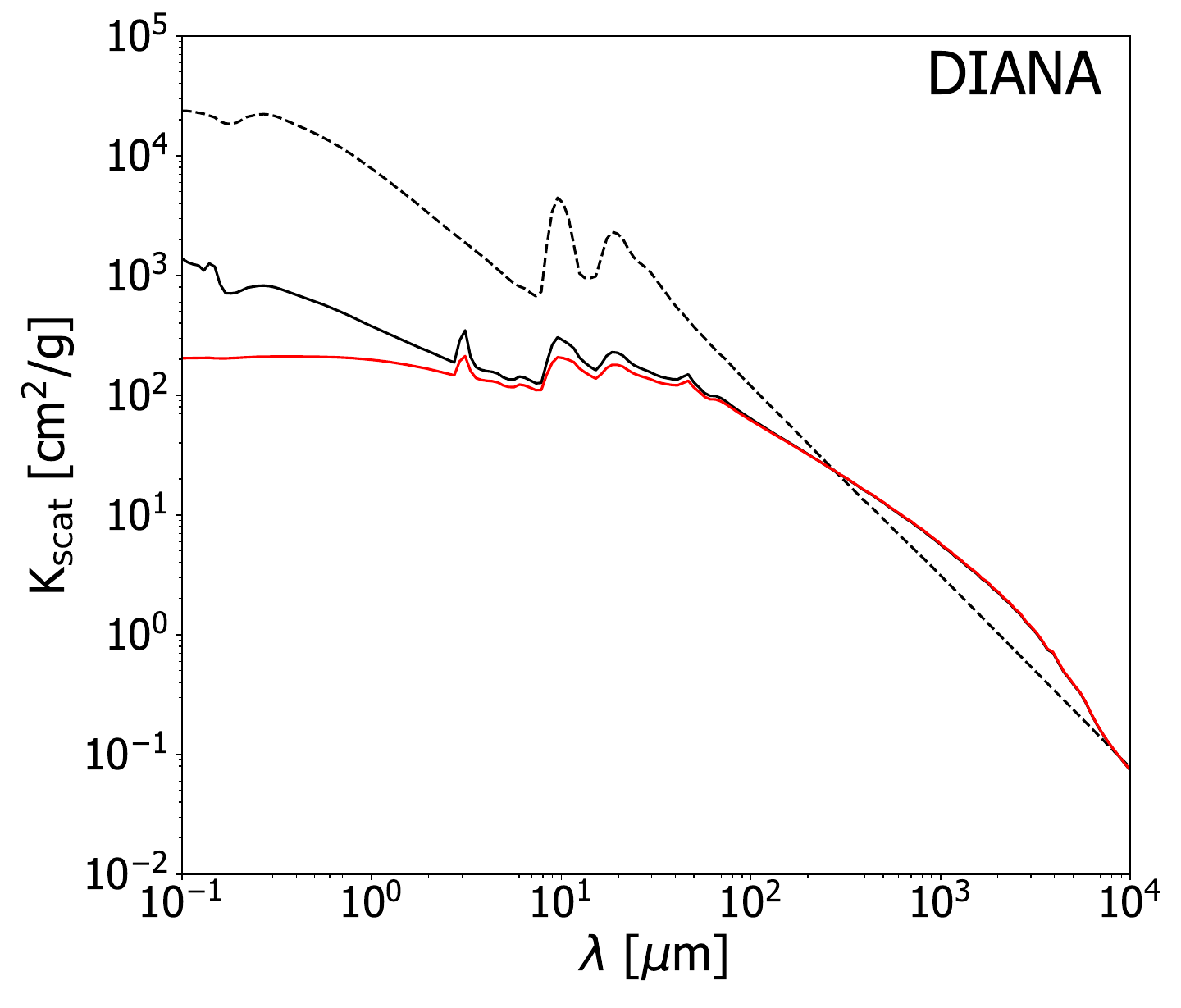}
  \caption{Absorption opacity (DIANA) \label{fig:scat_diana_amin}}
\end{subfigure}
\caption{Absorption opacity profiles from DSHARP (left) and DIANA standard opacities (right) for Model D. The red line represents the opacities of the large grains when a$_\mathrm{min, l}$ is set to 1~$\mu$m. The black lines are the standard choice of a$_\mathrm{min,l}$ = 0.005 $\mu$m.}
\label{fig:opa_amin}
\end{figure*}

\begin{figure*}
\begin{subfigure}{.5\linewidth}
  \centering
  \includegraphics[width=1.0\linewidth]{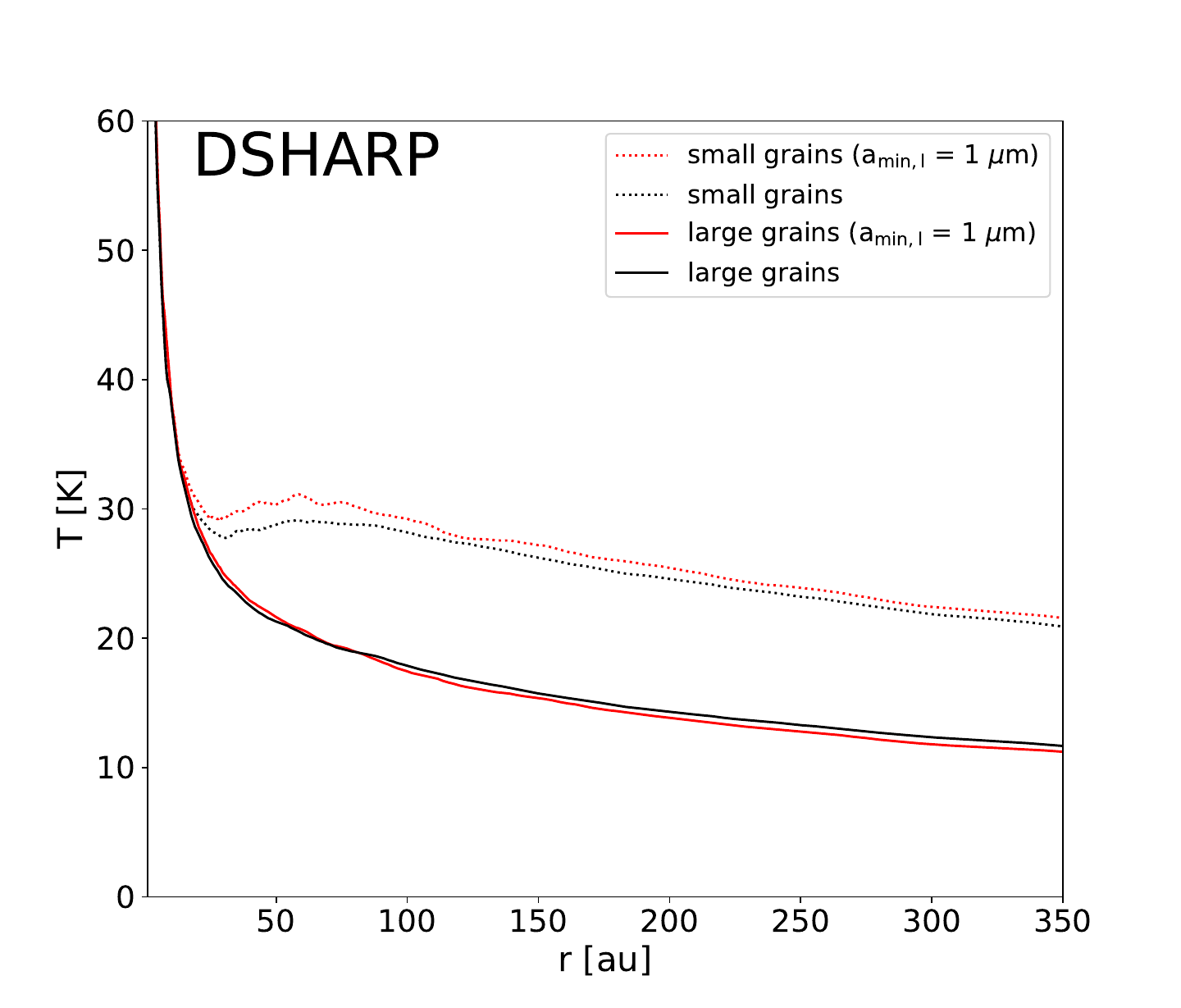}
  \caption{Dust temperature with the DSHARP opacities \label{fig:amina}}
\end{subfigure}
\begin{subfigure}{.5\linewidth}
  \centering
  \includegraphics[width=1.0\linewidth]{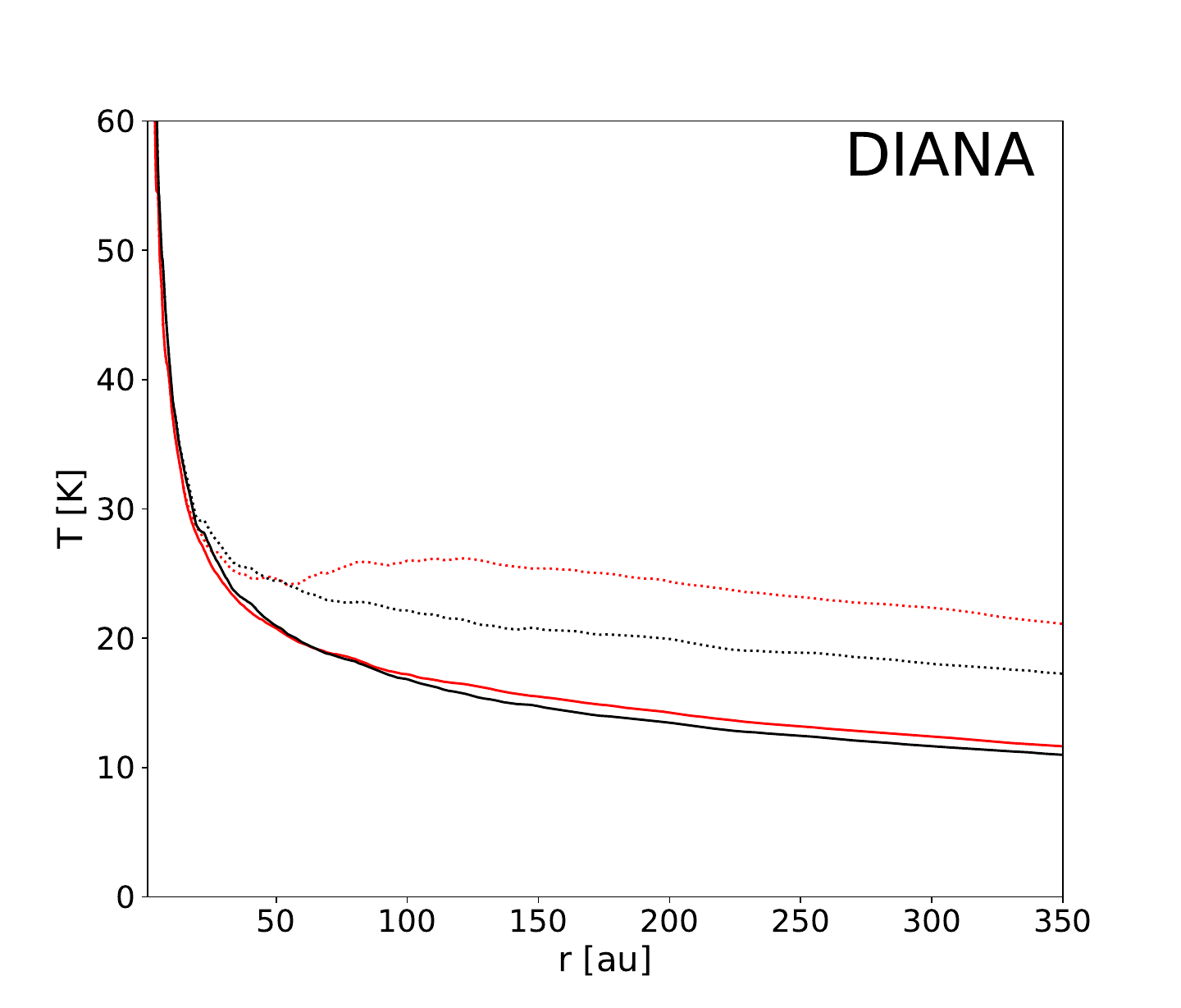}
  \caption{Dust temperature with the DIANA opacities \label{fig:aminb}}
\end{subfigure}
\caption{Disk midplane radial profiles of the dust temperatures. The parameters are that of Model D. The red lines show the temperatures when using a$_\mathrm{min,l}$ = 1 $\mu$m. For comparison, the black lines are the original temperatures using the standard sets (a$_\mathrm{min,l}$ = 0.005 $\mu$m). Left: DSHARP opacities. Right: DIANA standard opacities.}
\label{fig:amin}
\end{figure*}

However, a few other studies adopted the prescription where a$_\mathrm{min, l}$ equals a$_\mathrm{max,s}$ for the calculation of the opacities \citep[e.g.,][]{Du+Bergen_2014}, which should decrease a little the opacities at sub-millimeter wavelengths. Here, we test the effect of this prescription both with DSHARP and DIANA opacities on the thermal structure, that is by using the range [0.005 $\mu$m , 1 $\mu$m] for the small grains and [1 $\mu$m , 1 mm] for the large grains. The resulting absorption opacities are shown by the red lines in Fig.\,\ref{fig:opa_amin}. We see that a larger a$_\mathrm{min, l}$ value decreases the absorption opacities at visible and UV wavelengths by up to an order of magnitude. The resulting dust temperature profiles are shown by the red lines in Fig.\,\ref{fig:amin}. Interestingly, changing the large grain opacities mostly affects the small dust temperature rather than that of the large one, because the scattering albedo of the large grains is increased with the new size range (note that the scattering opacities are also slightly decreased in the same wavelength range but less than the absorption opacities). 

It is generally assumed that a higher albedo decreases the absorbed energy in the disk and therefore decreases the overall dust temperature. This result shows that this is not exactly correct when one considers a dust size distribution: the decrease of energy input for a grain species can imply an increase of energy input for another grain species.

We also notice that while the DIANA standard opacities do not raise a bump when a$_\mathrm{min, l}$ = a$_\mathrm{min,s}$, they do raise a bump and increase
the overall small grain temperatures when choosing a$_\mathrm{min, l}$ = a$_\mathrm{max,s}$. This highlights the importance of a careful selection of the optical properties. In particular, this questions the widely chosen prescription of a$_\mathrm{min, l}$ = a$_\mathrm{min,s}$ since this result shows that the value of a$_\mathrm{min, l}$ can have a significant impact on the dust temperature when polydispersion is considered, even for a dust distribution following a power-law.

\subsection{Disk mass}\label{app:mass}
\begin{figure*}
\begin{subfigure}{.5\linewidth}
  \centering
  \includegraphics[width=1.0\linewidth]{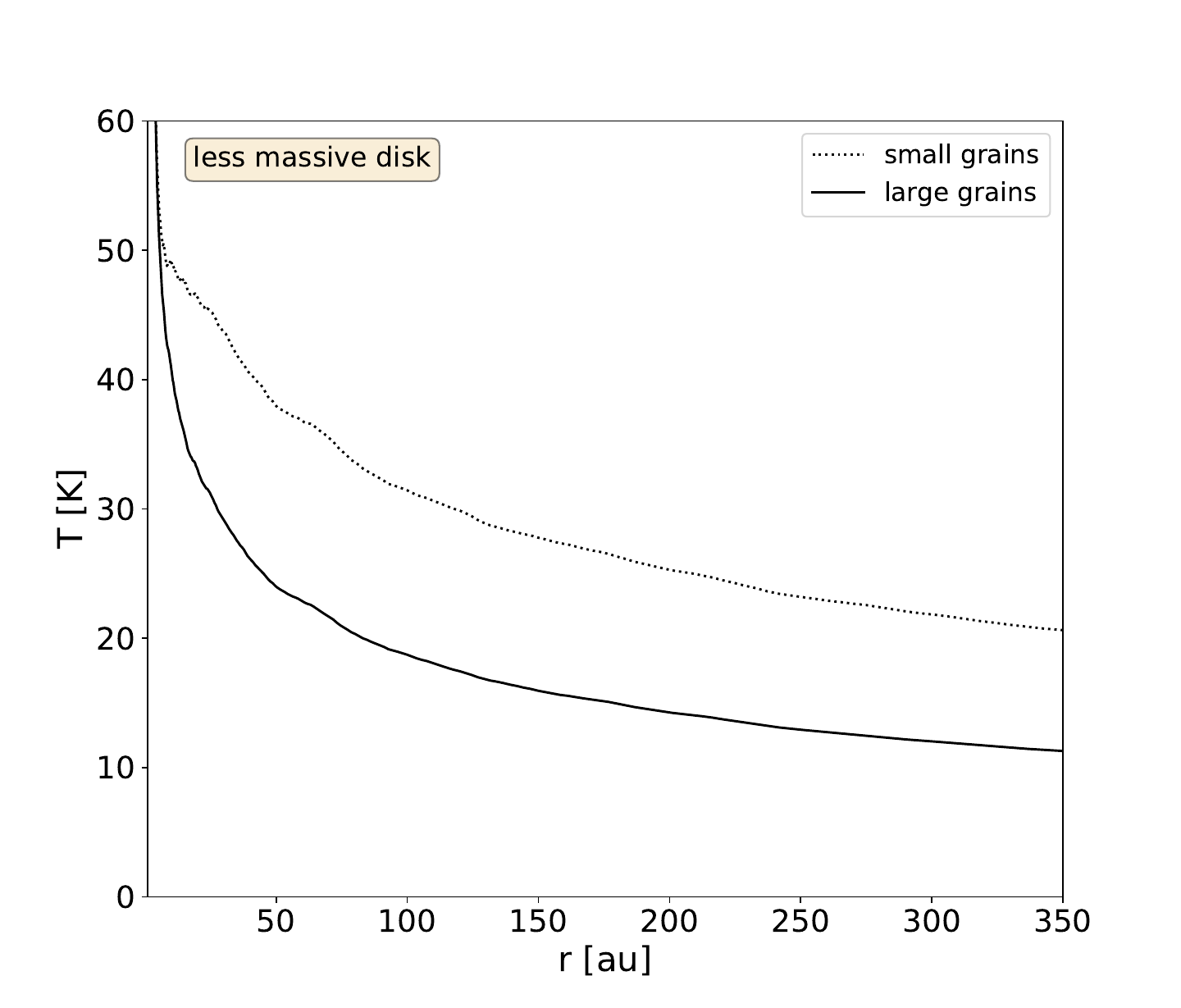}
  \caption{Less massive Model D ($M_\mathrm{disk} = 7.5\times10^{-4} M_\odot$)}
\end{subfigure}
\begin{subfigure}{.5\linewidth}
  \centering
  \includegraphics[width=1.0\linewidth]{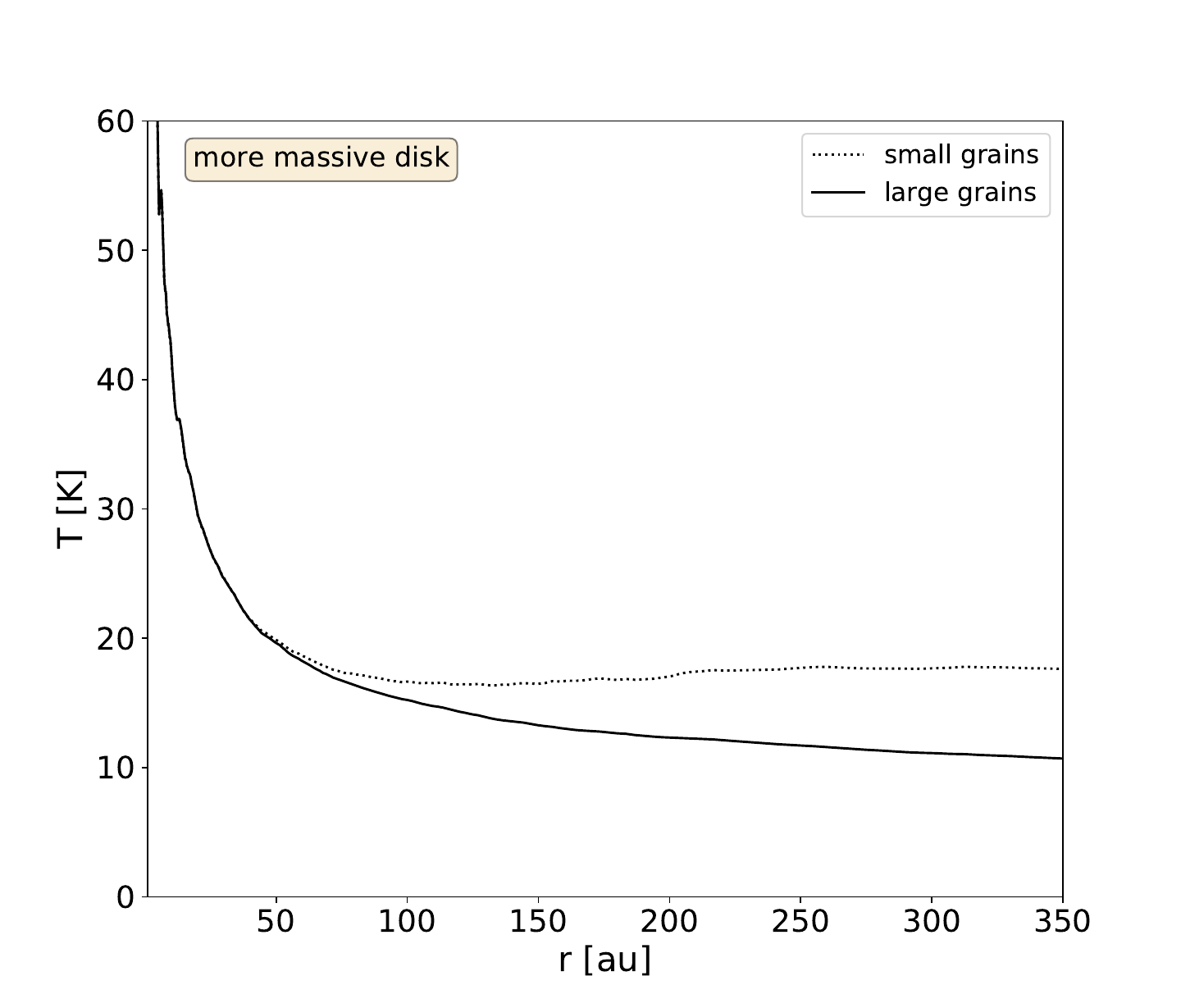}
  \caption{More massive Model D ($M_\mathrm{disk} = 7.5\times10^{-2} M_\odot$)}
\end{subfigure}
\caption{Disk midplane radial profiles of dust temperatures. The black curves are the temperature profiles of the small (dotted) and large (solid) grains. The parameters are that of Model D but with a different mass.}
\label{fig:masstest}
\end{figure*}

Using the same stellar luminosity (L$_\star$ = 1 L$_\odot$), we compare disks of different masses to investigate the impact of optical depth onto the dust thermal decoupling. We use Model D with the DSHARP opacities and change only its mass for two extreme cases: one model is 10 $\times$ less massive than Model D ($M_\mathrm{disk} = 7.5\times10^{-4} M_\odot$), whereas the other is 10 $\times$ more massive ($M_\mathrm{disk} = 7.5\times10^{-2} M_\odot$). The resulting dust temperatures in the midplane are shown in Fig.\,\ref{fig:masstest}.

In the less massive disk, a similar CO ice distribution as in Model D should be expected. The large grain temperature drops below 25 K inside 50 au and below 20 K at radius $\sim$ 100 au. The small grain temperature goes below 25 K at around 250 au and 
reach 20 K around 350 au. In the more massive and thus more optically thick disk model, the thermal decoupling occurs, as expected, at a larger distance than for fiducial Model D (outside 50 au). A bump is present for the small grains further away than in Model D but the temperature never goes above 20 K. In this model, we can expect that CO molecules preferably stick on the small grains almost everywhere in the disk.
However, this conclusion is only valid because we hold the stellar luminosity 
to the same value (L$_\star$ = 1 L$_\odot$), as appropriate for T\,Tauri stars. In practice, the more massive disks are found
around Herbig Ae stars, and those will have higher temperatures because of higher star
luminosity, possibly sufficient to bring back the bump temperature above 20 K.

\subsection{Impact of the turbulence}\label{app:turbulence}

\begin{figure*}
\begin{subfigure}{.5\linewidth}
  \centering
  \includegraphics[width=1.0\linewidth]{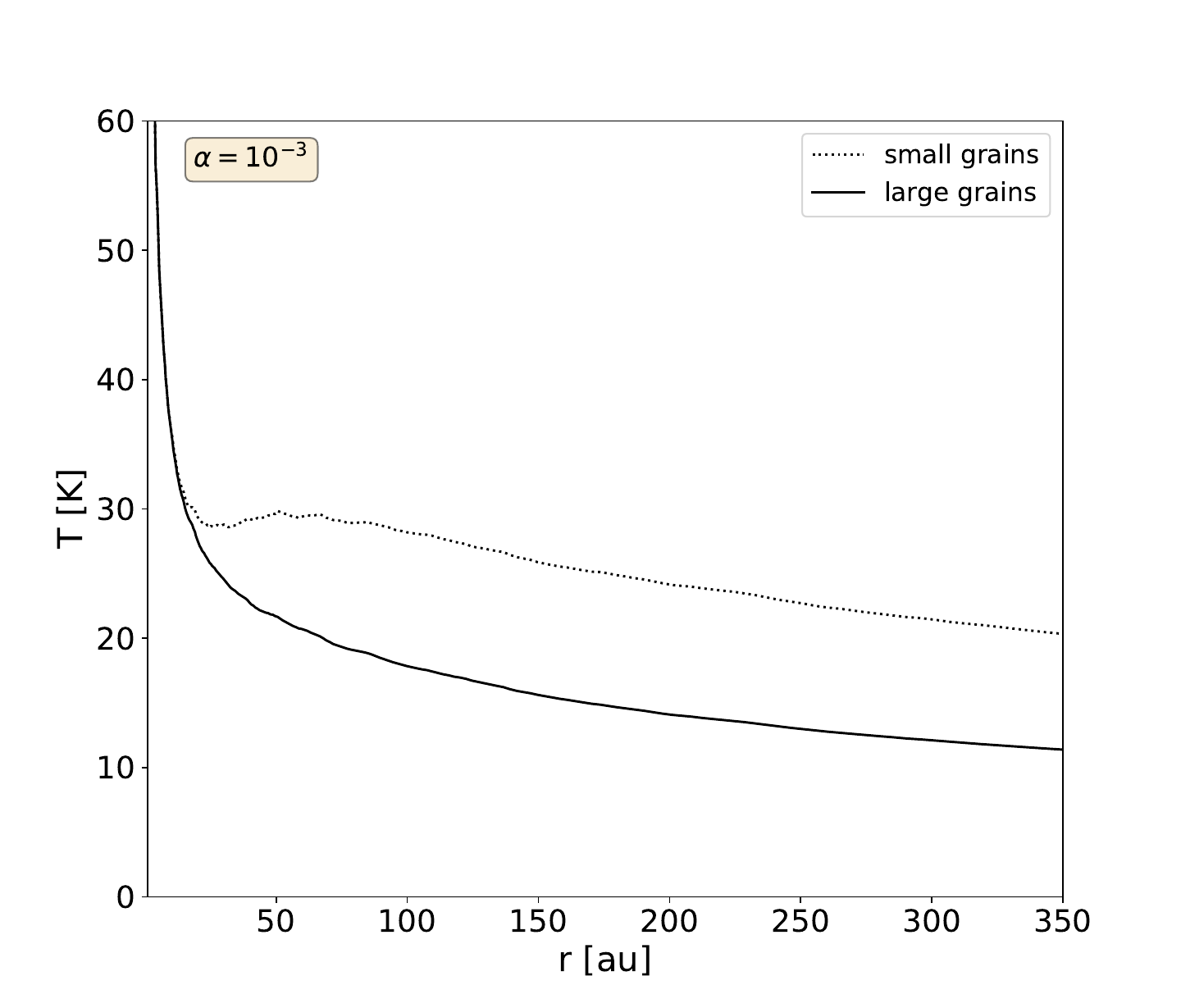}
  \caption{Model D with $\alpha = 10^{-3}$ \label{fig:alphatesta}}
\end{subfigure}
\begin{subfigure}{.5\linewidth}
  \centering
  \includegraphics[width=1.0\linewidth]{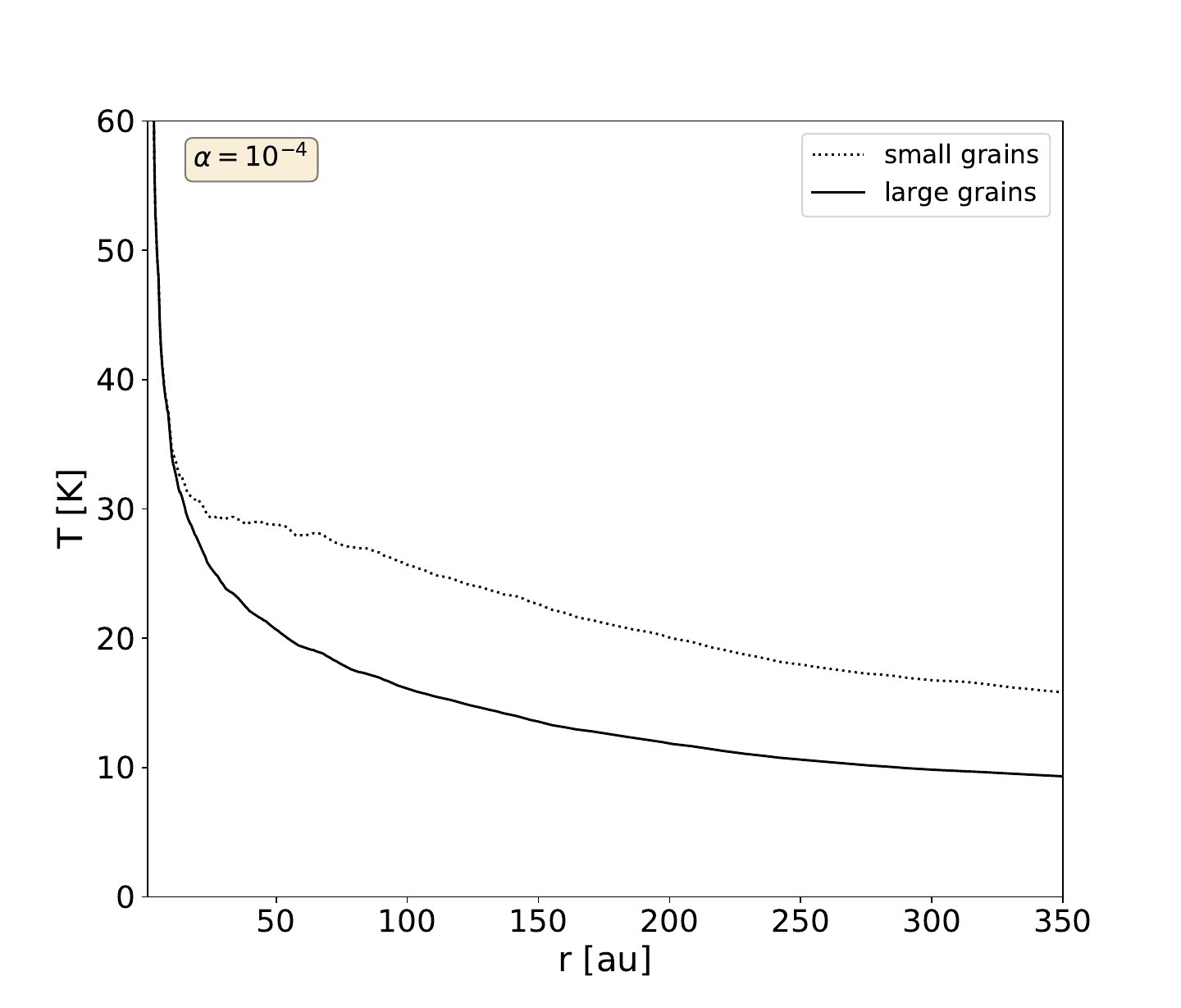}
  \caption{Model D with $\alpha = 10^{-4}$ \label{fig:alphatestb}}
\end{subfigure}
\caption{Disk midplane radial profiles of dust temperatures. The black curves are the temperature profiles of the small (dotted) and large (solid) grains. The parameters are that of Model D but with different $\alpha$ values.}
\label{fig:alphatest}
\end{figure*}

We also test different values of the $\alpha$ parameter (which affects our model only through dust settling) in order to see how varying the ratio of vertical concentration of solid particles to that of the gas may affect the bump shape and dust thermal profile. The fiducial model has an $\alpha$ value of $10^{-2}$. In Fig.\,\ref{fig:alphatest}, we show the resulting dust temperatures in the midplane of Model D (using the DSHARP opacities) with two other $\alpha$ values compatible with observations 
\citep{Cuzzi+etal_1993, Brauer+etal_2008}, the Model D being computed with $\alpha$ values of $10^{-3}$ and $10^{-4}$. In the first case ($\alpha=10^{-3}$), we see that the thermal structure in the midplane is roughly unchanged. For the smaller value of $\alpha$
($10^{-4}$), the bump appears flatter while the temperature spread is smaller, in particular at large radii. Overall, the CO spatial distribution should not vary much from the fiducial Model D because in all cases the small grain temperature remains higher than that of the large ones. 

\section{The bump effect: simulation artefacts or physical effects}\label{app:number}
\begin{figure*}
\begin{subfigure}{.5\linewidth}
  \centering
  \includegraphics[width=1.0\linewidth]{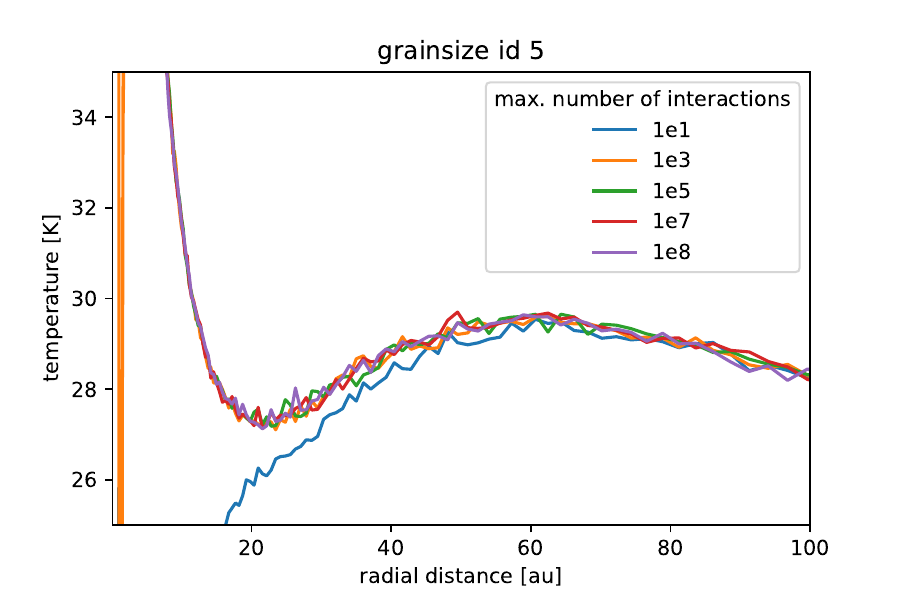}
  \caption{Maximum number of photon package interactions}
\end{subfigure}
\begin{subfigure}{.5\linewidth}
  \centering
  \includegraphics[width=1.0\linewidth]{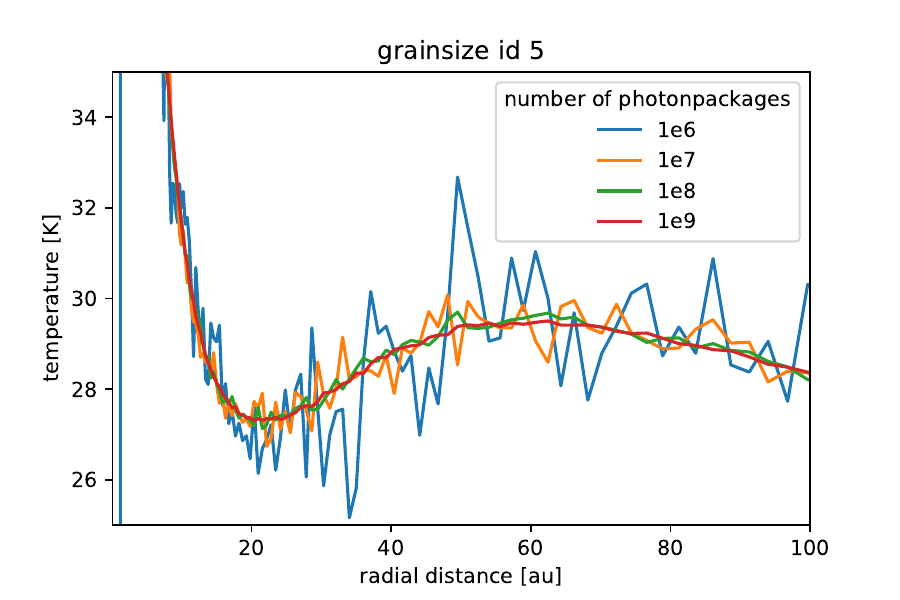}
  \caption{Number of photon packages}
\end{subfigure}
\caption{2D dust density maps in cm${-3}$ of the dust distribution.}
\label{fig:test}
\end{figure*}

One can rightfully argue that a dip in temperature in optically thick regions is very often the result of numerical effects. The bump effect would therefore not be a physical effect and would rather be caused by not using enough photon packages and/or by a too strong limitation of the maximum number of interactions a photon package is allowed to have before being removed during a simulation. In order to make sure the temperature bump is an actual physical effect, we perform additional test simulations where we compare various number of photon packages and various maximum number of interactions. The results are shown in Fig.\,\ref{fig:test}. The left panels shows the temperature profile of the size bin 5 from Model M16 for various maximum number of interactions from 10$^1$ to 10$^8$ interactions. The right panel shows the same grain population for various numbers of photon packages from 10$^6$ to 10$^9$. Both tests strongly suggest that the visible bump is a real physical effect and is not due to thermal simulation effects.

We also note that these simulation results have been cross checked and confirmed using the 3D Monte-Carlo radiative transfer code POLARIS \citep{Reissl+etal_2016}.

\end{appendix}

\end{document}